\documentclass{article}

\usepackage{arxiv}
\usepackage[utf8]{inputenc} % allow utf-8 input
\usepackage[T1]{fontenc}    % use 8-bit T1 fonts  
\usepackage{hyperref}       % hyperlinks
\usepackage{url}            % simple URL typesetting
\usepackage{booktabs}       % professional-quality tables
\usepackage{amsmath,amssymb,amsfonts}%
\usepackage{amsthm}%
\usepackage{mathrsfs}%
\usepackage{nicefrac}       % compact symbols for 1/2, etc.
\usepackage{microtype}      % microtypography
\usepackage{graphicx}
\usepackage[numbers]{natbib}
\usepackage{doi}
\usepackage{bm}

\usepackage[font=footnotesize,labelfont=bf]{caption}
\usepackage{subcaption}
\usepackage{makecell}
\usepackage{colortbl}
\usepackage[capitalise,nameinlink]{cleveref}
\usepackage{pdflscape}
\usepackage{rotating}
\usepackage{floatpag}
\usepackage{array}
\usepackage{ragged2e}
\usepackage{tabularray}
\usepackage{multirow}%
\usepackage{authblk}
\newcommand{\lb}[1]{\left( #1 \right)} % wrap functions in large brackets
\newcommand{\jdist}[2]{\text{#1}\left( #2 \right)} % distribution
\newcommand{\jut}[1]{^{\text{#1}}}
\newcommand{\jeqref}{Eq.~\eqref} %includes Eq. before equation labels

\title{Mapping the prevalence of cancer risk factors at the small area level in Australia}

\author{
\href{https://orcid.org/0000-0002-4666-5900}{\includegraphics[scale=0.06]{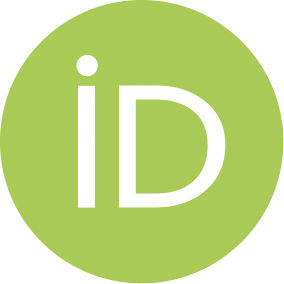}\hspace{1mm} James Hogg$^1$}\thanks{\href{mailto:james.hogg@hdr.qut.edu.au}{james.hogg@hdr.qut.edu.au}} 
\hspace{5mm}% 
\href{https://orcid.org/0000-0002-8161-0358}{\includegraphics[scale=0.06]{plots/orcid.png}\hspace{1mm}Jessica Cameron$^{1,2}$} 
\hspace{5mm}%
\href{https://orcid.org/0000-0001-9041-9531}{\includegraphics[scale=0.06]{plots/orcid.png}\hspace{1mm}Susanna Cramb$^{1,3}$} 
\hspace{5mm}% 
\href{https://orcid.org/0000-0001-8576-8868}{\includegraphics[scale=0.06]{plots/orcid.png}\hspace{1mm}Peter Baade$^{1,2}$} 
\hspace{5mm}% 
\href{https://orcid.org/0000-0001-8625-9168}{\includegraphics[scale=0.06]{plots/orcid.png}\hspace{1mm}Kerrie Mengersen$^{1}$}}
\date{%
    $^1$Centre for Data Science, Queensland University of Technology\\%
    $^2$Viertel Cancer Research Centre, Cancer Council Queensland\\%
    $^3$Australian Centre for Health Services Innovation, School of Public Health and Social Work, Queensland University of Technology\\[2ex]%
}

\renewcommand{\shorttitle}{Mapping cancer risk factors}

\hypersetup{
colorlinks,
allcolors=blue,
pdftitle={Mapping the prevalence of cancer risk factors at the small area level in Australia},
pdfsubject={},
pdfauthor={Hogg, Cameron, Cramb, Baade, Mengersen},
pdfkeywords={Bayesian statistics, small area estimation, disease mapping, cancer prevention},
}

\begin{document}
\maketitle

\begin{abstract}
Cancer is a significant health issue globally and it is well known that cancer risk varies geographically. However in many countries there are no small area level data on cancer risk factors with high resolution and complete reach, which hinders the development of targeted prevention strategies. Using Australia as a case study, the 2017-2018 National Health Survey was used to generate prevalence estimates for 2221 small areas across Australia for eight cancer risk factor measures covering smoking, alcohol, physical activity, diet and weight. Utilising a recently developed Bayesian two-stage small area estimation methodology, the model incorporated survey-only covariates, spatial smoothing and hierarchical modelling techniques, along with a vast array of small area level auxiliary data, including census, remoteness, and socioeconomic data. The models borrowed strength from previously published cancer risk estimates provided by the Social Health Atlases of Australia. Estimates were internally and externally validated. We illustrated that in 2017-18 health behaviours across Australia exhibited more spatial disparities than previously realised by improving the reach and resolution of formerly published cancer risk factors. The derived estimates reveal higher prevalence of unhealthy behaviours in more remote areas, and areas of lower socioeconomic status; a trend that aligns well with previous work. Our study addresses the gaps in small area level cancer risk factor estimates in Australia. The new estimates provide improved spatial resolution and reach and will enable more targeted cancer prevention strategies at the small area level, supporting policy makers, researchers, and the general public in understanding the spatial distribution of cancer risk factors in Australia. To help disseminate the results of this work, they will be made available in the Australian Cancer Atlas 2.0.
\end{abstract}

% keywords can be removed

\newpage
\section{Background}

In 2020, an estimated 19.3 million people were diagnosed with cancer worldwide \cite{sung2021global}, causing a huge health burden. Moreover, incidence of cancer has been shown to exhibit strong spatial disparities, which due to improved models and better data accessibility are now communicated to the public via interactive Atlas platforms. In Australia, a notable Atlas is the Australia Cancer Atlas (ACA) \cite{RN424}, which provides interactive maps of small area level estimates of incidence and relative survival rates for a wide range of cancer types. 

Whiteman \emph{et al.} \cite{RN165} suggest that at least one in every three cancers in Australia can be attributed to modifiable risk factors such as tobacco smoking, obesity, poor diet, insufficient physical activity, excessive sun exposure and alcohol consumption. Understanding the prevalence of cancer risk factors is pivotal to cancer prevention. 

To better assess how cancer risk factors vary by location and target interventions, many countries have generated small area estimates for their prevalence including Australia \cite{RN113}, the US \cite{RN426}, Canada \cite{RN395}, Iran \cite{RN119}, and Luxembourg \cite{RN492}. When generating small area estimates, practitioners must consider the \emph{reach} and \emph{resolution} of their results. Reach refers to the proportion of the small areas for which estimates are available, while resolution pertains to the small area population and geographical sizes. While the need for high resolution relates to minimizing outcome heterogeneity in larger areas and populations, the need for complete coverage (or high reach) ensures policy makers have complete spatial information. If small area estimates suffer from low reach or resolution the effectiveness of targeted interventions could be affected. 

%There is a strong individual level association between these modifiable behaviours and cancer incidence \cite{RN582, RN165, RN486, RN121}. 

In Australia, the Social Health Atlases of Australia (SHAA) \cite{RN113} is the major platform providing nationwide estimates for cancer risk factors at a small area level. The estimates were derived from the 2017-2018 National Health Survey (NHS). However the reach and resolution of the SHAA estimates could be improved. The larger areal units used in the SHAA combine heterogeneous sub-populations, resulting in estimates that are averages over different populations. The limitation regarding reach meant that no estimates are provided for very remote areas. Given that health disparities tend to widen with increasing remoteness \cite{RN581, RN338, RN608}, generating estimates for these areas is important for targeted public health initiatives in Australia. The modelled estimates provided by the SHAA use the best data source available, so the problem cannot be solved by using a different dataset or collecting better data; the solution is to use new methods of small area estimation (SAE) \cite{RN28}. 

SAE is a well-established survey method that leverages auxiliary data, such as census data, to estimate parameters of interest for small geographic areas with limited or no survey data. Model-based SAE methods, which borrow strength across areas \cite{RN37}, are particularly suitable for sparse survey data. These methods can be applied at either the area \cite{RN54} or individual level \cite{RN48}, with the latter requiring access to survey and census microdata.

Proportion area level models are commonly used \cite{RN461, RN408, RN476, RN146}; however, they become unsuitable when some of the input data (area level proportion estimates) are unstable, i.e. exactly zero or one \cite{RN533}. Sparse survey data and modelling rare or common population characteristics exacerbate this instability \cite{self_cite}. Solutions to instability include perturbing direct estimates prior to modelling \cite{RN400} or excluding unstable areas \cite{RN408}. Alternatively, modelling at the individual level, such as through multilevel regression and poststratification (MrP) \cite{RN403}, can be pursued. However, the use of individual level SAE models to derive proportion estimates is limited by the need for census microdata \cite{RN467}, which restricts the choice of covariates. Note that the modelling for the SHAA was conducted by the Australian Bureau of Statistics (ABS). Unfortunately, given that the published details of the ABS approach are modest \cite{RN27}, we can only infer the use of a individual level model. 

While individual and area level models have limitations, recent work supports the utility of two-stage SAE approaches, which involve separate modelling at both levels \cite{self_cite, gao2023_sma, Das2022, RN376}. Two-stage approaches have many benefits that are particularly relevant for this application as they can alleviate unstable direct estimates by smoothing individual level outcomes, accommodate even severely sparse survey data thanks to multi-stage smoothing, and utilize more auxiliary data (e.g. survey-only covariates), permitting more flexible models and better predictions. 

In this work, we generate small area level prevalence estimates for eight cancer risk factor measures using the Bayesian two-stage small area estimation methodology we developed for sparse survey data \cite{self_cite}. Our method considers a variety of data sources, including individual level survey data and area level auxiliary data such as census, remoteness and socioeconomic data. To assess the quality of our estimates, we used a dual validation approach whereby most SA2s are benchmarked to the sub-state level using fully Bayesian benchmarking \cite{RN30}, with the remaining SA2s (predominantly very remote areas) undergoing external validation. The results of this work will complement the current small area level estimates of cancer incidence and relative survival already available in the ACA \cite{RN424}.

%% Methods %%
\section{Data} \label{sec:data}

\subsection{Geographical areas}
Geographical location was defined according to the 2016 Australia Statistical Geography Standard (ASGS) \cite{RN348}. We generated prevalence estimates at the statistical area level 2 (SA2) level, which is the lowest level of the ASGS hierarchy for which detailed census population characteristics are publicly available. SA2s are recognized as achieving the optimal balance between privacy and resolution \cite{RN26}. Note that the SHAA provides estimates at a lower resolution, using population health areas (PHAs) which are constructed from single or multiple SA2s (40\% and 39\% of PHAs are constructed from one or two SA2s, respectively). In 2016 Australia had 1165 PHAs and 2310 SA2s \cite{RN133} with median population sizes of 7500 and 15000 for SA2s and PHAs, respectively.

Throughout this analysis, we also used statistical area level 3 (SA3) and statistical area level 4 (SA4). By virtue of the hierarchical nature of the ASGS, SA2s are nested within SA3s, and SA3s are nested within SA4s. There is a median of 6 and 22 SA2s nested within each SA3 (n = 333) and SA4 (n = 88), respectively.

Of the 2310 SA2s covering Australia, SA2s with no physical location (comprising ``Migratory–Offshore–Shipping'' and ``No usual address'' codes for each State and Territory) (n = 18), very remote island SA2s (Christmas Island, Cocos Island, Norfolk Island and Lord Howe Island) (n = 4), and SA2s with annual average population $\leq$ 10 (n = 67) were excluded. This left 2221 SA2s to use in the modelling. The remaining SA2s had a median (interquartile range (IQR)) population of 7859 (4483, 12753). Note that although Jervis Bay is classified as an ``Other Territory'' by the ABS, we included it as part of the state New South Wales. 

\subsection{Data sources}
\subsubsection{Survey data}
The individual level survey data and sampling weights were obtained from the 2017-18 National Health Survey (NHS), which is an Australia-wide population-level health survey conducted every 3-4 years by the ABS \cite{RN478, RN598}. This survey excluded very remote areas of Australia ($\approx 0.8$\% of 2016 population), discrete Aboriginal and Torres Strait Islander communities ($\approx 0.5$\% of 2006 population as per the ABS Community Housing and Infrastructure Needs Survey conducted only in 1999, 2001, and 2006 \cite{RN606}), and non-private dwellings ($\approx 2$\% of 2016 population \cite{RN605}). Non-private dwellings include hotels and motels, hostels, boarding schools and boarding houses, hospitals, nursing and convalescent homes, prisons, reformatories and single quarters of military establishments and short-stay caravan parks. The ABS highlights that these exclusions should only have a minor effect on aggregate estimates for the states and territories of Australia.

\begin{figure}
    \centering
    \includegraphics[width=\textwidth]{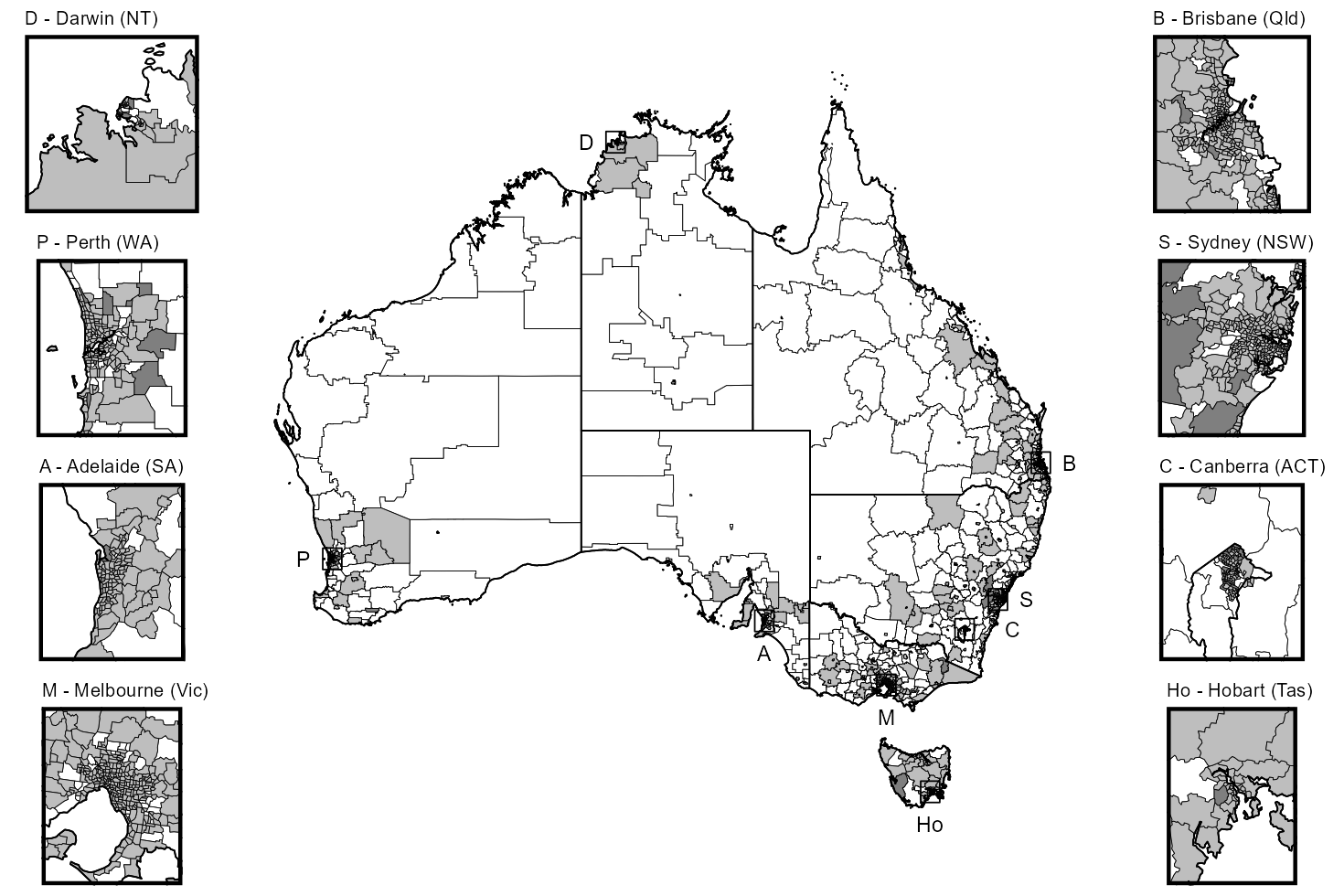}
    \caption{\small Map of 2221 SA2s in Australia with gray indicating an area with data from the 2017-2018 National Health Survey.}
    \label{fig:sampled}
\end{figure}

The 2017-18 NHS data consist of 17248 sampled persons 15 years and older, with 878 persons under the age of 18. The data cover 1694 (76\%) of the 2221 SA2s across Australia (see \cref{fig:sampled}) and provide a median (IQR) SA2 level sample size of 8 (5, 13). The median SA3 and SA4 level sample sizes were 42 (25, 65) and 154 (101, 226), respectively. The NHS was also used to obtain daily smoking rates at the SA4 level \cite{RN607}. Other sources of Australian health data are described in Section A of the Additional File.  

\subsubsection{Population data}
Estimated Resident Population data stratified by 5-year age groups (15 years and above), sex and SA2, were obtained from the ABS for both 2017 and 2018 \cite{RN650}. In this study the SA2 level population counts were derived by averaging across the two years. One of the risk factors (risky waist circumference) is only appropriate for ages 18+ and so modelling excluded persons under 18. Assuming that the single-year age distribution in this age group was uniform, we estimated that the population of 18-19-year olds was 40\% of the 15-19-year old population. 

For the SA2 level auxiliary data, we used data from the 2016 Australian census, represented as proportions (for categorical data) or averages (for continuous data) of individuals in each SA2. Census data for age, sex, non-school level education (higher education),  highschool completion status, occupation, labour force status, personal weekly income, religious affiliation, registered marital status, First Nations Australian status, and household composition were obtained from the ABS \cite{RN648}. These census factors made up 84 separate variables. Like Chidumwa \emph{et al.} \cite{RN618}, to reduce the dimension of these socioeconomic and demographic data we used Principal Components (PC) Analysis, where we retained the first six principal components as they accounted for approximately 62\% of the variation (see Section C.2 of the Additional File for more details). 

\subsubsection{Other data}
Australian research suggests that cancer burden \cite{RN121, RN26}, and the prevalence of cancer risk factors varies strongly by remoteness and socioeconomic status (SES) \cite{RN204, RN581}. Data on SA1 level remoteness were provided by ABS and based on the Accessibility and Remoteness Index of Australia (ARIA+) \cite{RN604}, and converted to SA2 using population proportions. Remoteness is divided into five groups - major cities, inner regional, outer regional, remote, and very remote - based on a measure of relative geographic access to services. Given that very remote areas of Australia were intentionally excluded during data collection for the 2017-18 NHS, we followed the approach of Das \emph{et al.} \cite{RN607}, and collapsed the outer regional, remote and very remote categories to a single remoteness group. Of the SA2s with sample data, 69\% were major cities. The SA2 sample sizes tended to be larger for outer regional to very remote areas (median of 11 and IQR of 6 to 21) than for major city areas (median of 8 and IQR of 4 to 12). 

SA2 level SES was sourced from the ABS Socio-Economic Indexes for Areas product \cite{RN560}. Like other Australian health studies \cite{RN204, RN582, RN607}, we used the Index of Relative Socio-Economic Disadvantage (IRSD). The IRSD is a general SES index constructed using principal components analysis that summarises the economic and social conditions of individuals and households within a given area in order to determine the area's overall relative disadvantage. A low IRSD score indicates a large proportion of relatively disadvantaged individuals in a given SA2 \cite{RN582}.  

In this work, we used IRSD national deciles as a categorical variable with 10 groups, where 1 represents the most disadvantaged or lowest SES group and was used as the reference group. Although the IRSD can be used as a continuous variable, it is recommended to use deciles \cite{RN560}, and this also gave superior model performance. There were  44 of the 2221 SA2s without an IRSD value provided, so these had the closest IRSD decile assigned according to their corresponding PC1 (principal component 1) values.

We also obtained prevalence estimates and measures of uncertainty for risky alcohol consumption (more than 2 standard drinks a day on average), adequate fruit intake, obesity, overweight, current smokers and inadequate physical activity from the SHAA \cite{RN113} at the Primary Health Network (PHN) and PHA level for adults. These data were downloaded as age-standardised rates per 100 people with 95\% confidence intervals. Definitions and details are available in Section C and D of the Additional File and the online SHAA platform \cite{RN113}. 

\subsection{Risk factors} \label{sec:rf}
Broad risk factor groups were selected by consulting three sources: a wide range of experts in the fields of public health, epidemiology and oncology; literature, specifically evidence for casual associations \cite{RN123} with, and population attributable fractions \cite{RN121, RN485, RN165} for, cancer incidence; and the availability of data in the 2017-18 NHS. In this work we selected the following five broad risk factor groups: tobacco smoking, alcohol, diet, weight and physical activity. According to the 2015 Australian Burden of Disease study \cite{RN121} these were attributable to 22.1\%, 4.5\%, 4.2\%, 7.8\% and 2.9\% of the total cancer burden, respectively. 

We explored a variety of possible measures and corresponding definitions for each of the five broad risk factor groups, placing priority on the definitions and recommendations used in the SHAA \cite{RN113}, the work by Whiteman \emph{et al.} \cite{RN165}, Cancer Council Australia \cite{RN600} and those provided by Australian government agencies such as Cancer Australia \cite{RN582}, the National Health and Medical Research Council (NHMRC), the Australian Institute of Health and Welfare (AIHW) \cite{RN513} and the Australian Department of Health and Aged Care (DOH).

\cref{table:aca_rf_defin} summarizes the five broad risk factor groups and the eight corresponding measures and definitions. \cref{table:national_direct} gives direct estimates for these measures stratified by the eight states and territories of Australia. The risk factor measures proposed are designed to be cross-sectional and strike a natural balance between being specific to cancer while maintaining applicability to a variety of other health conditions \cite{RN121}. Note that some risk factor groups, for example weight, required several differing measures. 

% Definitions for ACA risk factor measures
\begin{table}
\caption{\small Descriptions and definitions of the five cancer risk factor groups and the measures within each. More details are given in Section B of the Additional File.}
\label{table:aca_rf_defin}
    \centering
    \begin{tabular}{lll} 
    \noalign{\hrule height 1.5pt}
    Group & Measure & Measure definition \\ 
    \noalign{\hrule height 1.5pt}
    Smoking & \makecell[l]{Current\\smoking} & \makecell[l]{Those who reported to be current smokers (including daily, weekly\\or less than weekly), and had smoked at least 100 cigarettes in\\their life.} \\
     \hline\hline
    Alcohol & \makecell[l]{Risky alcohol\\consumption} & \makecell[l]{Those who exceeded the revised 2020 National Health and \\Medical Research Council (NHMRC) guidelines \cite{RN516} of up\\to 10 standard drinks/week and no more than 4 standard drinks\\in any day. Compliance with the guidelines were assessed\\using self-reported alcohol consumption during the last three\\drinking days from the proceeding seven days. } \\
     \hline\hline
    Diet & \makecell[l]{Inadequate\\diet} & \makecell[l]{Based on self-reported diet, those who did not\\meet both the fruit (2 serves/day) and\\vegetable (5 serves/day) 2013 NHMRC Australian \\Dietary guidelines \cite{RN520}.} \\ 
     \hline\hline
    \multirow{3}{*}{Weight} & Obese & Those with a measured BMI greater or equal to 30. \\
    \cline{2-3}
     & \makecell[l]{Overweight/\\obese} & Those with a measured BMI greater or equal to 25. \\
     \cline{2-3}
     & \makecell[l]{Risky waist\\circumference} & \makecell[l]{Those with a measured waist circumference measurements of\\$\geq$94cm (men) and $\geq$80cm (women). These cutoffs are only\\appropriate for adults, so we limited the survey dataset to\\all persons 18 years and older \cite{RN113}.} \\
    \hline\hline
    \multirow{2}{*}{\makecell[l]{Physical\\activity}} & \makecell[l]{Inadequate\\activity\\(leisure)} & \makecell[l]{Based on self-reported leisure physical activity, those\\who did not meet the 2014 Department of Health (DOH)\\Physical Activity guidelines \cite{RN517}, i.e. that each week adults\\(those between the ages of 18 and 64)\\should either do 2 $1/2$ to 5 hours of moderate-intensity\\physical activity or 1 $1/4$ to 2 $1/2$ hours of vigorous-intensity\\physical activity or an equivalent combination of both, plus\\muscle-strengthening activities at least 2 days each week.\\The DOH guidelines also provide specific recommendations\\for children (5 to 17 years), older persons (65 years and older)\\and pregnant women. In this work, the physical activity measures\\were derived from the ABS created variables that\\accommodated the guidelines\\across age groups.} \\
    \cline{2-3}
     & \makecell[l]{Inadequate\\activity (all)} & \makecell[l]{Although similar to inadequate activity (leisure), this measures is\\based on workplace and leisure self-reported physical activity.} \\
    \noalign{\hrule height 1.5pt}
    \end{tabular}
\end{table}

% National and state direct prevalence estimates
\begin{table}
\caption{\small Direct prevalence estimates for Australia and the eight states and territories for all eight cancer risk factor measures. Each cell of the table gives the direct prevalence estimate and corresponding 95\% confidence interval as percentages.}
\label{table:national_direct}
\centering
    \begin{tabular}{rcccc}
    \hline\hline
     & \makecell[c]{Current \\smoking} & \makecell[c]{Risky\\alcohol\\consumption} & \makecell[c]{Inadequate\\diet} & Obese \\
     \hline
    Australia & 15.1 (14.5, 15.7) & 28.5 (27.8, 29.3) & 46.9 (46.1, 47.7) & 30.6 (29.9, 31.4) \\
    \makecell[r]{New South\\Wales} & 14.4 (13.1, 15.7) & 27.2 (25.5, 28.8) & 45.4 (43.5, 47.2) & 30.7 (29.0, 32.4) \\
    Victoria & 15.1 (13.6, 16.6) & 26.0 (24.2, 27.8) & 45.3 (43.3, 47.4) & 30.9 (29.0, 32.8) \\
    Queensland & 15.8 (14.6, 17.1) & 28.8 (27.2, 30.4) & 47.0 (45.1, 48.8) & 32.0 (30.4, 33.7) \\
    South Australia & 14.1 (12.3, 16.0) & 28.0 (25.6, 30.3) & 49.3 (46.8, 51.9) & 31.6 (29.2, 33.9) \\
    \makecell[r]{Western\\Australia} & 13.7 (11.9, 15.4) & 31.3 (28.9, 33.7) & 46.0 (43.4, 48.6) & 28.0 (25.7, 30.3) \\
    Tasmania & 16.8 (14.9, 18.7) & 29.2 (26.9, 31.6) & 47.7 (45.1, 50.4) & 33.7 (31.3, 36.2) \\
    \makecell[r]{Northern\\Territory} & 20.5 (18.0, 23.0) & 33.4 (30.4, 36.4) & 48.4 (45.2, 51.5) & 28.8 (26.0, 31.6) \\
    \makecell[r]{Australian\\Capital Territory} & 11.2 (9.2, 13.2) & 28.2 (25.3, 31.1) & 49.6 (46.4, 52.7) & 25.6 (22.9, 28.2) \\
    \hline
     & \makecell[c]{Overweight/\\obese} & \makecell[c]{Risky\\waist\\circumference} & \makecell[c]{Inadequate\\activity\\(leisure)} & \makecell[c]{Inadequate\\activity\\(all)} \\
     \hline
    Australia & 65.7 (64.9, 66.4) & 63.6 (62.8, 64.4) & 85.2 (84.7, 85.8) & 83.5 (82.9, 84.1) \\
    \makecell[r]{New South\\Wales} & 64.9 (63.1, 66.6) & 62.8 (61.0, 64.7) & 84.8 (83.4, 86.1) & 82.9 (81.5, 84.4) \\
    Victoria & 66.5 (64.6, 68.5) & 64.1 (62.0, 66.1) & 85.3 (83.8, 86.7) & 83.4 (81.8, 84.9) \\
    Queensland & 65.0 (63.3, 66.7) & 63.9 (62.1, 65.7) & 87.0 (85.8, 88.3) & 85.3 (84.1, 86.6) \\
    South Australia & 68.6 (66.1, 71.0) & 66.0 (63.4, 68.6) & 85.7 (83.8, 87.5) & 84.3 (82.4, 86.2) \\
    \makecell[r]{Western\\Australia} & 65.1 (62.6, 67.6) & 62.0 (59.4, 64.7) & 84.1 (82.2, 86.0) & 82.2 (80.2, 84.2) \\
    Tasmania & 68.8 (66.4, 71.3) & 68.5 (65.9, 71.0) & 84.6 (82.6, 86.5) & 83.1 (81.1, 85.1) \\
    \makecell[r]{Northern\\Territory} & 63.1 (60.1, 66.2) & 58.2 (55.0, 61.5) & 86.2 (84.1, 88.4) & 84.8 (82.5, 87.1) \\
    \makecell[r]{Australian\\Capital Territory} & 62.5 (59.5, 65.6) & 60.6 (57.4, 63.8) & 82.1 (79.7, 84.6) & 79.9 (77.4, 82.5) \\
    \hline\hline
    \end{tabular}
\end{table}

We defined the risk factor measures as binary where a survey individual received a value of one if they did not meet guidelines, or were in the unhealthy category. Unlike the SHAA which provides age-standardised rates by PHAs \cite{RN113}, we used proportions (prevalence) due to their common use in both the literature \cite{RN28, RN625, RN461} and other digital Atlases \cite{RN426}. Furthermore, deriving age-standardised rates requires prevalence estimates by area \emph{and} age. This level of disaggregation is possible at the PHA level, but not feasible at the SA2 level. 

We provide further details and the motivation for the selected risk factor measure definitions in Section B of the Additional File. 

%% Statistical models %%
\section{Statistical models}
\subsection{Bayesian model}
Given the sparse nature of the available data for this SAE analysis, we used the Bayesian two-stage logistic normal (TSLN) model we proposed recently \cite{self_cite}. Our previous study shows that the TSLN model can outperform commonly used area \cite{RN408, RN476, RN552} and individual level \cite{RN44} models both in a simulation study focusing on sparse survey data and an application using the 2017-18 NHS data. The two-stage structure of the TSLN model includes an individual level stage 1 model, followed by an area level stage 2 model. 

The same TSLN model, with very similar components, was chosen to be applied to all eight risk factor measures. The selection of fixed and random effect structures for the two models was guided by the goal of achieving a balance between parsimony across risk factor measures and predictive performance. We followed the advice by Goldstein \cite{RN553} and initially used frequentist algorithms to select fixed and random effects, with fully Bayesian inference via Markov chain Monte Carlo (MCMC) for final model checking. Further details regarding model selection are given in Section E of the Additional File. 

Let $y_{ij} \in \{0,1\}$ be the binary value from the NHS for sampled individual $j = 1, \dots, n_i$ in SA2 $i = 1, \dots, m$, where $n_i$ is the sample size in SA2 $i$. Further, let $m = 1694$ and $M = 2221$ be the number of sampled and total number of SA2s, respectively. The goal of this analysis is to generate estimates of the true proportions of each risk factor measure, $\boldsymbol{\mu} = \lb{\mu_1, \dots, \mu_M }$. 

In this analysis, we used two versions of the survey weights, $w\jut{raw}_{ij}$, provided by the ABS \cite{RN508,RN44} to correct for sampling bias and promote design-consistency. The first, $w_{ij}$, was used for direct estimation and the second, $\tilde{w}_{ij}$, was used in the stage 1 model (see Section C.1 in the Additional File). Using the survey weights, small area proportion estimates can be computed using the Hajek \cite{RN571} direct estimator,

\begin{equation}
    \hat{\mu}^D_i = \frac{\sum_{j=1}^{n_i} w_{ij} y_{ij} }{n_i}, \label{eq:ht}
\end{equation}

\noindent with an approximate sampling variance of \cite{RN147, RN552},

\begin{eqnarray}
        \psi_i^D = \widehat{\text{v}} \lb{ \hat{\mu}_i^D }  = \frac{1}{n_i} \left( 1 - \frac{n_i}{N_i} \right) \left( \frac{1}{n_i - 1} \right) \nonumber
        \\
        \sum_{j=1}^{n_i} \left( w_{ij}^2 \left( y_{ij} - \hat{\mu}_i^D  \right)^2 \right). \label{eq:ht_var}
\end{eqnarray}

Direct estimators, such as \jeqref{eq:ht} and \eqref{eq:ht_var}, have low variance and are design-unbiased for $\mu_i$ when $n_i$ is large, but have high variance when $n_i$ is small \cite{RN37}. 

\subsubsection{Stage 1: Individual level model}

\begin{figure}
    \centering
    \includegraphics[width=0.8\textwidth]{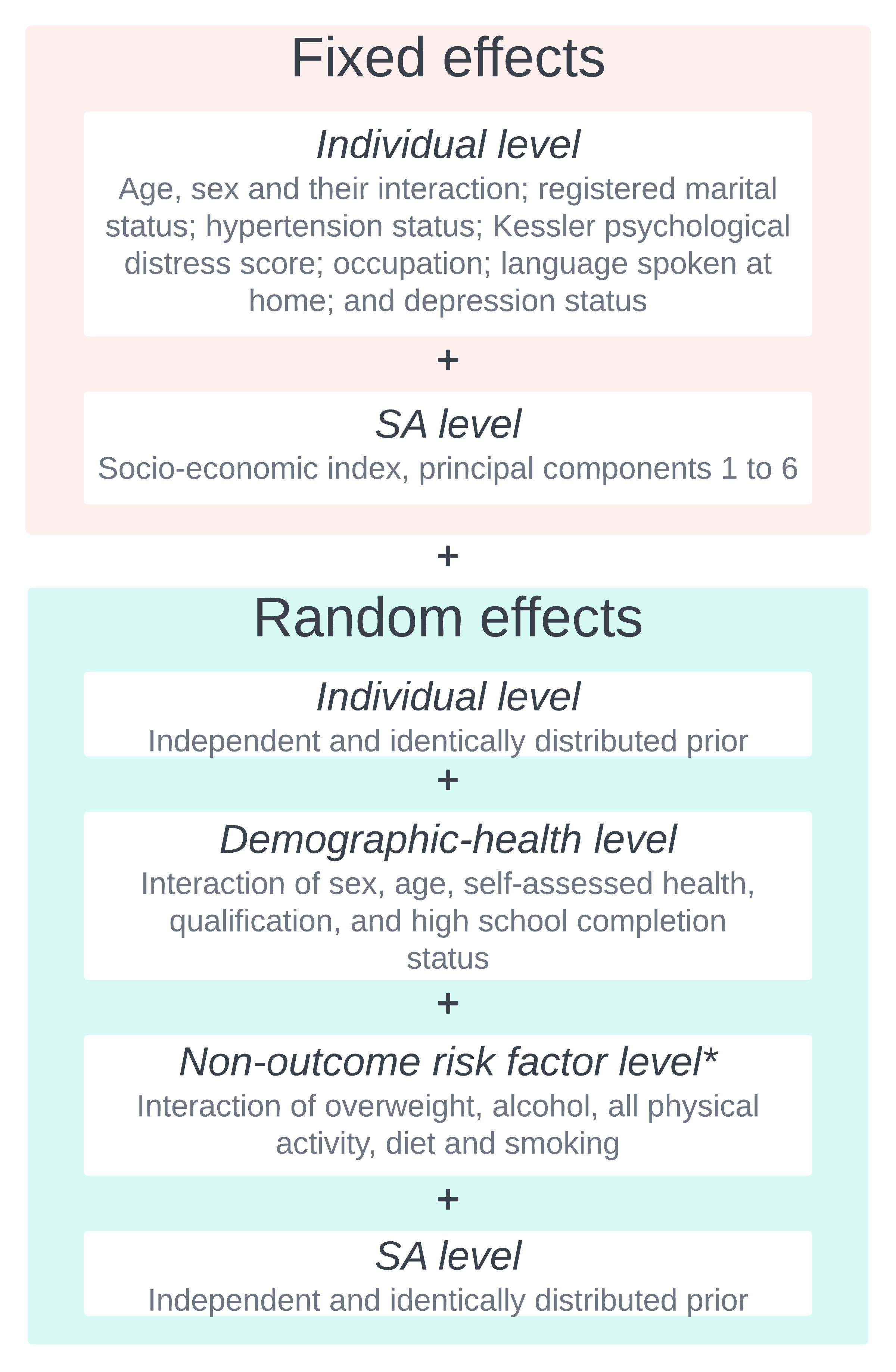}
    \caption{\small Schematic describing the components of the linear predictor for $\jdist{logit}{\pi_{ij}}$ in the stage 1 model. *The non-outcome risk factor categorical covariate was derived from the interaction of the binary risk factor outcomes not directly associated with the risk factor being modelled. For more details see Section C of the Additional File. IID: Independent and identically distributed; IRSD: Index of Relative Socio-Economic Disadvantage; PC: Principal component; SA2: Statistical area level 2.}
    \label{fig:s1mb}
\end{figure}

The stage 1 model is a Bayesian pseudo-likelihood logistic mixed model. Let $\pi_{ij}$ be the probability of $y_{ij} = 1$ for sampled individual $j$ in SA2 $i$. Following the notation of Parker \emph{et al.} \cite{RN44}, we represent the pseudo-likelihood for a probability density, $p(.)$, as $p\lb{y_{ij}}^{\tilde{w}_{ij}}$. Pseudo-likelihood is used to ensure the predictions from the logistic model are approximately unbiased under the sample design \cite{RN500, RN501}. Thus, the stage 1 model likelihood is given by,

\begin{equation}
    y_{ij} \sim \jdist{Bernoulli}{\pi_{ij}}^{\tilde{w}_{ij}}, \label{eq:model_s1}
\end{equation}

\noindent where $\jdist{logit}{\pi_{ij}}$ is modelled using a generic linear predictor that is application-specific. In this work, we used several unique components summarised in \cref{fig:s1mb}.
The linear predictor included eight individual level categorical covariates and seven area level covariates as fixed effects. Unstructured individual and SA2 level random effects were also applied. In addition to these, borrowing ideas from MrP \cite{RN439}, we included two hierarchical random effects based on categorical covariates that were themselves derived from the interaction of numerous individual level demographic and health covariates. A discussion of the priors used is given in \cref{sec:priors}. More details can be found in Section C of the Additional File.  

\subsubsection{Stage 2: Area level model}

After fitting the stage 1 model, the individual level predictions are aggregated to the area level, producing stage 1 (S1) proportion estimates $\hat{\mu}^{\text{S1}, (t)}_i$m using \jeqref{eq:ht}, and sampling variances, $\psi^{\text{S1}, (t)}_i = \widehat{\text{v}} \lb{ \hat{\mu}_i^D } + \widehat{\text{v}} \lb{ \hat{B}^{(t)}_i }$, for all posterior MCMC draws, $t=1, \dots, T$ \cite{self_cite}, where the function to compute the sampling variance, $\widehat{\text{v}}(.)$, is given in \jeqref{eq:ht_var} and $\hat{B}^{(t)}_i = n_i^{-1} \lb{ \sum_{j=1}^{n_i} w_{ij} \lb{ \pi^{(t)}_{ij} - y_{ij} } } $ quantifies the level of smoothing achieved by using $\pi_{ij}$ instead of $y_{ij}$. 

Using the common logistic transformation \cite{RN552, RN476}, let 

\begin{eqnarray}
    \hat{\theta}_i^{\text{S1}, (t)} & = & \jdist{logit}{ \hat{\mu}_i^{\text{S1}, (t)} } \label{eq:el_theta}
    \\
    \tau_i^{\text{S1}, (t)} & = & \psi_i^{\text{S1}, (t)} \left[ \hat{\mu}_i^{\text{S1}, (t)} \left( 1 - \hat{\mu}_i^{\text{S1}, (t)} \right) \right]^{-2}, \label{eq:el_tau}
\end{eqnarray}

\noindent thereby permitting the use of a Gaussian likelihood in the second stage model. Let $\bar{\tau}_i\jut{S1}$ be the empirical posterior mean of $\tau_i^{\text{S1}}$ and $\widehat{\text{v}} \lb{ \hat{\theta}_i\jut{S1} }$ be the empirical posterior variance of $\hat{\theta}_i\jut{S1}$. Finally, by selecting a random subset of the posterior draws, say $\widetilde{T}$, let $\hat{\boldsymbol{\theta}}\jut{S1}_i = \lb{ \hat{\theta}^{\text{S1}, (1)}_i, \dots, \hat{\theta}^{\text{S1}, (\widetilde{T})}_i }$.   

The stage 2 model is a Bayesian spatial Fay-Herriot \cite{RN54} model. Unlike previous two-stage approaches \cite{gao2023_sma, Das2022}, we accommodate some of the uncertainty inherent in fitting the stage 1 model by using the vector $\hat{\boldsymbol{\theta}}\jut{S1}_i$ as input to the stage 2 model. The stage 2 model likelihood for the posterior draws from the stage 1 model is,

\begin{equation}
    \hat{\boldsymbol{\theta}}_i\jut{S1} \sim \jdist{N}{ \theta_i , \bar{\tau}_i\jut{S1} + \widehat{\text{v}} \lb{ \hat{\theta}_i\jut{S1} } }
\end{equation}

\noindent where $\theta_i$ is modelled using a generic linear predictor that is problem specific. The final proportion/prevalence estimate for the $i$th SA2, denoted $\mu_i$, is given by the posterior distribution of $\text{logit}^{-1} \lb{\theta_i}$. To ensure that posterior uncertainty remains unaffected by the choice of $\widetilde{T}$, we downscale the likelihood contribution by $1/\widetilde{T}$.  

\begin{figure}
    \centering
    \includegraphics[width=0.8\textwidth]{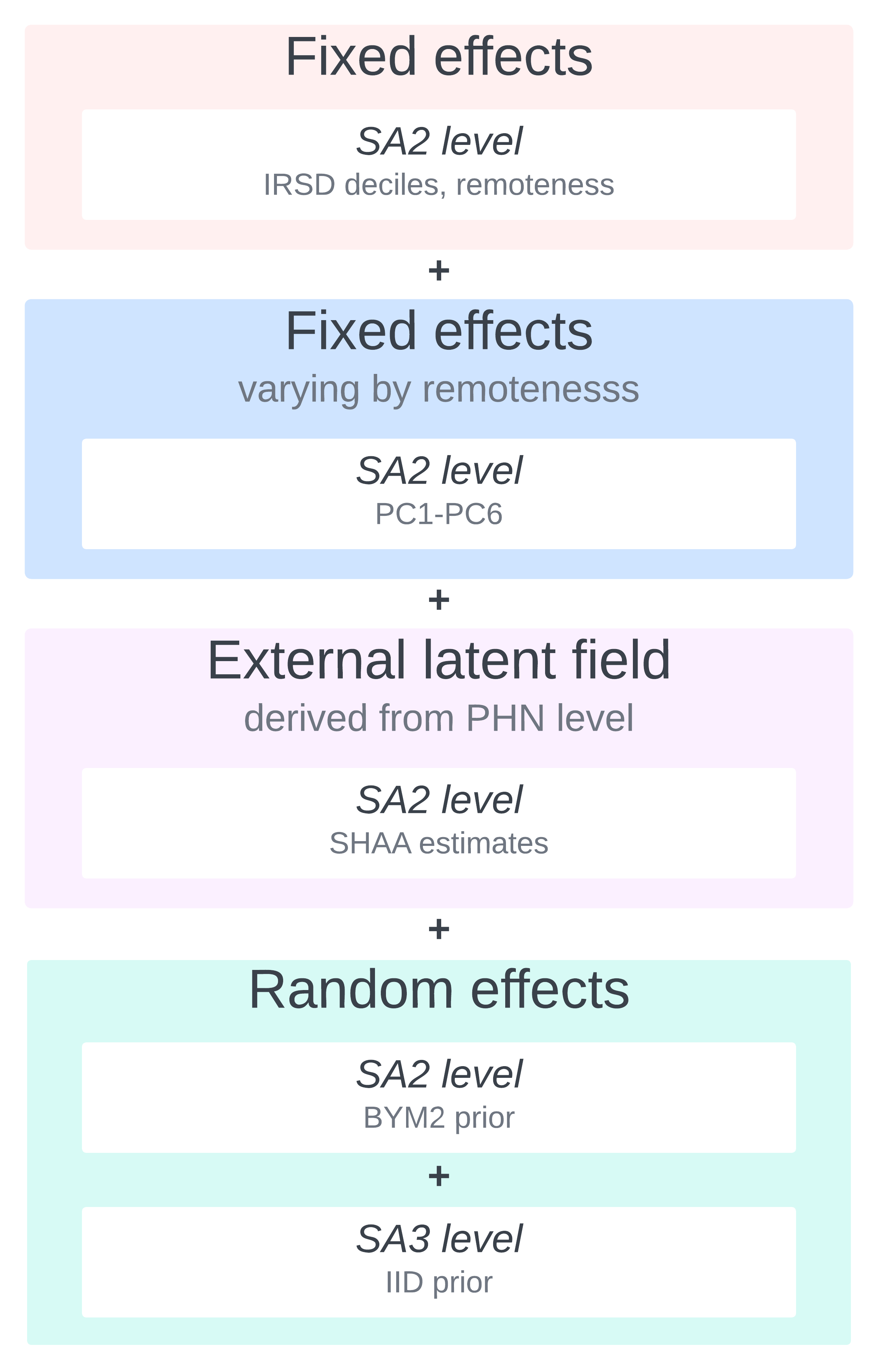}
    \caption{\small Schematic describing the components of the linear predictor for $\theta_i$ in the stage 2 model. For more details see Section C of the Additional File. IID: Independent and identically distributed; IRSD: Index of Relative Socio-Economic Disadvantage; PC: Principal component; PHN: Primary health network; SA2: Statistical area level 2; SA3: Statistical area level 3; SHAA: Social Health Atlases of Australia}
    \label{fig:s2mb}
\end{figure}

In this work, we used several unique components for the linear predictor of $\theta_i$ which are summarised in \cref{fig:s2mb}. The linear predictor included the SES index deciles and remoteness as standard fixed effects. In addition, PC1 to PC6 were used as fixed effects with coefficients varying according to remoteness. The linear predictor also included an external latent field constructed from the SHAA's estimates and a BYM2 spatial random effect \cite{RN394} at the SA2 level. Given we did not include SA3 level census covariates, an unstructured random effect at the SA3 level was employed. To smooth unstable variances we used the generalized variance function \cite{RN430, RN28, RN137} described in Section C.4.6 of the Additional File. More details can be found in Section C of the Additional File.

\subsection{Priors} \label{sec:priors}

The Bayesian models described above are completed by the specification of priors. Given the complexity of the two models, in this work generic weakly informative priors were adopted based on preliminary analysis of the data \cite{stan2023priorchoice}. In both models, all fixed effect coefficients were given $\jdist{N}{0, 2^2}$ priors with intercepts given a student-$t\lb{0, 2^2, \text{df} = 3}$. We used $\text{N}^{+}\lb{0, 1^2}$ and $\text{N}^{+}\lb{0, 2^2}$ priors for all standard deviation terms in the stage 1 and stage 2 models, respectively. The mixing parameter in the BYM2 \cite{RN394} random effect was given a $\jdist{Uniform}{0,1}$ prior (see Section C of the Additional File). 

We conducted sensitivity analysis by using more, $\jdist{N}{0, 1^2}$, and less, $\jdist{N}{0, 100^2}$, informative priors for the fixed effects in both models. We also experimented with using exponential priors with rates of 0.5 and 1 for standard deviation terms. Finally, we examined model fit when using an informative Beta prior for the mixing parameter. We found that the model fit and prevalence estimates were unaffected by these prior changes. The chosen priors gave superior sampling efficiency and convergence.

\subsection{Validation} \label{sec:validate}

For validation of the small area estimates, we adopted a dual approach, using both internal and external methods. See Section C.5 in the Additional File for details.   

\subsubsection{Internal benchmarking}

Internal validation involved a fully Bayesian benchmarking procedure \cite{RN30} that adjusts the results obtained in the stage 2 model by penalizing discrepancies between modelled and direct estimates. Unlike previous benchmarking approaches that adjust the point estimates only \cite{RN37, RN15}, Bayesian benchmarking adjusts the entire posterior --- automatically accounting for benchmarking-induced uncertainty. 

In this work we simultaneously enforced two benchmarks referred to as ``state'' and ``major-by-state''. The state benchmark had seven groups which were composed of the states and territories of Australia (except the Northern Territory, which was not benchmarked due to ABS instruction \cite{RN508}). 

%The median (minimum, maximum) sample sizes across the state benchmark groups was 1746 (1183,3546), with the direct estimates having coefficients of variation around 6\%.

The major-by-state benchmark had 12 groups, composed of the interaction of the states and territories of Australia (except the Northern Territory) and dichotomous remoteness (major city vs non-major city). Thus, for each state, apart from Tasmania (where all areas were non-major city), and the Australian Capital Territory (where 96\% of areas were major city), each SA2 was benchmarked differently depending on whether the area was in a major city or not. 

\subsubsection{External validation}

External validation was performed by comparing the estimates to those from the SHAA at the PHA level and the overall trends observed in the modelled results with the general findings from other Australian health surveys conducted on specific sub-populations, such as states \cite{RN631} or First Nations Australians \cite{RN628}. Although this validation affirmed the validity and reliability of our estimates in general, it was particularly helpful in assessing the credibility of estimates for areas that could not be benchmarked.

\subsection{Computation}
We used fully Bayesian inference using MCMC via the \texttt{R} package \texttt{rstan} Version 2.26.11 \cite{RN452}. Where possible we used the non-mean centered parameterization for random effects and the QR decomposition for fixed effects \cite{RN572}. The \texttt{stan} code for the stage 1 and stage 2 models can be found on GitHub \cite{hogg2023acariskfactors}. 

For the stage 1 model we used 1000 warmup and 1000 post-warmup draws for each of the four chains, feeding a random subset of 500 posterior draws from the stage 1 to the stage 2 model. For the stage 2 model we used 3000 warmup and 3000 post-warmup draws for each of four chains. For storage reasons we thinned the final posterior draws from the stage 2 model by 2, resulting in 6000 useable posterior draws. 

Convergence of the models was assessed using trace and autocorrelation plots, effective sample size and $\hat{R}$ \cite{RN499}. While convergence ranged slightly between risk factors, all the proportion parameters, $\boldsymbol{\mu} = \lb{\mu_1, \dots, \mu_M}$, had $\hat{R} < 1.03$, with 96\% having effective sample sizes $>1000$ and 99\% having $\hat{R} < 1.01$.  

\subsection{Summaries and visualisation} \label{sec:summ}

Estimates from the benchmarked stage 2 model were reported in a variety of forms, including absolute, relative and classification measures. For point estimates we used posterior medians and for uncertainty intervals we used 95\% highest posterior density intervals (HPDIs). We used the modelled proportions as the absolute indicator and odds ratios (ORs) as the relative indicator. The ORs for the $t$th posterior draw were derived as,

\begin{eqnarray}
    \text{OR}^{(t)}_i & = & \frac{\mu^{(t)}_i/(1-\mu^{(t)}_i)}{\hat{\mu}^D/(1-\hat{\mu}^D)} \label{eq:or}
\end{eqnarray}

with $\hat{\mu}^D$ being the overall direct estimate for the risk factor measure. An OR above one indicates that the SA2 has a prevalence higher than the national average. 

In addition to using point estimates and credible intervals to summarize the ORs, we also used the exceedence probability (EP) \cite{RN26, RN460, RN625}. 

\begin{equation}
    EP_i = \frac{1}{T} \sum_t \mathbb{I} \lb{\text{OR}^{(t)}_i > 1} \label{eq:dpp}
\end{equation}

Generally an EP above 0.8 (or below 0.2) is considered to provide evidence that the prevalence rate in the corresponding SA2 was substantially higher (or lower) than the national average, respectively \cite{RN396}. Note that the exceedance probabilities calculated using either ORs or prevalence are identical. 

To facilitate decision-making, we classified SA2s by assessing whether their individual and neighbor values (i.e. clusters \cite{RN568, RN617}) were significantly different to the national average. In this work, these classifications were called \emph{evidence classifications}. Any area classified as HC, H, L, or LC has an exceedance probability suggesting that the modelled prevalence is significantly different to the national average; HC or H denotes higher, while L or LC denotes lower. The difference between HC and H (or LC and L) is that the former provides an indication of clustering of areas, while the later only indicates significance of the area itself. If an area is not classified according to the criteria above (defined as None (``N'')) the modelled estimate is not sufficiently different to the national average. See details in Section D.3 of the Additional File.  

Code to produce subsequent plots is available on GitHub \cite{hogg2023acariskfactors}. 

%% Results %%
\section{Results} \label{sec:results}

\subsection{Prevalence}

\begin{figure}
    \centering
    \includegraphics[width=\textwidth]{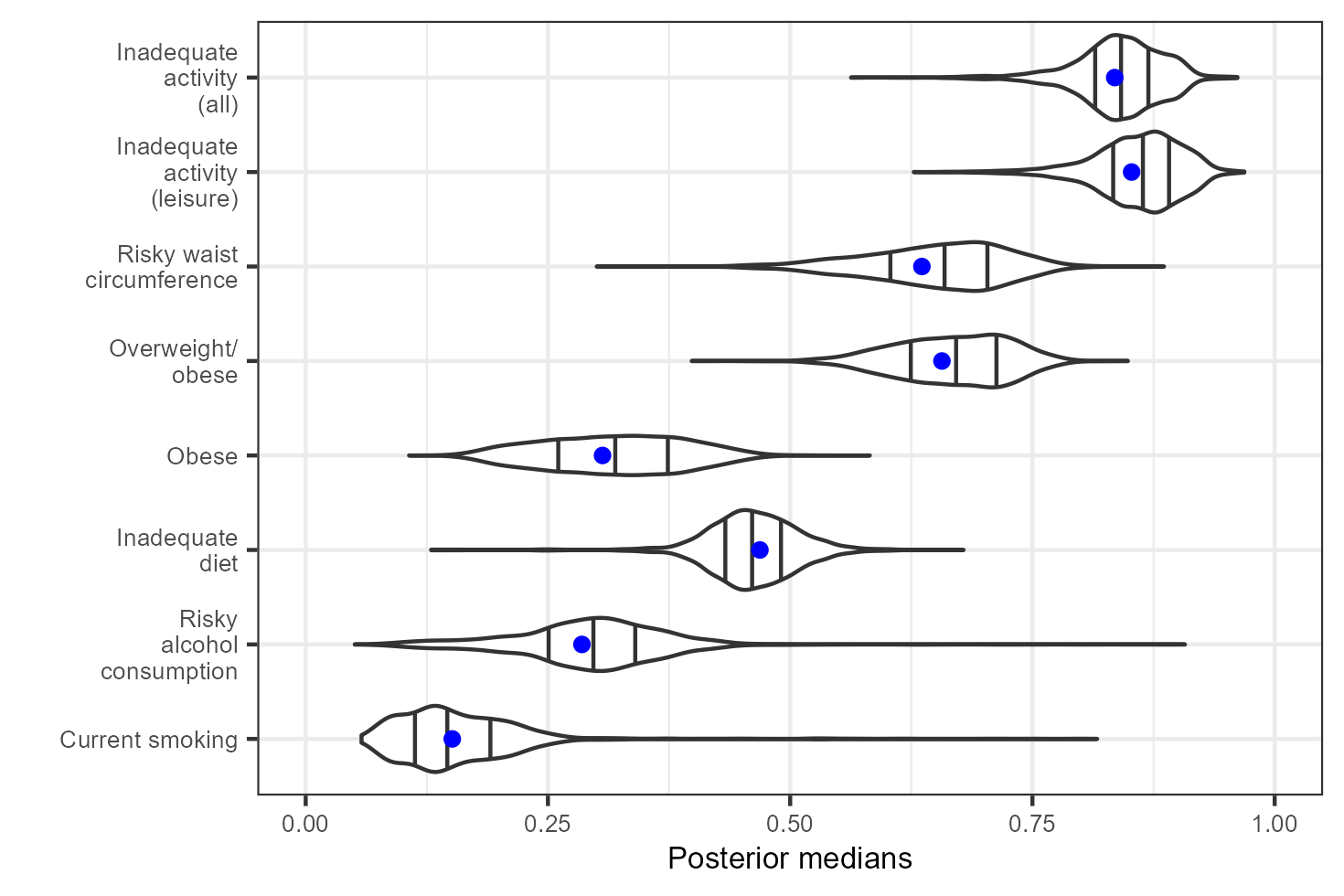}
    \caption{\small Violin plots describing the distribution of the posterior medians of the proportion estimates for each of the eight cancer risk factor measures. The width of each curve corresponds to the approximate frequency of the posterior medians similar to a density plot. The three vertical lines within the violins denotes the 25th, 50th and 75th quantiles of the posterior medians. The tails of each violin extend to the minimum and maximum values. The blue dots represent the nationwide direct estimates.}
    \label{fig:summary_violin}
\end{figure}

\begin{figure}
    \floatpagestyle{empty}
    \centering
    \includegraphics[width=0.8\textwidth]{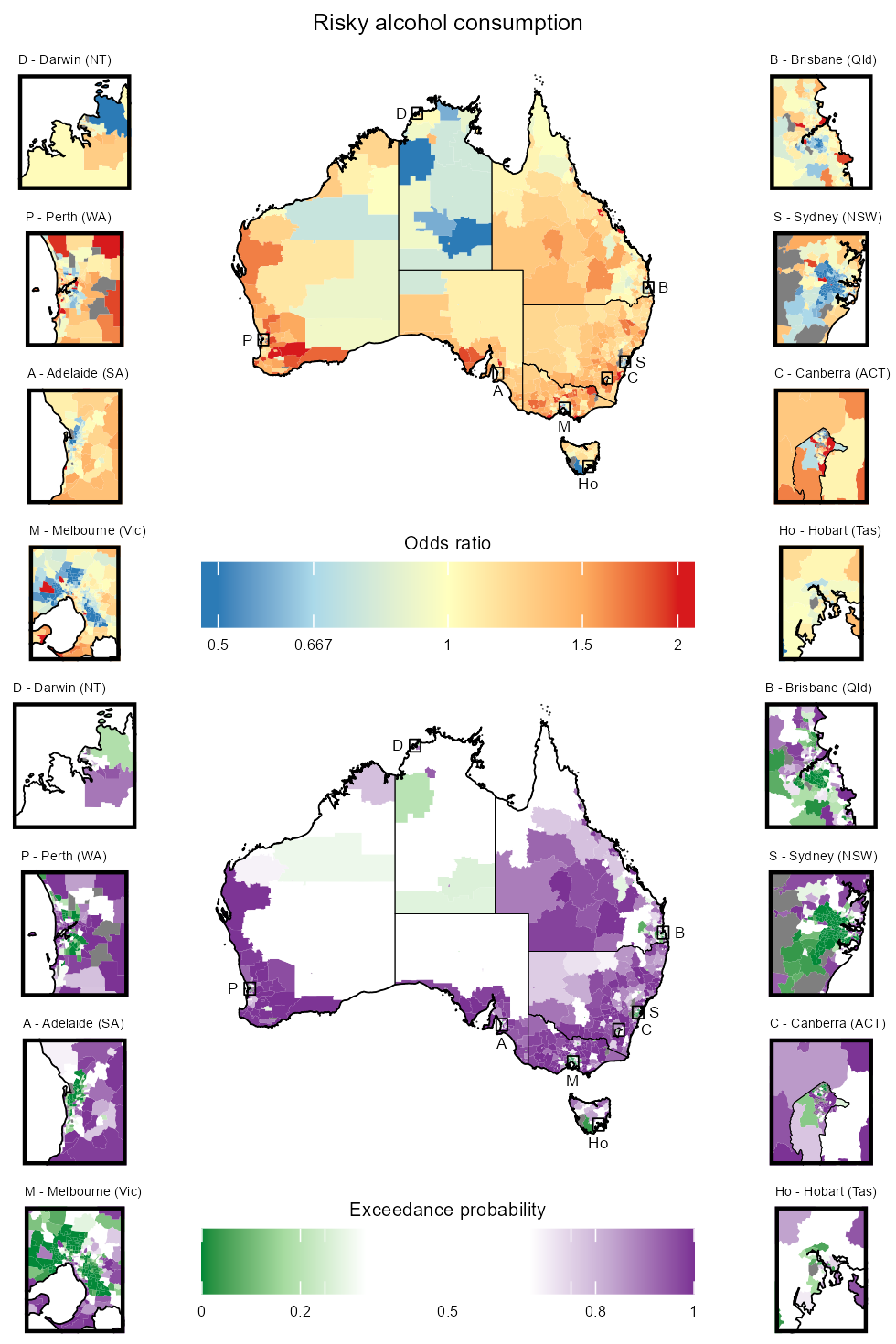}
    \caption{\small Choropleth maps displaying the results for risky alcohol consumption (see \cref{table:aca_rf_defin}) for 2221 SA2s across Australia. The top plot gives the posterior median of the odds ratios (OR). ORs above 1 indicate that the prevalence is higher than the national average. The bottom plot gives the exceedance probabilities (EPs) for the ORs described in \cref{sec:summ}. The map includes insets for the eight capital cities for each state and territory, with black boxes on the main map indicating the location of the inset. Note that some values are lower (or higher) than the range of color scales shown; for these values, the lowest (or highest) color is shown. Grey areas were excluded from estimation due to the exclusion criteria described in \cref{sec:data}. Black lines represent the boundaries of the eight states and territories of Australia.}
    \label{fig:alcohol}
\end{figure}

\begin{figure}
    \centering
    \includegraphics[width=0.8\textwidth]{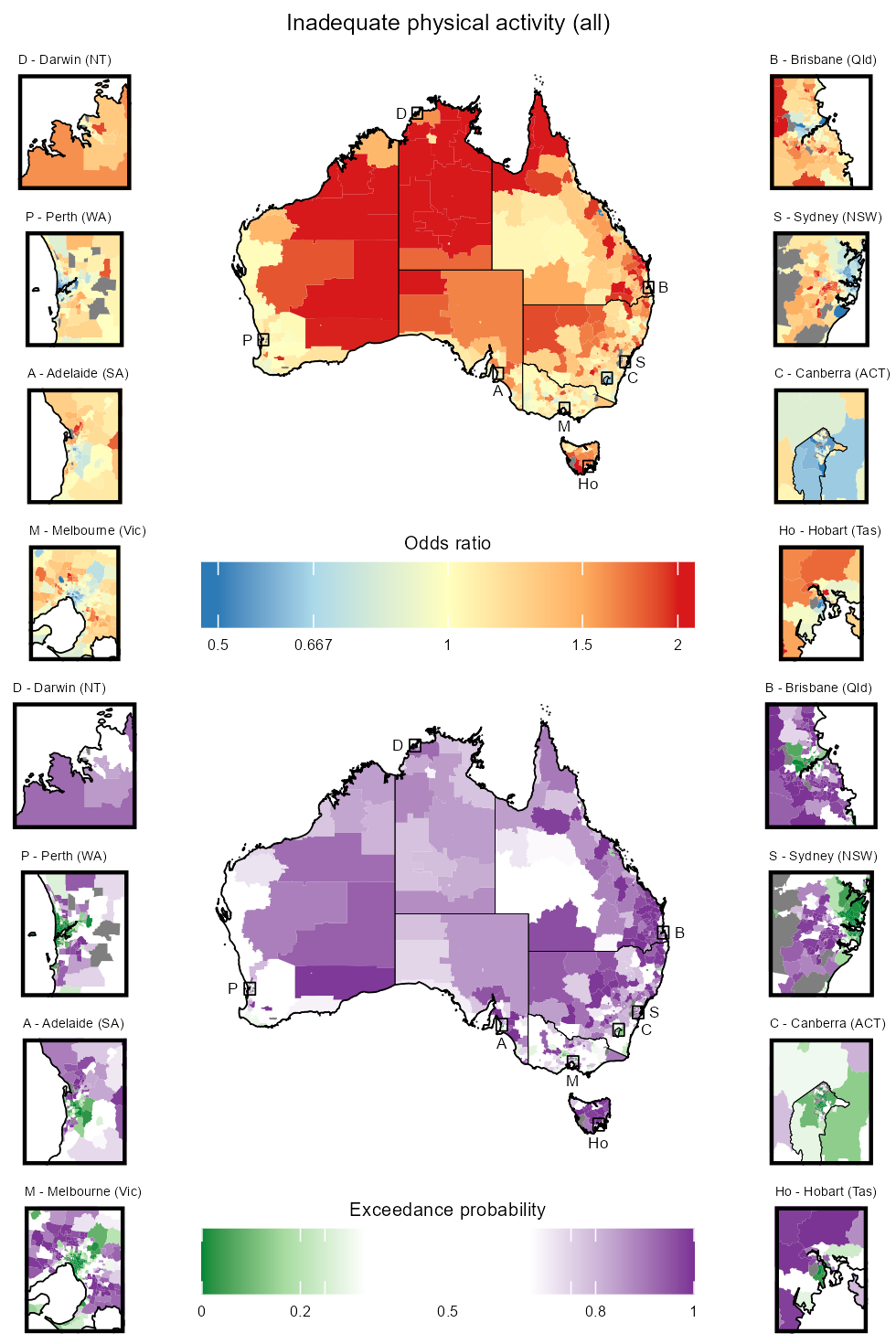}
    \caption{\small Choropleth maps displaying the results for inadequate physical activity (all) (see \cref{table:aca_rf_defin}). For more details see the caption for \cref{fig:alcohol}.}
    \label{fig:activityleiswkpl}
\end{figure}

Large spatial variation in the proportion of cancer risk factors across Australia can be clearly observed in \cref{fig:summary_violin,fig:alcohol,fig:activityleiswkpl} and Section H of the Additional File. Slightly more heterogeneity of the point estimates was observed within major cities as a result of the much greater socioeconomic variation within these areas. For example, the range of principal component 1 (a proxy for SES that is unique to the SES index) was largest in major cities and inner regional areas, but 50\% the size in remote and very remote areas. 

\begin{figure}
    \floatpagestyle{empty}
    \centering
    \includegraphics[width=0.8\textwidth]{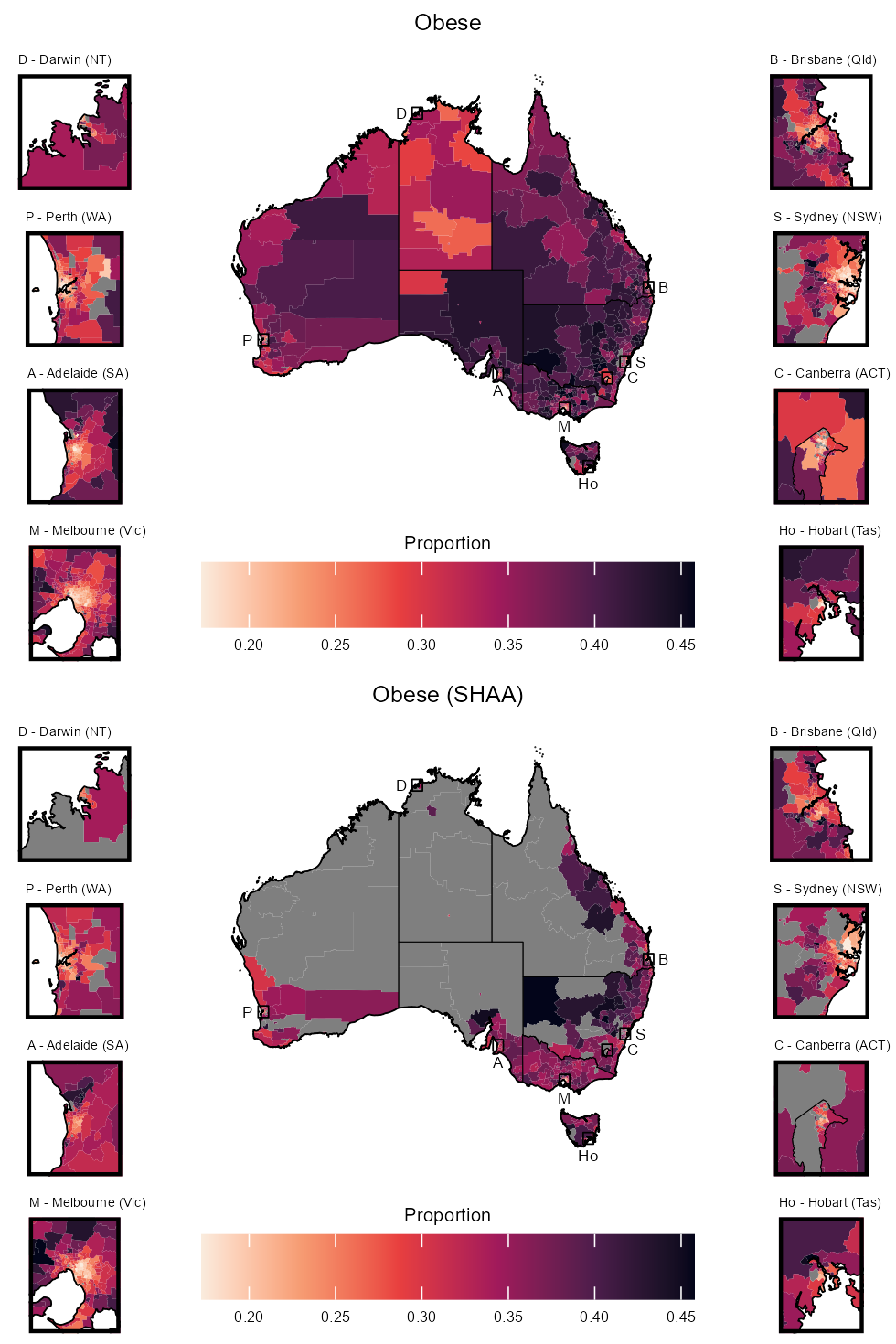}
    \caption{\small Choropleth maps of obesity prevalence at the (top) SA2 level from this work and (bottom) PHA level from the SHAA platform \cite{RN113}. The maps include insets for the eight capital cities in each state and territory, with black boxes indicating their location. Note that some values are lower (or higher) than the range of color scales shown; for these values, the lowest (or highest) color is shown. Grey areas represent no estimates, and black lines denote state and territory boundaries. Our estimates and SHAA's use similar but not identical definitions, with our values reported as proportions and SHAA's as age-standardized rates converted to proportions for comparison.}
    \label{fig:obesity}
\end{figure}

Stratifying by risk factor, the results highlight interesting patterns and trends. A more thorough discussion of the result is given in Section F of the Additional File. 
\begin{itemize}
    \item Current smoking (Section H.2 in the Additional File): Spatial patterns show lower prevalence in major cities and less disadvantaged areas. Although very high prevalence estimates are observed in the very remote regions in the middle of the country, these estimates come with substantial uncertainty. 
    \item Risky alcohol consumption (Section H.3 in the Additional File): The spatial patterns were inconsistent with the other factors, particularly in terms of the relationship between (higher) socioeconomic status and healthy behaviours. The results suggest that less disadvantaged areas have higher prevalence, which generally manifests in higher prevalence in major cities. Unlike other risk factors where prevalence estimates exhibit relative homogeneity within the SES index deciles and remoteness groups (see Section G of the Additional File), for risky alcohol consumption the estimates exhibit far greater heterogeneity for more disadvantaged areas in major cities. 
    \item Inadequate diet (Section H.4 in the Additional File): The spatial patterns suggest less dependence on the SES index and remoteness than the other risk factors. Inadequate diet exhibits the lowest heterogeneity of the risk factors considered in this work.  
    \item Body weight (Sections H.5 to H.7 in the Additional File): Similar spatial patterns are observed for the three measures. The prevalence was very strongly tied to remoteness with substantially lower prevalence almost exclusively occurring in major cities. Furthermore, the most notable differences in patterns between the estimates for obese and overweight/obese are found in major cities. 
    \item Physical activity (Sections H.8 to H.9 in the Additional File): Similar spatial patterns are observed for the two measures. Lower prevalence of inadequate activity is observed in major cities and less disadvantaged areas.
\end{itemize}

The estimates demonstrate reliability, as around 97\% of them possess coefficients of variation (CV) below 25\% --- a widely accepted threshold for reliability \cite{RN27}. Furthermore, the modelled estimates show considerable stability improvements over the SA2 direct estimates with a reduction in variability (measured by standard deviation) across Australia by an average factor of 3.3. The estimate uncertainty varied by risk factor, with current smoking having the highest median CV (17.4) and inadequate activity (leisure) having the lowest (2.6). The distribution of CV also varied by remoteness; the median CV for major cities (62\% the survey data) was, on average, 1.8 to 3.1 times smaller than that for very remote areas. 

To investigate the impact of the finer resolution, we derived PHA level CVs by taking the population weighted mean of the SA2 level estimates. The CVs of point estimates at the SA2 level range from 5\% to 34\% larger than point estimates at the PHA level across the risk factors. Similarly, by calculating and summarising the heterogeneity of SA2s within each PHA, we find that across the risk factors, the median PHA CV is between 1.5\% to 9.1\%. Of the PHAs composed of multiple SA2s, 10\% have CVs $>15$\%. The large CVs indicate that the corresponding PHAs were highly heterogeneous, highlighting the benefits of using higher resolution estimates. Given the similar definitions for the obese risk factor measure, \cref{fig:obesity} compares the estimates used in this work and that of the SHAA, indicating strong agreement.  

Section G of the Additional File provides more plots describing the modelled results, including how they vary by the SES index and remoteness. An interactive exploration of the modelled results will be made available in the Australian Cancer Atlas 2.0 \cite{RN424}, planned for release in late 2023. 

\subsection{Evidence classifications} \label{sec:ec}

\begin{figure}
    \centering
    \includegraphics{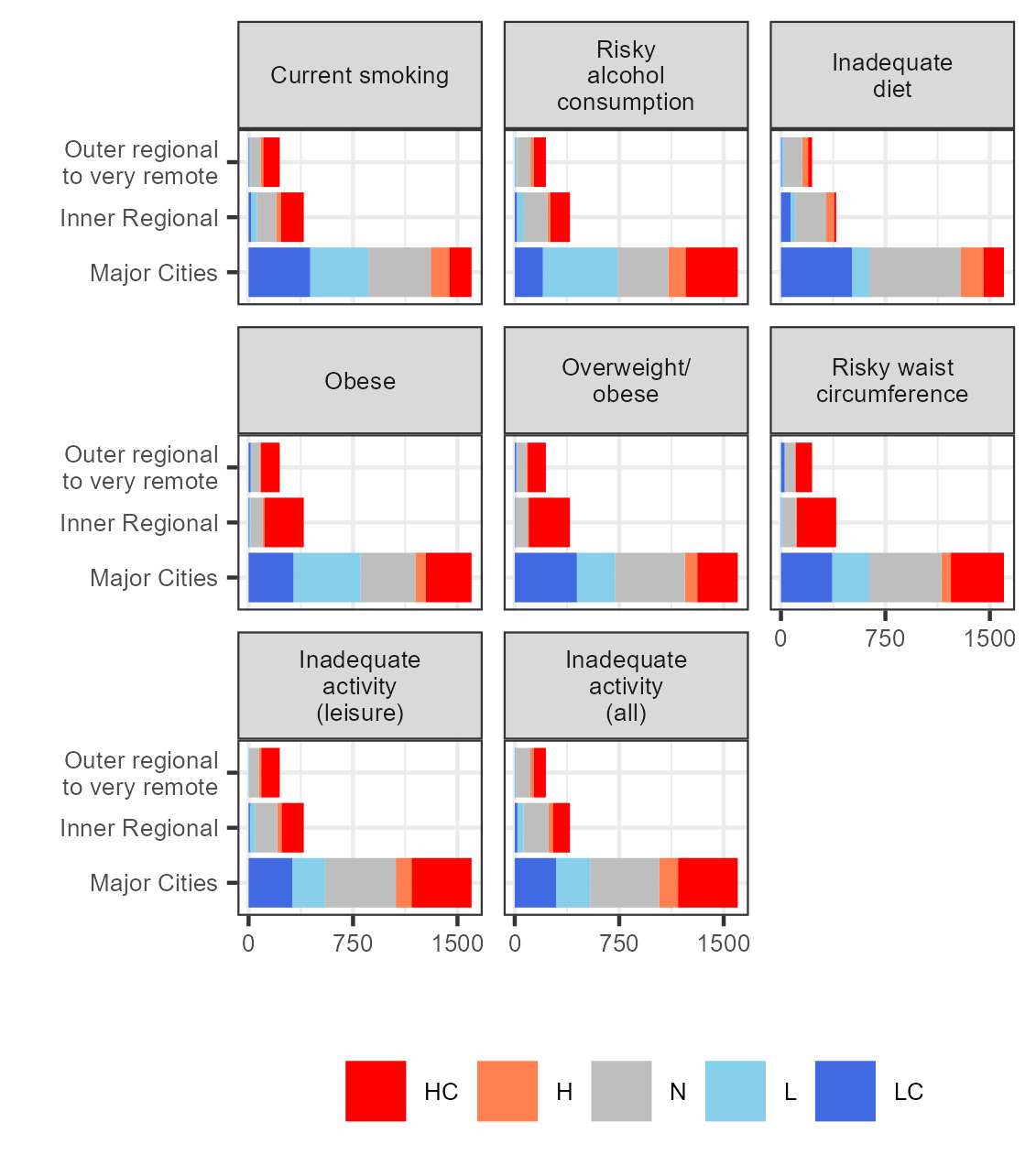}
    \caption{\small Distribution of the evidence classifications (HC, H, N, LC, and L) by remoteness and risk factor. The $x$-axis is the weighted number of SA2s using the 2017-2018 ERP as weights.}
    \label{fig:ec_ra_barchart_pw_cec}
\end{figure}

\cref{table:ec} summarises the number of evidence classifications for each risk factor measure. \cref{fig:ec_ra_barchart_pw_cec} stratifies these by remoteness. A similar stratified plot for the SES index is found in Section G of the Additional File.

% Distribution of evidence classifications
\begin{table}
\caption{\small Distribution of evidence classifications by risk factor measure (excluding ``N'' category). The values in the table are population weighted counts.}
\label{table:ec}
\centering
\begin{tabular}{rrrrrr} 
%\cline{2-6}
 & Total & HC & H & L & LC \\ 
\hline\hline
Current smoking & 1469 & 442 & 154 & 408 & 466 \\ 
\hline
\makecell[r]{Risky alcohol\\consumption} & 1482 & 603 & 137 & 523 & 219 \\ 
\hline
\makecell[r]{Inadequate\\diet} & 1155 & 194 & 221 & 155 & 586 \\ 
\hline
Obese & 1663 & 745 & 89 & 487 & 342 \\ 
\hline
Overweight/obese & 1539 & 718 & 94 & 267 & 460 \\ 
\hline
\makecell[r]{Risky waist\\circumference} & 1523 & 783 & 76 & 267 & 397 \\ 
\hline
\makecell[r]{Inadequate\\activity (leisure)} & 1458 & 724 & 148 & 262 & 324 \\ 
\hline
\makecell[r]{Inadequate\\activity (all)} & 1411 & 637 & 177 & 278 & 318 \\
\hline\hline
\end{tabular}
\end{table}

Across most risk factors many more HC areas are identified than LC areas. For example, for risky waist circumference, around 783 SA2s are classed as HC, while only around 397 are classed as LC. We observed that HC or H evidence classifications are generally found in the most disadvantaged areas, while L or LC areas are more likely in the least disadvantaged areas in major cities. 

The evidence classifications revealed several interesting trends. For the physical activity risk factor measures, a larger proportion of the areas in major cities were classified as HC or H as opposed to LC or L. For inadequate physical activity, the HC classifications favour the most disadvantaged areas. The weight risk factor measures exhibited different trends with a relatively even distribution of evidence classifications in major cities. Furthermore, almost all areas classified as LC or L occurred in less disadvantaged areas. As mirrored in the maps, the evidence classifications for smoking suggest a very strong correlation with remoteness and SES; almost all the LC or L classifications occur in major cities and less disadvantaged areas. Inadequate diet has the smallest number of evidence classifications (1155 out of 2221), with the largest proportion of them being LC areas in major cities and less disadvantaged areas. The results for risky alcohol consumption suggest that less disadvantaged areas have higher proportions of risky alcohol consumption; a trend unique to this risk factor measure. 

%% Discussion %%
\section{Discussion} \label{sec:discussion}

This work improves the spatial resolution and reach of previously published cancer risk factor estimates in Australia. While the estimates highlight broadly similar findings as those from the SHAA, they provide greater resolution and reach allowing for more granular exploration of the spatial disparities (see \cref{fig:obesity}). This is particularly pertinent due to the heterogeneity of the component SA2s within each PHA in terms of population size, socioeconomic status and remoteness. 

By improving the reach of the previously published cancer risk factor estimates, the estimates in this work uniquely enable the exploration of spatial disparities in very remote areas of Australia. As expected, the very remote areas have far greater uncertainty than those in major cities (CVs greater than 3 times higher). Nevertheless, by utilising the estimates and their uncertainty measures policy makers will have the capability to more effectively allocate health interventions and resources to these disadvantaged areas and triage areas where more data should be collected in the future to improve the quality of small area estimates. 

The cancer risk factor estimates generated in this work reveal substantial spatial disparities in cancer risk behaviours across Australia, with higher prevalence of high risk behaviours generally occurring in more remote areas. While the prevalence of most cancer risk factors is higher in areas of lower SES, the spatial patterns for risky alcohol consumption demonstrated the opposite effect. Point estimates for risky alcohol consumption and current smoking exhibited the most heterogeneity across Australia, while those from the physical activity measures exhibit the least. The distribution of the point estimates are mostly consistent across states and territories of Australia. 

Although generating prevalence estimates and their uncertainty intervals are useful in a variety of applications, using them to visualize which areas are substantially different to the national average can be difficult as the two components must be considered jointly. By further classifying the estimates according to their posterior probabilities, we were able to streamline this process. Classifications, such as those used in this work, are pivotal in developing targeted interventions as they enable policymakers to quickly identify areas, or groups of areas, with substantially higher (or lower) prevalence. 

Our Bayesian methodology, along with its associated exceedance probabilities and evidence classifications, provides insights that cannot easily be attained via the estimates from the SHAA. Although the spatial patterns of the evidence classifications (see \cref{sec:ec}) vary by risk factor, a consistent pattern was that areas with lower than average prevalence of risk factors (classified as LC or L) were almost exclusively located in major cities. Although there were areas with higher than average prevalence (HC or H) in major cities these were often less common, except for the physical activity risk factors where about half were higher and lower than the average prevalence. 

Although this applied work represents a significant step in the ongoing improvements in cancer prevention in Australia, it has some limitations. Firstly and most critically, like previous research \cite{RN485, RN184}, most of the risk factor measures used were based on data derived from self-reports which are highly susceptible to various biases \cite{RN472}. Furthermore, some 2017-18 NHS questions focused on behaviour from the previous week (e.g. alcohol, physical activity), while others on a usual week (e.g. fruit and vegetables consumption, smoking). 

Given the nature of the survey questions, caution must be exercised in using the risk factor measures presented herein. While the estimates provide insights into the spatial variation, due to the ecological fallacy \cite{RN369} and the often varying lag time between exposure (to a risk factor) and a cancer diagnosis \cite{RN165}, the estimates here cannot be used to establish individual level associations between risk factors and cancer incidence. Moreover, as these estimates are derived from cross-sectional data, they do not enable inference into lifetime risky behaviour or causal relationships with cancer. 

The second limitation, relevant to any spatial analysis of lattice data, is the modifiable areal unit problem (MAUP) \cite{Openshow1979}. The MAUP refers to the sensitivity of estimates to the specified definition of a \emph{small area} (e.g. choice of partitioning and resolution). While we have presented our estimates for SA2s, which offers wide applicability, we acknowledge that this particular partitioning of Australia represents just one of countless possible configurations, each yielding unique results. Thus, the conclusions drawn from our estimates are inherently entwined with the choice of partitioning and resolution of the small areas we employed \cite{RN31}.    

Thirdly, the accuracy of our estimates are conditional on the 2017-18 NHS exclusions (very remote areas, discrete Aboriginal and Torres Strait Islander communities and non-private dwellings \cite{RN508}). Without data for these sub-populations, there is currently no way to assess the impact of these exclusions on modelled estimates from this survey. 

Next, while the SHAA provides estimates by sex \cite{RN113}, our study, constrained by the sparsity of the survey data at the SA2 level, did not allow for a similar disaggregation. Given the evidence that health behaviours can depend on sex, the non sex-specific estimates generated in this work may suffer from inadvertent smoothing toward the mean. 

The final limitation is that the quality, in terms of both bias and variance, of small area estimates can always be improved by using larger surveys. Although we used the best survey data available, in the future, linkage of multiple surveys could provide much larger sample sizes across Australia, enabling the production of higher resolution estimates. 

In terms of future research directions, one approach could involve developing distinct models for each of the eight risk factor measures. That is the linear predictor for each risk factor measure could have different sets of covariates, random effect structures or even include non-linear relationships via splines. Alternatively, future work could model the numerous risk factors jointly by leveraging univariate stage 1 models, followed by a multivariate spatial stage 2 model \cite{RN610}.

%% Conclusions %%
\section{Conclusions}

Using a Bayesian two-stage small area estimation model we have, for the first time, generated and validated point estimates of the prevalence of eight cancer risk factors, and measures of their uncertainty, at the SA2 level across Australia. By aggregating the estimates, we have shown that they are very similar to those given by the SHAA \cite{RN113}, external surveys \cite{RN635, RN630, RN633, RN645, RN646, RN514} and previous research on how area level socioeconomic status and remoteness relate to healthy behaviours \cite{RN204}. The new estimates provide improved spatial resolution and reach and will enable more targeted cancer prevention strategies at the small area level. Furthermore, by including the results in the next release of the Australian Cancer Atlas \cite{RN424}, this work promises to provide a more comprehensive picture of cancer in Australia. Since the health factors used in this study are also common risk factors for other diseases, the prevalence estimates generated here may be useful in other disease modelling applications both in Australia and internationally.  

\subsection*{Acknowledgements}

This study has received ethical approval from the Queensland University of Technology Human Research Ethics Committee (Project ID: 4609) for the project entitled ``Statistical methods for small area estimation of cancer risk factors and their associations with cancer incidence''. Ethics approval was received for the inclusion of the modelled estimates from the 2017-18 NHS into the Australian Cancer Atlas (Griffith University Human Research Ethics Committee (EC00162) Ref:2018/052). Approval was received to access the secure ABS Datalab for analyses of the NHS data (Project ID: 2021-033 QUT). 

We thank the Australian Bureau of Statistics (ABS) for designing and collecting the National Health Survey data and making it available for analysis in the DataLab. The views expressed in this paper are those of the authors and do not necessarily reflect the policy of QUT, CCQ or the ABS.

\subsection*{Competing interests}
The authors declare that they have no competing interests.

\subsection*{Funding}
JH was supported by the Queensland University of Technology (QUT) Centre for Data Science and Cancer Council QLD (CCQ) Scholarship. SC receives salary and research support from a National Health and Medical Research Council Investigator Grant (\#2008313).

\subsection*{Additional Files}
The Additional File (appended to this paper) contains additional material including further details of the data and model, and more plots, maps, and results. 

\bibliographystyle{unsrtnat}
\bibliography{ref}  %%% Uncomment this line and comment out the ``thebibliography'' section below to use the external .bib file (using bibtex) .

\end{document}

\maketitle
\tableofcontents

% Use Letters for the 4 sections
\setcounter{section}{0}
\setcounter{subsection}{0}
\renewcommand{\thesection}{\Alph{section}}
\renewcommand{\thesubsection}{\thesection.\arabic{subsection}}

\newpage
\section{Other sources of Australian health data} \label{supp:other_aus_data}

% https://www.stylemanual.gov.au/accessible-and-inclusive-content/inclusive-language/aboriginal-and-torres-strait-islander-peoples

By evaluating the extent to which our estimates at the SA2 level aligned with those from the SHAA and the broader socioeconomic status (SES), and remoteness trends observed in external surveys, we were able to validate the credibility and generalizability of our results. In this section, we provide details about these external surveys and their appropriateness with respect to the validation procedure carried out in this work. Note that our analyses did not involve any form of record linkage. 

\subsection{State surveys}

Although the National Health Survey is the optimal survey for this study given its national coverage and breadth of health data collected, each state and territory of Australia also independently conduct surveys for health surveillance purposes. 

\begin{figure}[H]
    \centering
    \includegraphics[width=\textwidth]{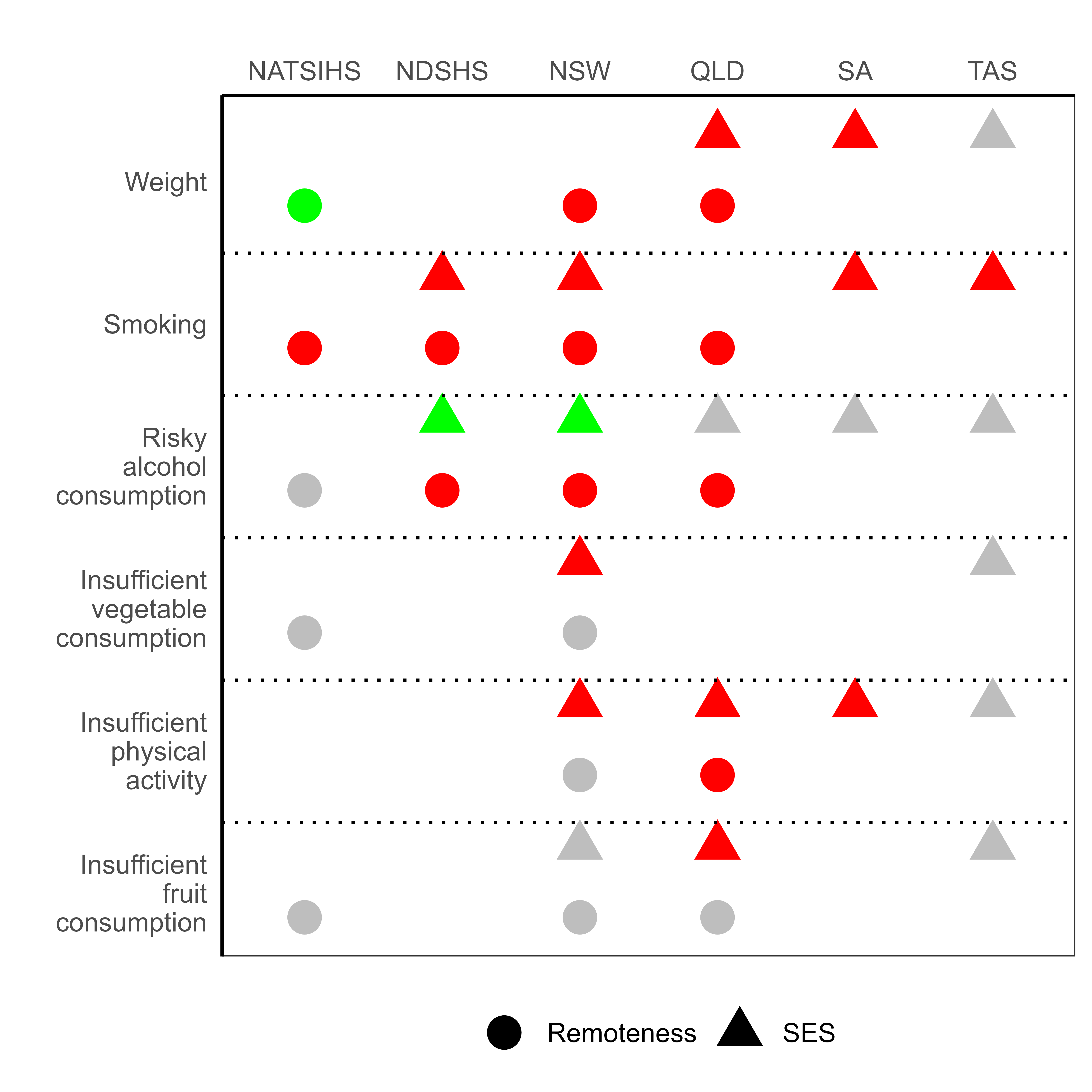}
    \caption{\small Schematic summarizing the key findings from external health surveys conducted by state governments (New South Wales (NSW) \cite{RN635}, Queensland (QLD) \cite{RN630}, South Australia (SA) \cite{RN633} and Tasmania (TAS) \cite{RN645}) and the Australian Bureau of Statistics (The National Aboriginal and Torres Strait Islander Health Survey (NATSIHS) \cite{RN646} and National Drug Strategy Household Survey (NDSHS) \cite{RN514}). More details can be found in \cref{supp:other_aus_data} and \ref{supp:data_details}. Each column corresponds to a specific survey, while rows represent different cancer risk factors, each with varying definitions across the surveys. Within each risk factor, we provide a summary of trends based on remoteness and socioeconomic status (SES), represented by distinct shapes. The colors of the shapes indicate the direction of the trends: red indicates a higher prevalence for remote areas compared to urban areas, and for socioeconomic disadvantage compared to socioeconomic advantage. For instance, a red triangle signifies a higher prevalence of the risk factor in more remote areas, while gray indicates insignificant or trivial differences between remoteness categories. Conversely, green denotes a lower prevalence of the risk factor in more remote areas. The schematic only includes trends supported by available data from digital platforms or publicly available reports.}
    \label{fig:state_summary}
\end{figure}

\paragraph{New South Wales (NSW)}
NSW, the state with the largest 2016 adult population in Australia, conducts annual population health surveys by telephone. These surveys generally aim for a sample size of 13,000 people from the state \cite{RN631}. The NSW Ministry of Health provides data on cancer risk factors stratified by year, remoteness, and SES status via their interactive dashboard, HealthStats NSW \cite{RN635}. As shown in \cref{fig:state_summary}, trends in cancer risk factors from NSW contribute a large proportion of the data we used to externally validate our modelled estimates.

\paragraph{Victoria (VIC)}
VIC, the second largest state in terms of population, also conducts annual health surveys, which are administered using computer assisted telephone interviews \cite{RN632}. In 2017, the survey collected data on 33,654 adults. Unlike other states, the Victorian Agency for Health Information provides cancer risk factor prevalence estimates for small areas called Local Government Areas (LGAs) \cite{RN604}, which are non-ASGS boundaries that continue to be used by the Australian government. Unfortunately data are not reported by remoteness or SES. 

\paragraph{Queensland (QLD)}
QLD, the second largest state in terms of population, conducts annual health surveys, which are also administered by telephone. Around 12,500 adults participate each year \cite{RN630}. Although not as flexible as the NSW data system, Queensland Health provide an online tool called the Queensland survey analytic system that provides prevalence estimates for cancer risk factor by LGAs, PHNs, remoteness and SES. Although LGA estimates are available, for the regions we would be interested in --- far north Queensland for example --- data are not available or not releasable. %Results from Queensland were vital in terms of validating estimates for First Nations Australians. Around 8.5\% of SA2s in Queensland in 2016 had greater than 10\% of the residents identifying as First Nations Australians. This is second only to the NT, where this figure is 57.6\%. Unfortunately the online tool does not provide data stratified by First Nations Australian status. 

\paragraph{Western Australia (WA)}
WA also conducts annual health surveys administered using computer assisted telephone interviews. In 2018, the sample included 5,750 adults aged 16 years and over and boasted an average participation rate of approximately 90\% \cite{RN629}. These data would have been important for validation as 5\% of the states SA2s are very remote. Unfortunately there is no publicly available dashboard summarising the Western Australian survey results. In published reports, cancer risk factor data are not stratified by small area, remoteness or SES, which limits their usefulness for validation purposes. 

% These data would have been important for validation as 5\% of the states SA2s are very remote and 4.2\% had over 20\% of their residents identifying as First Nations Australians. 

\paragraph{South Australia (SA)}
SA collects health data monthly via telephone interviews. In one year, around 7,000 South Australians are interviewed for the survey \cite{RN633}. Unfortunately there is no publicly available dashboard summarising the SA survey results. In published reports prevalence of cancer risk factors are sporadically stratified by remoteness, and SES.

\paragraph{Northern Territory (NT)}
The first state-wide health survey in the NT was due to occur in 2022 with the goal of sampling 2000 residents \cite{RN636}. At the time of writing, the most relevant health data for the state is available in the NHS \cite{RN598} (6.7\% of participants were from the NT) or National Aboriginal and Torres Strait Islander Health Survey \cite{RN628} (18\% of participants were from the NT).  

\paragraph{Tasmania (TAS)}
TAS has been collecting health data triennially since 2009 via computer assisted telephone interviews. The 2019 Tasmanian Population Health Survey sampled 6300 adults. The published report presented detailed prevalence estimates by remoteness, and SES \cite{RN645}.

\paragraph{Australian Capital Territory (ACT)}
The ACT has been conducting an annual General Health Survey since 2007 via computer assisted telephone interviewing \cite{RN666}. In 2019, the survey collected a wide range of health data from 2002 adults. Given its population (only 1.7\% of the 2016 adult population) and geographical size (less than 0.03\% of Australia's landmass), data from the ACT were not informative for external validation in this research. The ACT is also relatively homogeneous in terms of remoteness and SES. Of its 114 SA2s, 110 are major cities and 102 have SES index deciles above 5 (denoting lower levels of disadvantage). %and 106 have the proportion of First Nations Australians under 3\%.  

\subsection{Other surveys}

Other nationwide surveys have health data for either a subset of the population (like the National Aboriginal and Torres Strait Islander Health Survey (NATSIHS) \cite{RN628}) or a subset of variables (such as the National Drug Strategy Household Survey (NDSHS) \cite{RN514}). 

The NATSIHS is a hexennial survey which collects social and health data from First Nations Australians with the goal of reporting statistics at the national and remoteness level \cite{RN628}. The 2015 survey collected data on 11,178 First Nations Australians living in private dwellings across Australia. These surveys collect information on all the risk factors considered in this research.  

The NDSHS is the leading survey of licit and illicit drug use in Australia. In 2019, the household-based survey collected data from 22,274 people aged 14 and over \cite{RN514}. Relevant to this research, the NDSHS provides data on alcohol consumption and smoking. 

%% NEW SECTION %% --------------------------------------------------------------------------------------------
\newpage
\section{Risk factor details} \label{supp:rf_details}

This section provides further details for the eight risk factor measure definitions used in this work. A summary is provided in Table 1 in the main paper. 

\subsection{Smoking}
According to Whiteman \emph{et al.} \cite{RN165} and the 2015 Australian Burden of Disease study \cite{RN121} smoking contributes the highest proportion of total cancer burden in Australia. Similar to the definition used in the SHAA \cite{RN113}, \emph{current smoking} was defined as those who reported to be daily, weekly or less than weekly current smokers, and had smoked at least 100 cigarettes in their life. Even though ex-smokers experience greater risk than non-smokers for some cancers \cite{Tindle2018}, including ex-smokers in the definition would create a measure of lifetime smoking which deviates from the cross-sectional nature of the other risk factor measures. 

\subsection{Alcohol}
The revised 2020 guidelines by the NHMRC stipulate that persons should drink no more than 10 standard drinks a week and no more than 4 standard drinks on any one day \cite{RN516}. The previous 2009 NHMRC guidelines stipulated that adults should drink no more than 2 standard drinks on any day \cite{RN165}. Both versions of the guidelines emphasize that the recommendations do not represent a ``safe'' or ``no risk'' level of alcohol consumption \cite{RN512}. 

Although Cancer Australia \cite{RN582} and overseas governments, such as the US \cite{RN586}, recommends alcohol consumption in line with the 2009 NHMRC guidelines, Cancer Council Australia \cite{RN599} recommends the more recent 2020 guidelines. In this work, we defined \emph{risky alcohol consumption} as persons who did not meet the 2020 guidelines.

In the 2017-18 NHS, detailed alcohol consumption data were based on self-reports. Participants were asked about their alcohol consumption for the most recent three days (from the preceding 7 days) that they drank alcohol. Results cannot indicate lifetime alcohol behaviour.

\subsection{Diet}
Although a variety of diet-related factors, such as fruit, vegetables, meat, fibre and wholegrains, have been found to contribute to the risk of developing cancer \cite{RN123, RN121, RN165}, the 2017-18 NHS collected diet information for fruit and vegetables only \cite{RN508}. Participants of the survey were asked to report the number of serves of fruit and serves of vegetables they usually ate each day. 

Both Whiteman \emph{et al.} \cite{RN165}, the SHAA \cite{RN113} and Cancer Australia \cite{RN582} use the 2013 NHMRC guidelines \cite{RN520} for fruit and vegetables, which stipulates two serves (equivalent to 300g) of fruit and five serves (equivalent to 375g) of vegetables per day.

While we acknowledge the benefits in producing separate maps for fruit and vegetable consumption, we found that only 7.5\% of Australians met the guidelines for vegetable intake \cite{RN478}. Modelling these sparse data could give very unstable prevalence estimates \cite{self_cite}, a possible reason why the SHAA does not provide estimates for vegetable consumption alone. To address this issue, we jointly modelled whether participants met the guidelines for fruit or vegetables. Specifically, we assigned a value of one, indicating \emph{inadequate diet}, to persons who did not meet either guideline.

\subsection{Weight}
According to Whiteman \emph{et al.} \cite{RN165}, 3.4\% of cancer diagnoses in Australia are attributable to being overweight or obese. Following the SHAA, we used both weight-based risk factor measures according to the conventional Body Mass Index (BMI) \cite{RN521, RN511}, and a measure related to waist circumference \cite{RN600}. 

BMI is calculated using a participant's height and weight. While self-reported height and weight are commonplace \cite{RN472, RN625} and useful to approximate BMI \cite{RN426, RN71}, the reported values can be subject to bias \cite{RN509}. Measurement of weight, height and waist circumference was a voluntary section of the NHS interview. To increase the applicability (reduce the number of missing values) of these weight-based measurements, any participant who refused to be measured had their weight, height and waist circumference measurements imputed using a hot-decking method where they received weight, height and waist circumference measurements from another very similar participant based on demographic characteristics, such as age, sex, state, self-perceived body mass, exercise, cholesterol and self-reported BMI \cite{RN508}. Although approximately 40\% of the measured data were imputed, measured values can have significantly less bias than self-reported measures \cite{RN472}. In this work, we used the measured data.

\emph{Overweight/obese} and \emph{obese} was defined as persons with a BMI greater than or equal to 25 and 30, respectively.  

Following Cancer Australia \cite{RN582} and the SHAA \cite{RN113}, we defined \emph{risky waist circumference} as measurements of 94cm and 80cm or more for men and women, respectively \cite{RN512}. Note that these waist circumference cutoffs are only appropriate for adults, so for this risk factor we limit the dataset to all persons 18 years and older. Assuming that the single-year age distribution in any of the age group was uniform, we estimated that the population of 18-19-year olds was 40\% of the 15-19-year old population. 

\subsection{Physical activity}
The national Department of Health (DOH) guidelines for physical activity, published in 2014 \cite{RN517}, closely mirror those given by the World Health Organization \cite{RN519}. The DOH guidelines stipulate that each week adults (those between the ages of 18 and 64) should either do 2 $1/2$ to 5 hours of moderate-intensity physical activity or 1 $1/4$ to 2 $1/2$ hours of vigorous-intensity physical activity or an equivalent combination of both. In addition, the guidelines recommend muscle-strengthening activities at least 2 days each week. The DOH guidelines also provide specific recommendations for children (5 to 17 years), older persons (65 years and older) and pregnant women. 

The benefits of physical activity on overall health, including cancer, are best described by considering the total volume of activity, which is calculated by considering the frequency, duration and intensity of exercise \cite{RN518}. In a systematic review of physical activity and cancer outcomes, McTiernan \emph{et al.} \cite{RN406} found that most studies related leisure physical activity to cancer, whereas non-leisure (or work-related) physical activity was not classed as physical activity. Elsewhere \cite{RN587}, physical activity has been defined as any movement that results in energy expenditure, which includes deliberate exercise or sport, incidental movement or work-related activity.

To capture both leisure only and all activity, we opt to allow workplace physical activity (if of sufficient intensity) to count toward meeting the 2014 DOH guidelines. \emph{Inadequate activity (leisure)} was defined as those who did not meet the guidelines based on their reported leisure physical activity alone in the week before the survey interview. \emph{inadequate activity (all)} was defined as those who did not meet the guidelines based on their reported leisure and workplace physical activity in the week before the survey interview. In this work, the physical activity measures were derived from the ABS created variables that accommodated the guidelines across age groups.  

%% NEW SECTION %% --------------------------------------------------------------------------------------------
\newpage
\section{Model details} \label{supp:model_details}

This section provides details of the two-stage model described in Section 3 of the main paper. Figures 2 and 3 (in the main paper) provide graphically summaries of the components of the two-stage model. Initially we describe the principal components analysis that was carried out before modelling. 

\subsection{Survey weights}
To help correct for sampling bias and promote design-consistency, survey practitioners provide survey weights. In this analysis, we used two versions of the survey weights, $w\jut{raw}_{ij}$, provided by the ABS \cite{RN508,RN44}. 

\begin{equation*}
    w_{ij} = n_i w \jut{raw}_{ij} \lb{ \sum_{j=1}^{n_i} w\jut{raw}_{ij} }^{-1}
\end{equation*}

was used for direct estimation, and 

\begin{equation*}
    \tilde{w}_{ij} = n w \jut{raw}_{ij} \lb{  \sum_{i=1}^m \sum_{j=1}^{n_i} w\jut{raw}_{ij} }^{-1}
\end{equation*}

was used in the stage 1 model.

\subsection{Principal components analysis} \label{sec:pca}
As mentioned in Section 2.2.2 in the main paper, principal components analysis was conducted on 84 continuous census covariates represented as proportions or averages. The input data were scaled and centered prior to the analysis. Principal component 1 (PC1) to PC6 were retained as they accounted for approximately 62\% of the variation. PC1 to PC6 captured approximately 24\%, 11\%, 11\%, 7\%, 5\% and 4\% of the variation, respectively. All remaining PCs contributed less than 3\% of the total variation each with 83\% of these contributing less than 1\%. 

PC1 was very strongly correlated (Pearson correlation of 0.86) with the SES index deciles (see \cref{fig:pc1_vs_irsd}), obtained from the ABS Socio-Economic Indexes for Areas product. Despite the correlation, model performance (assessed using metrics described in \cref{supp:mb}) was improved when both were included. Including both was pragmatic as the SES index is constructed from 16 purposely-selected census measures of relative disadvantage, while PC1 to PC6 captured a wider range of census data. Moreover, the SES index is included as a categorical variable with ten groups, whilst the PC1 to PC6 variables are included as continuous quantities. A further benefit of including the SES index (which captured 43\% of the variation) is that it accommodates some information not captured by the principal components analysis including characteristics of dwellings (internet connection, number of cars, etc), disability and equivalised income \cite{RN560}. 

\subsection{Stage 1: Individual level model} \label{sec:s1}

In this section we describe the components of the stage 1 linear predictor, 

\begin{equation*}
    \jdist{logit}{\pi_{ij}} = \mathbf{X}_{ij} \boldsymbol{\beta} + e_i + \delta_{r[ij]} +  \xi_{d[ij]} + \epsilon_{ij},
\end{equation*}

\noindent for the individual level probability of $y_{ij} = 1$ for sampled individual $j = 1, \dots, n_i$ in SA2 $i = 1, \dots, m$. Details on how the model was selected are given in \cref{sec:mb_s1}. 
   
\paragraph{Fixed effects}
The fixed effects were included via the individual level design matrix, $\mathbf{X}_{ij}$, and corresponding coefficients, $\boldsymbol{\beta}$. We used the following individual level categorical covariates in the stage 1 models: age, sex and their interaction; registered marital status; hypertension status; Kessler psychological distress score; occupation; language spoken at home; and depression status. See \cref{table:categorical_vars} for the categories of the individual level covariates and the reference groups used. Along with the individual covariates, we also included seven SA2 level fixed effects; SES index deciles (IRSD) as a categorical covariate with 10 groups and the first six principal components (PC) (derived from the 2016 census data, see \cref{sec:pca}). 

\paragraph{Random effects}
Unstructured individual level random effects ($\epsilon_{ij}$) and area (SA2) level random effects ($e_i$) were applied. In addition to these, by borrowing ideas from MrP \cite{RN439}, we included two hierarchical random effects based on categorical covariates that were themselves derived from the interaction of numerous individual level demographic and health covariates. We derived a demographic-health (DH) categorical covariate from the interaction of sex, age, self-assessed health, qualification, and high school completion status. See \cref{table:categorical_vars} for the categories of these covariates. The median (IQR) samples size in each DH group was 7 (3, 20). 
The random effect for DH group $d = 1, \dots, 1050$ is denoted by $\xi_{d[ij]}$, where $d[ij] \in \{1, \dots, 1050\}$ indexes the DH group for sampled individual $j$ in SA2 $i$. 

We also derived a categorical covariate ($\delta_{r[ij]}$ with $r = 1, \dots, 16$ levels) from the interaction of the binary risk factor outcomes not directly associated with the risk factor being modelled. See \cref{table:norf_concor} for more details on this covariate, termed the \emph{non-outcome risk factor (NORF)} covariate. For example, when modelling waist circumference, $\boldsymbol{\delta} = \lb{ \delta_{1}, \dots, \delta_{16} }$ was constructed using risky alcohol consumption, inadequate activity (all), inadequate diet and current smoking. This variable was identical for risky waist circumference, obese and overweight/obese, while for other outcomes, overweight was one of the binary risk factors included.  

The probability that $y_{ij} = 1$ for sampled individual $j$ in SA2 $i$ is denoted by $\pi_{ij}$. Pseudo-likelihood was used to ensure the predictions from the logistic model were approximately unbiased under the sample design \cite{RN500, RN501}. The remaining notation is described in Section 3.1 of the main paper. 

\paragraph{Stage 1 model}
The stage 1 model used for all risk factor measures is,  

\begin{eqnarray}
    y_{ij} & \sim & \jdist{Bernoulli}{\pi_{ij}}^{\tilde{w}_{ij}} \label{eq:model_s1}
    \\
    \jdist{logit}{\pi_{ij}} & = & \mathbf{X}_{ij} \boldsymbol{\beta} + e_i + \delta_{r[ij]} +  \xi_{d[ij]} + \epsilon_{ij} \nonumber
    \\
    e_i & \sim & \jdist{N}{0, \sigma^2_e}  \nonumber
    \\
    \delta_{r[ij]} & \sim & \jdist{N}{0, \sigma^2_{\delta}}  \nonumber
    \\
    \xi_{d[ij]} & \sim & \jdist{N}{0, \sigma^2_{\xi}}  \nonumber
    \\
    \epsilon_{ij} & \sim & \jdist{N}{0,2^2},   \nonumber
\end{eqnarray}

\noindent where we represent the pseudo-likelihood for a probability density, $p(.)$, as $p\lb{y_{ij}}^{\tilde{w}_{ij}}$ \cite{RN44}. Note the fixed variance for $\epsilon_{ij}$; an explanation is given in \cref{supp:mb}. Details on the priors used are given in Section 3.2 in the main paper. 

\subsection{Stage 2: Area level model}

In this section we describe the components of the stage 2 linear predictor,

\begin{equation*}
    \theta_i = \mathbf{Z}_i \boldsymbol{\Lambda} + \alpha \gamma_{i} + \mathbf{G}_i \boldsymbol{\Gamma}_{r[i]} + \zeta_i + \eta_{h[i]},
\end{equation*} 

\noindent for SA2 $i = \dots, M$, where $M = 2221$ is the total number of areas. Details on how the model was selected are given in \cref{sec:mb_s2}. 

\subsubsection{Fixed effects}
The design matrix of fixed effects ($\mathbf{Z}_i$) for the second stage model with corresponding coefficients ($\boldsymbol{\Lambda}$) included two categorical covariates: the SES index deciles (IRSD) and remoteness. Apart from the six principal components (introduced below), no other specific census variables were found to improve model fit. 

\subsubsection{Fixed effects with varying coefficients}
The second component of the linear predictor for $\theta_i$ was the six continuous covariates, PC1 to PC6. Australia is highly decentralised, meaning that area level statistical relationships may be very different in major cities as opposed to very remote areas. To incorporate this, we allowed the fixed effect regression coefficients for PC1 to PC6 to vary according to remoteness (major cities, inner regional and outer regional to very remote). 

The principal components values for SA2 $i$ are denoted by $\mathbf{G}_i$, a row-vector of length six. The regression coefficients specific to the $r$th remoteness category for the $i$th SA2 are denoted by $\boldsymbol{\Gamma}_{r[i]}$, a column-vector of length six. Although we explored partial pooling for the regression coefficients, because we only used three remoteness groups we found better performance and convergence when using independent priors; treating them as fixed effects. As described in Section 3.2 in the main paper, fixed effects were given generic weakly informative priors. Thus, $\mathbf{\Gamma}_{r[i]} \sim \jdist{N}{0, 2^2}$. 

\subsubsection{Estimates from the SHAA}
The third component of the stage 2 linear predictor was modelled prevalence estimates from the SHAA at the PHN level (see \cref{supp:data_details}). The data from the SHAA is provided as age-standardised rates per 100 people with 95\% confidence intervals. We used these data to derive proportion estimates and their associated standard errors, before transforming the estimates to the unconstrained scale. 

The logistic transformed SHAA estimate and variance for SA2 $i$ are denoted by $\hat{\gamma}_{i}$ and $\widehat{\text{v}} \lb{ \hat{\gamma}_{i} }$, respectively. Note that $\hat{\gamma}_{i}$ and $\hat{\gamma}_{k}$ (and corresponding variance) were identical if SA2 $i$ and $k$ were within the same PHN.

To accommodate the variance of the estimates from the SHAA, we assumed classical measurement error \cite{RN549}, whereby the model variance (or error) is independent of the true value. Thus,

\begin{eqnarray*}
    \hat{\gamma}_{i} & \sim & \jdist{N}{\gamma_{i}, \widehat{\text{v}} \lb{ \hat{\gamma}_{i} }}
    \\
    \gamma_{i} & \sim & \jdist{N}{0,2^2}
\end{eqnarray*}

\noindent where $\gamma_{i}$ is assumed to be the true logistic transformed SHAA estimate. Note that other work models the true values using spatial priors \cite{RN343, RN547, RN545}. However, because there are few PHNs compared to SA2s, we used an independent weakly informative prior, $\jdist{N}{0,2^2}$. 

The true SHAA estimate, $\gamma_{i}$, was included as an external latent field in the linear predictor for $\theta_i$, with $\alpha$ (the coefficient for $\gamma_{i}$) controlling the influence of the external latent field on the modelled estimates. \cref{table:shaa_rf_concor} shows which of the variables from the SHAA was used as the external latent field for each of our risk factor measures.  

\subsubsection{Random effects} \label{sec:re}
The final components of the stage 2 model were random effects. Many studies have illustrated the benefits of accommodating the spatial structure of the small areas in SAE models \cite{RN458, RN147, RN343, gao2023_sma, RN533}. Although others have used conditional autoregressive (CAR) \cite{RN363} or simultaneous autoregressive (SAR) priors only, Gomez-Rubio \emph{et al.} \cite{RN35} argues that including a structured and unstructured random effect provides a useful compromise between accurate small area estimates and their variances. 

We used the BYM2 spatial prior \cite{RN394} ($\zeta_i$) with mixing parameter, $\rho \in [0,1]$, scaling factor, $\kappa$, variance parameter, $\sigma_\zeta^2$, and $M \times M$ adjacency matrix, $\mathbf{W}$. The BYM2 prior is a linear combination of a unit-scale intrinsic conditional autoregressive (ICAR) prior \cite{RN363} and a standard normal. The BYM2 prior, denoted as $\jdist{BYM2}{ \mathbf{W}, \kappa, \rho, \sigma_{\zeta}^2 }$, is 

\begin{eqnarray}
    \zeta_i & = & \sigma_{\zeta} \left(s_i \sqrt{\rho/\kappa} + v_i \sqrt{1-\rho}\right) \label{eq:bym2}
    \\
    s_i & \sim & N\left( \frac{\sum_{k=1}^{M} W_{ik} s_k}{\sum_{k=1}^{M} W_{ik}}, \frac{1}{\sum_{k=1}^{M} W_{ik}} \right) \nonumber
    \\
    v_i & \sim & N(0,1). \nonumber
\end{eqnarray}

As is common in disease mapping \cite{RN8}, we used the binary contiguous specification for $\mathbf{W}$ where $W_{ik} = 1$ if SA2 $i$ and SA2 $k$ are neighbors and zero otherwise. To reduce the complexity of our model we made several manual changes to the weight matrix.
\begin{itemize}
    \item We ensured the neighborhood structure was fully connected \cite{RN394} by treating the eastern and western SA2s at the top of Tasmania as neighbors of the furthest south SA2s in Victoria. 
    \item Donut SA2s, or areas with only one neighbor (n = 87), were also assigned the neighbors of their neighbor. The final weight matrix assigned 93\% of SA2s more than 2 neighbors. 
    \item Although Jervis Bay is classified as an ``Other Territory'' by the ABS, we altered its SA2 code to include it as part of NSW. 
\end{itemize}

Following the recommendations by Gomez-Rubio \emph{et al.} \cite{RN35}, Mohadjer \emph{et al.} \cite{RN433} and Banerjee \emph{et al.} \cite{RN390}, the ICAR prior for $\mathbf{s} = (s_1, \dots, s_M)$ was declared for all areas and thus the $s_i$'s for non-sampled areas are implicitly imputed during MCMC. We used the Stan implementation of the ICAR prior given by Morris \emph{et al.} \cite{RN397}.  

Given that we did not export, explore or include SA3 level census covariates, we employed a random effect at the SA3 level as well \cite{RN35}. Let $\eta_{h[i]}$ be the random effect for SA3 $h$. We found no discernible improvement in model fit when using spatial priors at the SA3 level, so we reverted to a standard normal prior $\jdist{N}{ 0, \sigma_{\eta}^2 }$ instead.

\subsubsection{Stage 2 model}
The stage 2 model used for all risk factor measures was,

\begin{eqnarray}
    \hat{\boldsymbol{\theta}}_i\jut{S1} & \sim & \jdist{N}{ \theta_i , \bar{\tau}_i\jut{S1} + \widehat{\text{v}} \lb{ \hat{\theta}_i\jut{S1} } }^{1/\tilde{T}} \label{eq:model_s2}
    \\
    \theta_i & = & \mathbf{Z}_i \boldsymbol{\Lambda} + \alpha \gamma_{i} + \mathbf{G}_i \boldsymbol{\Gamma}_{r[i]} + \zeta_i + \eta_{h[i]} \nonumber
    \\
    \hat{\gamma}_{i} & \sim & \jdist{N}{ \gamma_{i}, \widehat{\text{v}} \lb{ \hat{\gamma}_{i} } } \nonumber
    \\
    \zeta_i & \sim & \jdist{BYM2}{ \mathbf{W}, \kappa, \rho, \sigma_{\zeta}^2 } \nonumber
    \\
    \eta_{h[i]} & \sim & \jdist{N}{ 0, \sigma_{\eta}^2 }, \nonumber
\end{eqnarray}

\noindent where the prevalence estimate for the $i$th SA2 was given by the posterior distribution of $\mu_i = \text{logit}^{-1} \lb{\theta_i}$. The remaining notation and details on the priors used are given in Section 3.2 in the main paper. 

\subsubsection{Generalized variance functions} 
As discussed in our previous work \cite{self_cite}, the stage 1 sampling variances, $\psi_i\jut{S1}$, can be unrealistically low for unstable areas. To correct for this we used generalized variance functions (GVF), which are commonly used in area level SAE modelling to smooth or impute unstable sampling variances \cite{RN430, RN28, RN137}. The GVF used in this work is a Bayesian linear model, fitted to the stable areas, and is a generalisation of the GVF used by Das \emph{et al.} \cite{Das2022} to a fully Bayesian framework \cite{RN536}.

Let $\mathbf{L}$ be the design matrix and $\boldsymbol{\omega}$ the corresponding regression coefficients for the linear model. We used the log of the SA2 sample size, the log of the SA2 population, PC1 and the posterior median of $\hat{\theta}_i\jut{S1}$ as covariates. The GVF is,   

\begin{eqnarray}
    \jdist{log}{\sqrt{\psi_i\jut{S1}}} & \sim & \jdist{N}{ \mathbf{L}_i \boldsymbol{\omega}, \sigma_{\text{gvf}}^2}. \label{eq:gvf}
\end{eqnarray}

During MCMC we impute values for the unstable S1 sampling variances via $\lb{\jdist{exp}{ \mathbf{L}_i \boldsymbol{\omega} + 0.5\sigma_{\text{gvf}}^2}}^2$ \cite{Das2022}.

\subsection{Validation}

\subsubsection{Bayesian benchmarking}

Let $\widehat{C}_k^D$ and $\widehat{\text{v}}\left( \widehat{C}_k^D \right)$ be the direct Hajek \cite{RN571} estimate and sampling variance for benchmark $k = 1, \dots, K$. The goal of internal benchmarking in this work was to ensure that the population-weighted modelled estimate,

\begin{equation}
    \widetilde{C}_k = \frac{ \sum_{i \in S_k} \mu_i N_i }{\sum_{i \in S_k} N_i}  \label{eq:bench}
\end{equation}

\noindent was in approximate agreement with $\widehat{C}_k^D$. Note that $S_k$ denotes the set of SA2s in benchmark group $k$. Fully Bayesian benchmarking takes the form,

\begin{equation}
    \widetilde{C}_k \sim N\lb{ \widehat{C}_k^D, \lb{p \times \sqrt{ \widehat{\text{v}}\lb{ \widehat{C}_k^D }}}^{2} } \label{eq:bay_bench}
\end{equation}

\noindent where $p > 0$ was a discrepancy measure used to assert the desired level of concordance. We set $p = 0.5$ for both the state and major-by-state benchmarks. Note that benchmarking was not used during model selection and was only applied once a final model had been selected.

\subsubsection{Validation for very remote regions and the Northern Territory} \label{sec:remote_valid}

The ABS warns that direct estimates for the Northern Territory (NT) could be inaccurate. This is because 20\% of the population of the NT live in very remote or discrete First Nations Australian communities which were purposely excluded from the sampling frame \cite{RN508}. According to data from 2006, the NT had the highest proportion of First Nations Australian people residing in discrete communities, 41,681 (45\%) \cite{RN606}. Of the 40 SA2s across Australia with populations that were composed of over 25\% First Nations Australian people, 30 (75\%) were very remote and 18 (45\%) were in the NT, respectively. 

Furthermore, given the following warning from the ABS \cite{RN508},  
\begin{quote}
    \emph{\dots the estimates from the survey, do not (and are not intended to) match estimates of the total Australian estimated resident population (which include persons living in Very Remote areas of Australia and persons in non-private dwellings, such as hotels) obtained from other sources}
\end{quote} 
we could not use internal validation (benchmarking) for very remote SA2s.

\subsubsection{External validation}

External validation was performed by comparing the estimates to those from the SHAA at the PHA level and the overall trends observed in the modelled results with the general findings from other Australian health survey. It was not possible to match our definitions and age subset to those from the external data. Thus, comparisons were general in nature, assuming that general trends in risk factor prevalence would be evident.  

The graphical summary presented in Section A of the Additional File provides an overview of the broader associations between the prevalence of cancer risk factors, the SES index, and remoteness. The figure summarises the trends available from published reports and/or digital platforms ranging across six selected Australian health surveys. 

Due to the scope of the 2017-18 National Health Survey (NHS) \cite{RN508}, internal validation would not have been valid for very remote SA2s and those in the Northern Territory (n = 103) (see \cref{sec:remote_valid} above). Although these SA2s only accounted for approximately 1.5\% of the 2017-2018 adult population, they were of particular interest in this work since no estimates currently exist for many of these very remote areas.

\subsubsection{Results}

\begin{figure}[h]
    \centering
    \includegraphics[width=0.8\textwidth]{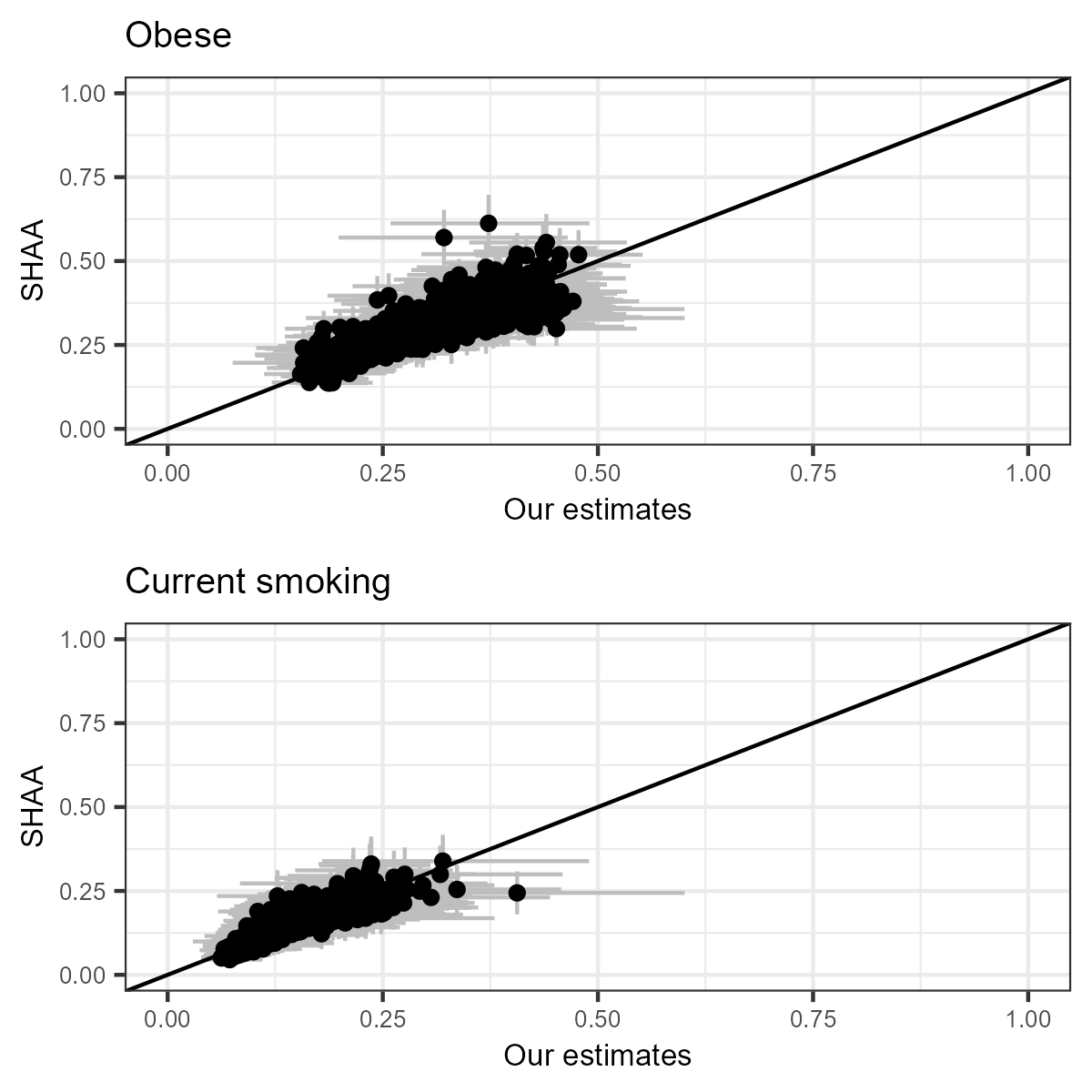}
    \caption{\small Scatter plots for obese and current smoking, where the posterior median PHA-level prevalence estimates from this work ($x$-axis) are compared to the corresponding PHA-level estimates from the SHAA ($y$-axis). Each point also has a 95\% highest density interval and 95\% confidence interval from this work and the SHAA, respectively. The black diagonal line represents when the two axis are equal. Note that the definitions used for our estimates and those from the SHAA are similar but not identical. Furthermore, we report proportions, while the SHAA reports age-standardised rates which have been converted to proportions for comparison.}
    \label{fig:scattersha}
\end{figure}

Our SA2 level modelled estimates corroborated well with the trends reported from the external surveys and SHAA estimates (\cref{supp:other_aus_data}). The strongest evidence of agreement was for current smoking and obese as the risk factor measure definitions from this work and those from the SHAA were similar (see \cref{fig:scattersha}). Moreover, the trends by socioeconomic status and remoteness from the external surveys agreed strongly with the modelled estimates.

\begin{figure}[h]
    \centering
    \includegraphics[width=0.8\textwidth]{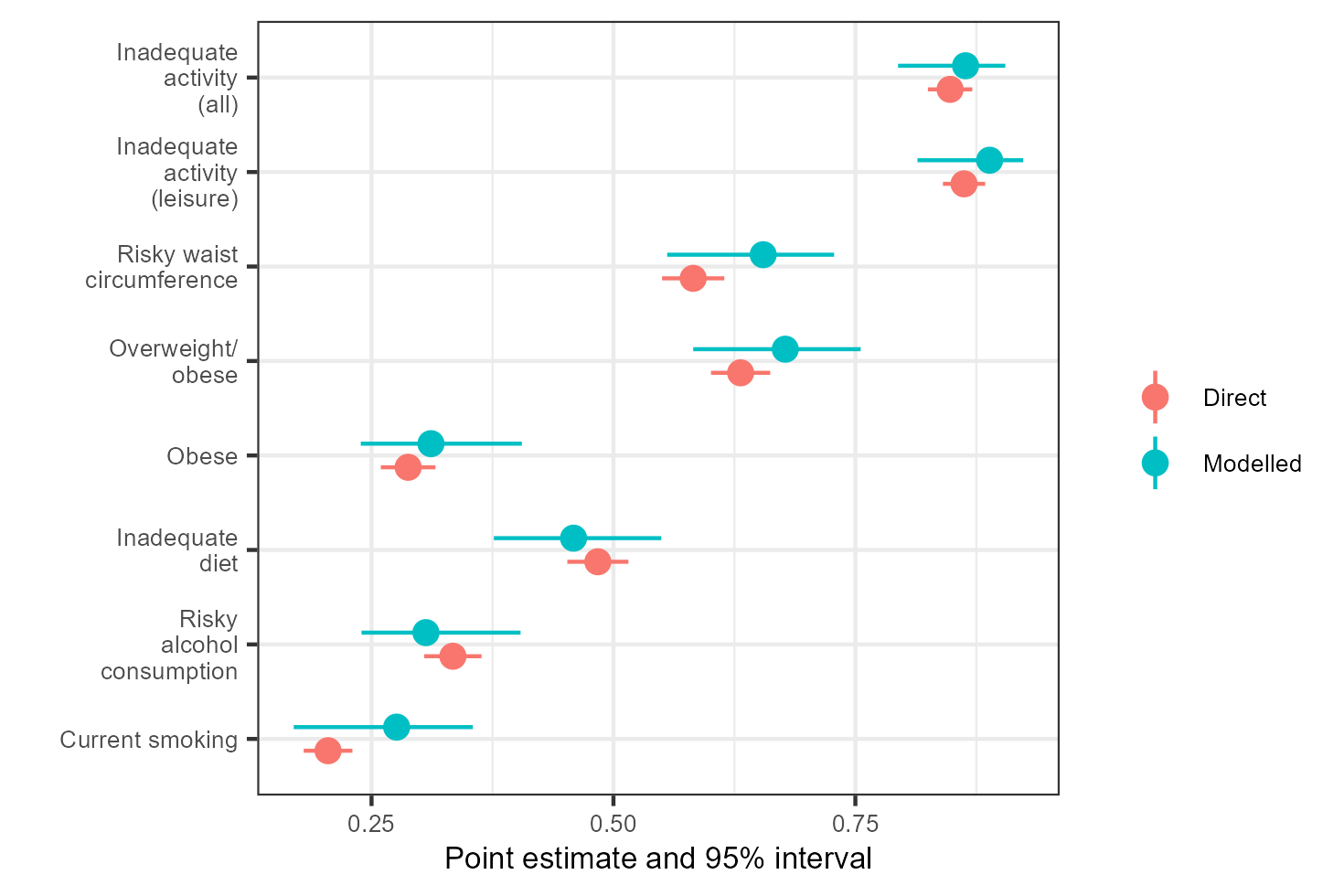}
    \caption{\small Comparison of the aggregated direct and modelled estimate for the non-benchmarked areas. For the modelled points, the posterior median and 95\% highest posterior density interval (HPDI) is used, while for the direct estimates the direct estimate and 95\% confidence interval is used.}
    \label{fig:benchmarkcomp}
\end{figure}

\begin{figure}[H]
    \centering
    \includegraphics[width=0.8\textwidth]{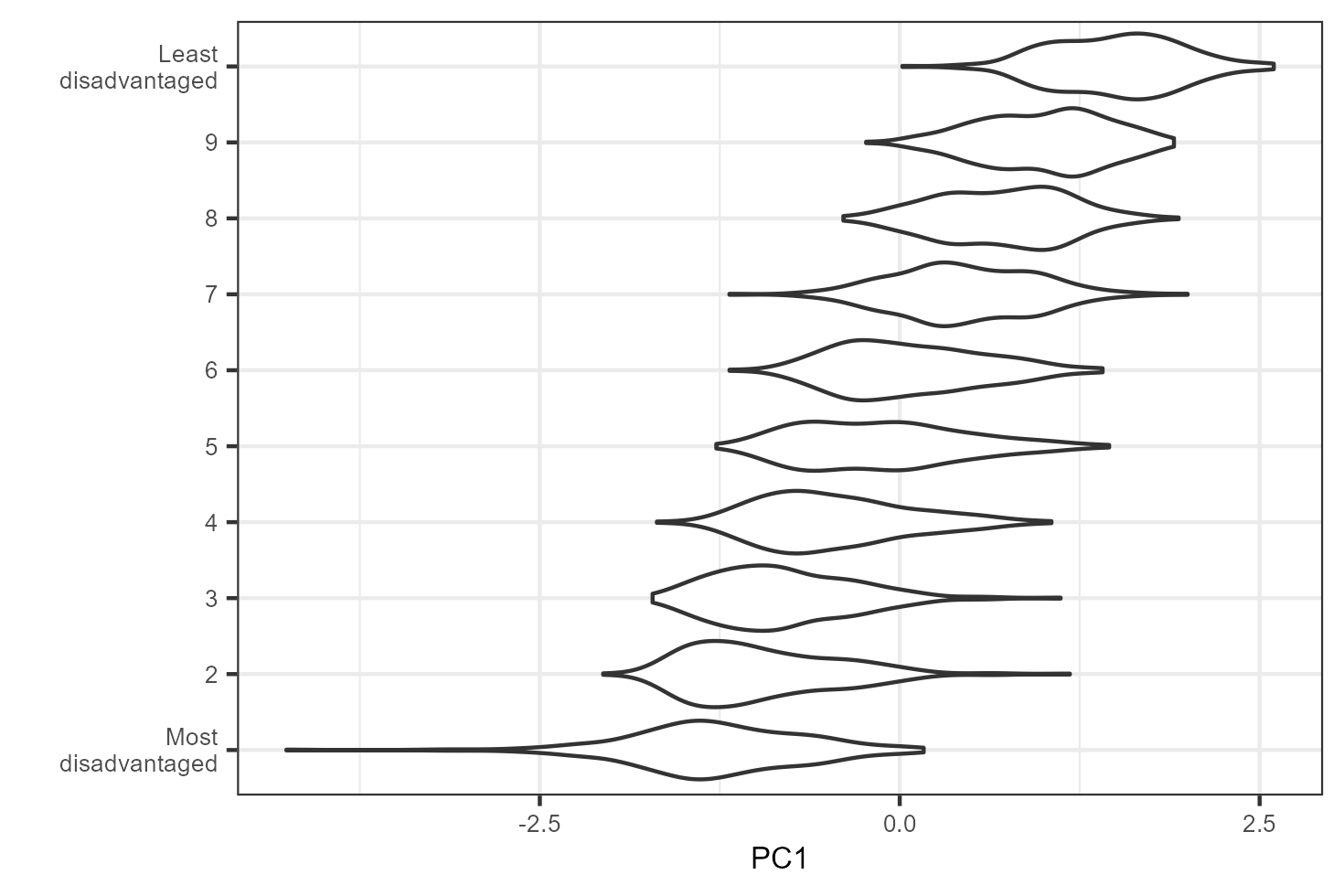}
    \caption{Distribution of principal component 1 (PC1) by SES index quintiles (IRSD).}
    \label{fig:pc1_vs_irsd}
\end{figure}

The level of agreement with external data was generally similar for benchmarked and non-benchmarked SA2s, with non-benchmarked SA2s having considerably higher uncertainty on average as expected (see \cref{supp:add_plots}). This was particularly true when deriving a single estimate for the non-benchmarked SA2s (see \cref{fig:benchmarkcomp}). Our model-based estimate of this quantity was generally very different (often higher) and had far greater uncertainty than the direct estimate which, due to the scope of the survey, is advertised as non-robust \cite{RN508}. This disparity was expected as the direct estimate did not capture any very remote areas, with most data coming from the capital of the Northern Territory, Darwin. In contrast, our model-based estimate captured SA2 estimates from Darwin \emph{and} very remote Australia where population health is generally poorer \cite{RN647}. 

The fully Bayesian benchmarking approach performed well, with minimal changes in point estimates. We also observed a relatively even spread of increased and decreased posterior uncertainty, ranging from about a 13\% increase to a 6\% decrease in the width of HPDIs when the benchmarks were accommodated (see \cref{table:benchmark_comp}). 

% Affects of benchmarking
\begin{table}[H]
\caption{Descriptive statistics comparing the SA2 level prevalence estimates from the benchmarked and non-benchmarked models. The first column summarises the mean absolute relative difference (MARD) between the posterior median proportions from the benchmarked and non-benchmarked models. The second column of the table compares the width of the 95\% highest posterior density intervals (HPDI) by dividing the non-benchmarked intervals sizes by the benchmarked interval sizes. Thus, a value greater than 1 indicates that the benchmarked model provides narrower HPDIs. We derived the ratio for all SA2 areas, but only display the median and interquartile range in the table.}
\label{table:benchmark_comp}
\centering
\begin{tabular}{rrr} 
 & \makecell[r]{MARD\\$\times 10$} & \makecell[r]{Median (IQR) of\\relative width of HPDIs} \\ 
\hline\hline
\makecell[r]{Inadequate\\activity (leisure)} & 0.13 & 1.01 (0.95, 1.07) \\
\hline
\makecell[r]{Inadequate\\activity (all)} & 0.15 & 1.06 (1.01, 1.12) \\
\hline
\makecell[r]{Risky\\alcohol\\consumption} & 0.60 & 1.00 (0.96, 1.05) \\
\hline
\makecell[r]{Inadequate\\diet} & 0.38 & 1.05 (1.02, 1.09) \\
\hline
Obese & 0.51 & 0.95 (0.90, 1.01) \\
\hline
\makecell[r]{Overweight/\\obese} & 0.23 & 0.96 (0.92, 1.02) \\
\hline
Current smoking & 0.70 & 0.95 (0.90, 1.01) \\
\hline
\makecell[r]{Risky waist\\circumference} & 0.38 & 0.87 (0.81, 0.94) \\
\hline\hline
\end{tabular}
\end{table}

%% NEW SECTION %% --------------------------------------------------------------------------------------------
\newpage
\section{Additional data details} \label{supp:data_details}

\subsection{Population Health Areas (PHA)}
There is substantial heterogeneity of the component SA2s within each PHA in terms of population size, socioeconomic status and remoteness. For example, 71 (6\%) and 491 (42\%) PHAs are composed of SA2s with different remoteness categories and different SES index deciles, respectively. The average standard deviation of SA2 population sizes within each PHA is 3254. The affect of aggregating these heterogeneous SA2s into PHAs is a loss of information, where the extent to the loss is somewhat dependent on the number of SA2s within each PHA. In 2016, there were 451 (39\%) and 242 (21\%) PHAs with 2 and more than 2 component SA2s, respectively. 

\subsection{Estimates from the SHAA}
As described in Section 2.2.3 of the main paper, we also obtained prevalence estimates and measures of uncertainty for six risk factors from the SHAA \cite{RN113} at the Primary Health Network (PHN) and PHA level for adults. There are 31 PHNs across Australia, with each comprising multiple SA2s (median of 72 and IQR of 53 to 96) and varying population sizes (median 505000 and IQR of 401000 to 808000, averaged over the 2017-2018 adult population). The SHAA's estimates and their respective uncertainties were included in the models and thus each SA2 was assigned to a PHN using ABS concordance files. To obtain a single value for any SA2 that overlapped several PHNs, we took the weighted mean of the PHN estimates using the given concordance ratios. Note that the SHAA does not provide estimates for the Western Queensland PHN. Based on its similar remoteness characteristics, estimates for this PHN were copied from those from the Northern Territory PHN. 

% Groups for the individual level categorical covariates
\begin{table}[H]
\caption{Categories for the individual level covariates used in this research (see \cref{sec:s1}). Bolded categories denote the reference group. Most of the categories for these covariates were derived by the Australian Bureau of Statistics \cite{RN598}. For details of the definitions, we refer the reader to publicly available data dictionaries \cite{abs_census_dictionary_2016}.}
\label{table:categorical_vars}
    \centering
    \begin{tabular}{>{\RaggedLeft\hspace{0pt}}m{0.3\linewidth}>{\hspace{0pt}}m{0.038\linewidth}>{\hspace{0pt}}m{0.6\linewidth}} 
    \cline{2-3}
    & \# & Categories \\ 
    \hline\hline
    Sex & 2 & \textbf{Male}, Female \\
    Age & 9 & \textbf{15-19}, 20-24, 25-34, 35-44, 45-54, 55-64, 65-74, 75-84, 85+ \\
    Registered marital status & 5 & \textbf{Married}, Never married, Widowed, Divorced, Separated \\
    Hypertension & 3 & Has hypertension, \textbf{Does not have hypertension} \\
    &  & Not applicable \\
    \makecell[r]{Language spoken\\at home} & 2 & \textbf{English}, Other \\
    Depression status & 4 & All or most of the time, Some or a little of the time \\
    &  & \textbf{None of the time}, Not available \\
    Kessler score & 3 & High/very high, \textbf{Low/moderate}, Not available \\
    Occupation & 10 & Community and personal service workers \\
    &  & \textbf{Unemployed or not in labourforce} \\
    &  & Clerical and administrative workers \\
    &  & Professsionals, Technicians and trades workers \\
    &  & Labourers, Sales workers \\
    &  & Machinery operators and drivers \\
    &  & Managers, Other \\
    \makecell[r]{Self-assessed\\health} & 5 & Poor, Fair, Good, Very good, \textbf{Excellent} \\
    Qualification & 4 & Certificate, \textbf{No non-school qualification or not determined} \\
    &  & Bachelor/Diploma, Postgraduate \\
    \makecell[r]{High school\\completion status} & 4 & 
    \textbf{Year 12 or equivalent}, Year 11 or equivalent \\
    &  & Year 10 or equivalent, At most year 9 or equivalent \\
    \hline\hline
    \end{tabular}
\end{table}

% ACA risk factor measures used to construct NORF variable
\begin{table}[H]
\caption{Description of the cancer risk factor measures used to construct the non-outcome risk factor (NORF) categorical variable, which was uniquely defined for each cancer risk factor measure. The NORF variable was used in the stage 1 model described in \cref{sec:s1}.}
\label{table:norf_concor}
    \centering
    \begin{tabular}{lllllll}
    &  & \multicolumn{5}{l}{Measures included in NORF categorical covariate} \\
    \cline{3-7}
     &  & \makecell[l]{Current\\smoking} & \makecell[l]{Risky\\alcohol\\consumption} & \makecell[l]{Inadequate\\diet} & Overweight & \makecell[l]{Inadequate\\activity (all)} \\ 
    \hline\hline
    \multirow{8}{0.05\linewidth}{\rotatebox[origin=c]{90}{Risk factor measures}} & Current smoking &  & \checkmark & \checkmark & \checkmark & \checkmark \\
     & \makecell[l]{Risky\\alcohol\\consumption} & \checkmark &  & \checkmark & \checkmark & \checkmark \\
     & \makecell[l]{Inadequate\\diet} & \checkmark & \checkmark &  & \checkmark & \checkmark \\
     & Obese & \checkmark & \checkmark & \checkmark &  & \checkmark \\
     & \makecell[l]{Overweight/\\obese} & \checkmark & \checkmark & \checkmark &  & \checkmark \\
     & \makecell[l]{Risky\\waist\\circumference} & \checkmark & \checkmark & \checkmark &  & \checkmark \\
     & \makecell[l]{Inadequate\\activity\\(leisure)} & \checkmark & \checkmark & \checkmark & \checkmark &  \\
     & \makecell[l]{Inadequate\\activity\\(all)} & \checkmark & \checkmark & \checkmark & \checkmark &  \\
    \hline\hline
    \end{tabular}
\end{table}

% Definitions for SHAA risk factor measures
\begin{table}[H]
\caption{Definitions for the six cancer risk factor measures used from the SHAA \cite{RN113}.}
\label{table:shaa_rf_defin}
    \begin{tabular}{>{\RaggedLeft\hspace{0pt}}m{0.248\linewidth}>{\hspace{0pt}}m{0.692\linewidth}} 
    \hline\hline
    SHAA variable & Definition \\ 
    \hline\hline
    Current smokers & Those who were classed as current smokers. \\
    Risky alcohol consumption & Those who consumed more than two standard alcoholic drinks per day on average. \\
    Adequate fruit intake & Those who consumed at least 2 serves of fruit per day on average. \\
    Obese & Those with a BMI greater or equal to 30. \\
    Overweight & Those with a BMI that was between 25 and less than 30. \\
    \makecell[r]{Inadequate physical\\activity} & Those who undertook low, very low or no exercise in the week prior to the survey. \\
    \hline\hline
    \end{tabular}
\end{table}

% Concordance between ACA and SHAA risk factor measures
\begin{table}[H]
\caption{Table summarising which of the variables from the SHAA was used as fixed effects in the stage 2 model (\cref{sec:s1}) for each of the cancer risk factor measures.}
\label{table:shaa_rf_concor}
\centering
    \begin{tabular}{llllllll}
     &  & \multicolumn{6}{>{\centering\arraybackslash\hspace{0pt}}m{0.589\linewidth}}{SHAA variables} \\ 
    \cline{3-8}
     &  & Smoking & Alcohol & Fruit & Obese & Overweight & Exercise \\ 
    \hline\hline
    \multirow{8}{0.05\linewidth}{\rotatebox[origin=c]{90}{Risk factor measures}} & \makecell[l]{Current\\smoking} & \checkmark &  &  &  &  &  \\
     & \makecell[l]{Risky\\alcohol\\consumption} &  & \checkmark  &  &  &  \\
     & \makecell[l]{Inadequate\\diet} &  &  & \checkmark &  &  &  \\
     & Obese &  &  &  & \checkmark  &  &  \\
     & \makecell[l]{Overweight/\\obese} &  &  &  &  & \checkmark &  \\
     & \makecell[l]{Risky\\waist\\circumference} &  &  &  &  & \checkmark &  \\
     & \makecell[l]{Inadequate\\activity\\(leisure)} &  &  &  &  &  & \checkmark \\
     & \makecell[l]{Inadequate\\activity\\(all)} &  &  &  &  &  & \checkmark \\
    \hline\hline
    \end{tabular}
\end{table}

\subsection{Evidence classifications}

We followed the work by Gramatica \emph{et al.} \cite{RN568} and Congdon \emph{et al.} \cite{RN617}, who used an adaption of the LISA (Local Indicator of Spatial Association) clustering approach \cite{RN11} to identify clusters of areas with significantly high (or low) prevalence. The LISA clusters were computed from the posterior draws by first deriving the deviation of the prevalence from the national average, $z^{(t)}_i = \mu^{(t)}_i - \hat{\mu}^D$, and then calculating the spatial lag of the deviation, $\mathbf{L}^{(t)} = \mathbf{z}^{(t)} \lb{ \mathbf{W}^{*} }^T$, where $\mathbf{z}^{(t)} = \lb{ z^{(t)}_1, \dots, z^{(t)}_M }$ was a $M$-dimensional row vector and $\mathbf{W}^{*}$ the row-standardized version of the adjacency neighborhood matrix described in \cref{sec:re}. 

To derive the evidence classification measure, SA2s were classified into one of four exclusive groups \cite{RN617}:
\begin{itemize}
    \item High-cluster (``HC'') if both $\frac{1}{T} \sum_t \mathbb{I} \lb{z^{(t)}_i > 0} > 0.8$ and $\frac{1}{T} \sum_t \mathbb{I} \lb{L^{(t)}_i > 0} > 0.8$,
    \item High (``H'') if $\frac{1}{T} \sum_t \mathbb{I} \lb{z^{(t)}_i > 0} > 0.8$ and $\frac{1}{T} \sum_t \mathbb{I} \lb{L^{(t)}_i > 0} \leq 0.8$
    \item Low (``L'') if both $\frac{1}{T} \sum_t \mathbb{I} \lb{z^{(t)}_i > 0} < 0.2$ and $\frac{1}{T} \sum_t \mathbb{I} \lb{L^{(t)}_i > 0} \geq 0.2$, 
    \item Low-cluster (``LC'') if both $\frac{1}{T} \sum_t \mathbb{I} \lb{z^{(t)}_i > 0} < 0.2$ and $\frac{1}{T} \sum_t \mathbb{I} \lb{L^{(t)}_i > 0} < 0.2$.
\end{itemize}

%% NEW SECTION %% --------------------------------------------------------------------------------------------
\newpage
\section{Model building} \label{supp:mb}

As described in Section 3.1 in the main paper, a single model specification was chosen to be applied to all eight risk factor measures.

\subsection{Stage 1 model} \label{sec:mb_s1}

To guide model selection for the stage 1 model, we used the smoothing ratio ($SR$) and area linear comparison ($ALC$) metrics, both of which indicate the level of concordance between the observed and smoothed data from the stage 1 model \cite{self_cite}. The $SR$ and the $ALC$ smoothing metrics achieve this by comparing the observed individual level data to the predicted probabilities and the area level direct estimates to the stage 1 estimates, respectively. Higher values of both are preferred. We used the posterior median of the $SR$, which for posterior draw $t$ is given by, 

\begin{equation}
    SR^{(t)} = 1- \frac{
    \sum_{i=1}^m \left| \frac{\sum_{j = 1}^{n_i} w_{ij} \left( y_{ij} - \pi^{(t)}_{ij} \right) }{n_i} \right|}{
    \sum_{i=1}^m \left| \frac{\sum_{j = 1}^{n_i} w_{ij} \left( y_{ij} - \hat{\mu}^D \right) }{n_i} \right|}, \label{eq:sr}
\end{equation}

\noindent where $\hat{\mu}^D$ is the overall prevalence. The $ALC$ is equal to the regression coefficient when we regress the posterior median of $\hat{\mu}\jut{S1}_i$ on $\hat{\mu}^D_i$ with weights $1/\psi_i^D$.

For survey-only and census fixed effects, a frequentist weighted logistic model with no random effects was used for variable selection. The primary focus in selecting these was to maximize both the $ALC$ and the $SR$, with AIC and BIC considered as secondary criteria. When validating frequentist decisions with Bayesian inference we used leave-one-out cross-validation (LOOCV) via Pareto-importance sampling \cite{RN116}. In general the selected variables were consistently good predictors across all the risk factor measures. See \cref{table:model_building_s11} and \ref{table:model_building_s12} for model metrics for the stage 1 model for all risk factor measures. 

To select the variables to include in the demographic-health categorical covariate, we explored a large range of possible interactions by fitting the resulting categorical variable using frequentist weighted logistic mixed models. The selected set of covariates showed persistent benefits, in terms of the $ALC$ and the $SR$, across all risk factor measures. Similar to our previous work \cite{self_cite}, we found that fitting the NORF categorical variable as a random effect improved model fit across the board (see \cref{supp:data_details}). 

The fixed standard deviation of the residual error ($\sigma_e$) was utilized to address stage 1 models in cases where the $SR$ or the $ALC$ was too low. In previous work we found that optimal performance of the area level MRRMSE and MARB generally occurred when the $0.55 < ALC < 0.75$ and the $0.4 < SR < 0.7$, with performance atrophy outside these bounds. We found that values in the lower end of these bounds generally provided narrower and more reliable credible intervals. 

As discussed in our previous work \cite{self_cite}, although the stage 1 model is designed to smooth the observed data, oversmoothing (e.g. when the $SR$ or the $ALC$ are close to zero) can significantly affect the model performance. Thus, the residual error scale was used to purposely overfit the first-stage model in order to improve the predictions from the second-stage model. In this work, we set $\sigma_e = 2$, which provided a range of the $SR$ from 0.37 to 0.45 and the $ALC$ from 0.55 to 0.70 across the risk factor measures. 

\subsection{Stage 2 model} \label{sec:mb_s2}
Unlike other SAE applications where predictive accuracy can be reasonably assessed via scoring rules \cite{RN533} or leave-one-out cross validation \cite{RN116}, given the unstable nature of the direct SA2 level estimates, we assessed performance of our stage 2 model by comparing direct and modelled estimates at two aggregated levels; the SA4 and major-by-state benchmark level. At these levels of aggregation, the direct estimates were plausibly treated as ground truth. For example, approximately 73\% and 100\% of the direct smoking estimates at the SA4 and major-by-state benchmark level had coefficients of variation below 25\%. Direct estimates for other risk factors exhibited similar or superior certainty.  

Under the assumption that the SA4 and major-by-state benchmark level direct estimates were the truth, we used the Bayesian analogue of mean absolute relative bias (MARB) and mean relative root mean square error (MRRMSE) to assess model performance. Below we give details at the major-by-state benchmark level using the notation given in Section 3.3.1 of the main paper. 

\begin{eqnarray}
    \text{MARB} & = & \frac{1}{K} \sum_{k=1}^{K} \left| \frac{ \frac{1}{T} \sum_{t=1}^T \lb{ \widehat{C}_k^D - \widetilde{C}^{(t)}_k} }{\widehat{C}_k^D} \right| \label{met:b_marb}
    \\
    \text{MRRMSE} & = & \frac{1}{K} \sum_{k=1}^K \frac{\sqrt{  \frac{1}{T} \sum_{t=1}^T \lb{ \widehat{C}_k^D - \widetilde{C}^{(t)}_k }^2  }}{\widehat{C}_k^D} \label{met:b_mrrmse}
\end{eqnarray} 

In addition to MARB and MRRMSE, we also derived the interval overlap probability (IOP) for each SA4 and major-by-state benchmark estimate. An IOP of 1 indicated that the modelled 95\% highest posterior density interval (HPDI) was entirely contained within the direct estimate 95\% confidence interval; the optimal situation. To summarizethe IOPs, we used the mean IOP (MIOP). Given that SA4 level direct estimates were still relatively unstable, we placed greater priority on the MIOP than the MARB and MRRMSE during model selection. See \cref{table:model_building_s21} and \ref{table:model_building_s22} for model metrics for the stage 2 model for all risk factor measures. 

% Model building: Stage 1.1
\newpage
\pagestyle{empty}
\begin{table}[H]
\caption{The progression of the performance metrics (smoothing ratio (SR), area linear comparison (ALC) and leave-one-out cross validation (LOOCV)) as components of the linear predictor for the stage 1 model are included. For each risk factor, the first row gives the performance metrics when only the intercept was included, the second the performance metrics when \emph{both} the intercept and fixed effects are added and so forth. The final row gives the performance metrics for the final stage 1 model. NORF: Non-outcome risk factor, DH: Demographic health, SA2: Statistical area level 2}
\label{table:model_building_s11}
\centering
\begin{tabular}{rrccc} 
    \cline{3-5}
     &  & ALC & SR & LOOCV \\ 
    \hline\hline
    \multirow{7}{*}{\makecell[r]{Risky\\alcohol\\consumption}} & Intercept only & 0.00 & 0.00 & -10218.90 \\
     & Fixed effects (FE) & 0.25 & 0.11 & -9343.83 \\
     & \makecell[r]{NORF random\\effects (RE)} & 0.27 & 0.12 & -9168.88 \\
     & DH RE & 0.28 & 0.13 & -9253.30 \\
     & SA2 RE & 0.59 & 0.30 & -9458.02 \\
     & \makecell[r]{Residual error\\(sd = 1)} & 0.62 & 0.35 & -9891.93 \\
     & \makecell[r]{Residual error\\(sd = 2)} & 0.69 & 0.45 & -12373.01 \\ 
    \hline
    \multirow{7}{*}{\makecell[r]{Inadequate\\diet}} & Intercept only & 0.00 & 0.00 & -11898.00 \\
     & FE & 0.09 & 0.04 & -11583.33 \\
     & NORF RE & 0.10 & 0.05 & -11413.50 \\
     & DH RE & 0.14 & 0.07 & -11463.79 \\
     & SA2 RE & 0.52 & 0.23 & -11670.57 \\
     & \makecell[r]{Residual error\\(sd = 1)} & 0.57 & 0.30 & -12253.55 \\
     & \makecell[r]{Residual error\\(sd = 2)} & 0.66 & 0.41 & -15639.11 \\ 
    \hline
    \multirow{7}{*}{Obese} & Intercept only & 0.00 & 0.00 & -10758.03 \\
     & FE & 0.26 & 0.10 & -9914.78 \\
     & NORF RE & 0.26 & 0.10 & -9850.31 \\
     & DH RE & 0.31 & 0.13 & -9883.91 \\
     & SA2 RE & 0.60 & 0.27 & -10126.73 \\
     & \makecell[r]{Residual error\\(sd = 1)} & 0.63 & 0.33 & -10677.55 \\
     & \makecell[r]{Residual error\\(sd = 2)} & 0.70 & 0.43 & -13525.59 \\ 
    \hline
    \multirow{7}{*}{Current smoking} & Intercept only & 0.00 & 0.01 & -7440.70 \\
     & FE & 0.25 & 0.12 & -6564.64 \\
     & NORF RE & 0.27 & 0.14 & -6314.59 \\
     & DH RE & 0.31 & 0.17 & -6381.14 \\
     & SA2 RE & 0.56 & 0.32 & -6626.02 \\
     & \makecell[r]{Residual error\\(sd = 1)} & 0.57 & 0.36 & -6893.80 \\
     & \makecell[r]{Residual error\\(sd = 2)} & 0.62 & 0.45 & -8467.00 \\
    \hline\hline
\end{tabular}
\end{table}

% Model building: Stage 1.2
\begin{table}[H]
\caption{See description in \cref{table:model_building_s11}}
\label{table:model_building_s12}
\centering
\begin{tabular}{rrccc} 
    \cline{3-5}
     &  & ALC & SR & LOOCV \\ 
    \hline\hline
    \multirow{7}{*}{\makecell[r]{Inadequate\\activity (leisure)}} & Intercept only & 0.00 & 0.00 & -7153.47 \\
     & Fixed effects (FE) & 0.16 & 0.06 & -6799.33 \\
     & \makecell[r]{NORF random\\effects (RE)} & 0.18 & 0.07 & -6657.59 \\
     & DH RE & 0.21 & 0.09 & -6670.94 \\
     & SA2 RE & 0.47 & 0.22 & -6864.23 \\
     & \makecell[r]{Residual error\\(sd = 1)} & 0.48 & 0.28 & -7050.93 \\
     & \makecell[r]{Residual error\\(sd = 2)} & 0.56 & 0.39 & -8433.78 \\ 
    \hline
    \multirow{7}{*}{\makecell[r]{Inadequate\\activity (all)}} & Intercept only & 0.00 & 0.00 & -7622.71 \\
     & FE & 0.15 & 0.06 & -7238.44 \\
     & NORF RE & 0.17 & 0.06 & -7101.44 \\
     & DH RE & 0.20 & 0.08 & -7114.28 \\
     & SA2 RE & 0.45 & 0.21 & -7293.53 \\
     & \makecell[r]{Residual error\\(sd = 1)} & 0.47 & 0.27 & -7487.59 \\
     & \makecell[r]{Residual error\\(sd = 2)} & 0.55 & 0.37 & -8983.52 \\ 
    \hline
    \multirow{7}{*}{\makecell[r]{Overweight/\\obese}} & Intercept only & 0.00 & 0.00 & -11030.30 \\
     & FE & 0.21 & 0.12 & -9929.64 \\
     & NORF RE & 0.21 & 0.13 & -9893.96 \\
     & DH RE & 0.24 & 0.15 & -9931.15 \\
     & SA2 RE & 0.49 & 0.26 & -10128.58 \\
     & \makecell[r]{Residual error\\(sd = 1)} & 0.52 & 0.32 & -10521.22 \\
     & \makecell[r]{Residual error\\(sd = 2)} & 0.61 & 0.42 & -13075.34 \\ 
    \hline
    \multirow{7}{*}{\makecell[r]{Risky waist\\circumference}} & Intercept only & 0.00 & 0.00 & -10503.49 \\
     & FE & 0.26 & 0.13 & -9400.15 \\
     & NORF RE & 0.26 & 0.14 & -9354.67 \\
     & DH RE & 0.29 & 0.16 & -9388.27 \\
     & SA2 RE & 0.60 & 0.30 & -9617.15 \\
     & \makecell[r]{Residual error\\(sd = 1)} & 0.62 & 0.36 & -9983.51 \\
     & \makecell[r]{Residual error\\(sd = 2)} & 0.70 & 0.46 & -12439.92 \\
    \hline\hline
\end{tabular}
\end{table}

% Model building: Stage 2.1
\begin{table}[H]
\caption{The progression of the performance metrics (given in \cref{sec:mb_s2}) as components of the linear predictor for the stage 2 model are included. For each risk factor, the first row gives the performance metrics when only the intercept was included, the second the performance metrics when \emph{both} the intercept and fixed effects are added and so forth. The final row gives the performance metrics for the stage 2 model used to derive the results given in the main paper. Both the mean absolute relative bias (MARB) and mean relative root mean squared error (MRRMSE) are given as $\times100$. SA2: Statistical area level 2, SA3: Statistical area level 3, SA4: Statistical area level 4}
\label{table:model_building_s21}
\centering
\begin{tabular}{rr|rrr|rrr}
     &  & \multicolumn{3}{c}{SA4} & \multicolumn{3}{c}{Major-by-state} \\ 
    \cline{3-8}
     &  & MARB & MRRMSE & MIOP & MARB & MRRMSE & MIOP \\ 
    \hline\hline
    \multirow{7}{*}{\makecell[r]{Risky\\alcohol\\consumption}} & Intercept only & 24.95 & 25.17 & 0.64 & 8.81 & 9.31 & 0.60 \\
     & \makecell[r]{Fixed\\effects (FE)\\(non-varying)} & 22.62 & 23.37 & 0.64 & 6.76 & 7.62 & 0.53 \\
     & FE (varying) & 13.98 & 16.10 & 0.74 & 5.53 & 7.06 & 0.65 \\
     & \makecell[r]{External\\latent\\field} & 13.24 & 15.50 & 0.76 & 4.66 & 6.51 & 0.73 \\
     & \makecell[r]{SA2 random\\effect (RE)} & 13.22 & 15.54 & 0.76 & 4.60 & 6.54 & 0.72 \\
     & SA3 RE & 13.19 & 15.61 & 0.76 & 4.59 & 6.55 & 0.72 \\
     & Benchmarking & 12.88 & 14.86 & 0.79 & 1.25 & 2.37 & 1.00 \\ 
    \hline
    \multirow{7}{*}{\makecell[r]{Inadequate\\diet}} & Intercept only & 10.02 & 10.18 & 0.81 & 3.38 & 3.73 & 0.75 \\
     & \makecell[r]{FE\\(non-varying)} & 9.50 & 10.03 & 0.81 & 3.56 & 4.38 & 0.71 \\
     & FE (varying) & 8.92 & 10.31 & 0.76 & 2.74 & 4.06 & 0.79 \\
     & \makecell[r]{External\\latent\\field} & 8.92 & 10.40 & 0.76 & 2.76 & 4.14 & 0.82 \\
     & SA2 RE & 8.92 & 10.44 & 0.76 & 2.74 & 4.20 & 0.82 \\
     & SA3 RE & 8.89 & 10.46 & 0.77 & 2.72 & 4.18 & 0.83 \\
     & Benchmarking & 8.82 & 9.94 & 0.81 & 1.14 & 1.77 & 1.00 \\ 
    \hline
    \multirow{7}{*}{Obese} & Intercept only & 23.63 & 23.76 & 0.55 & 12.16 & 12.44 & 0.23 \\
     & \makecell[r]{FE\\(non-varying)} & 16.45 & 17.20 & 0.67 & 5.71 & 6.47 & 0.65 \\
     & FE (varying) & 13.59 & 15.48 & 0.78 & 5.75 & 6.93 & 0.58 \\
     & \makecell[r]{External\\latent\\field} & 13.60 & 15.68 & 0.78 & 5.64 & 6.89 & 0.59 \\
     & SA2 RE & 13.51 & 15.67 & 0.78 & 5.52 & 6.86 & 0.61 \\
     & SA3 RE & 13.45 & 15.69 & 0.79 & 5.47 & 6.82 & 0.62 \\
     & Benchmarking & 12.42 & 14.58 & 0.80 & 1.17 & 2.16 & 1.00 \\ 
    \hline
    \multirow{7}{*}{\makecell[r]{Current\\smoking}} & Intercept only & 42.22 & 42.61 & 0.65 & 18.06 & 18.41 & 0.36 \\
     & \makecell[r]{FE\\(non-varying)} & 26.23 & 27.86 & 0.78 & 8.19 & 9.71 & 0.68 \\
     & FE (varying) & 25.70 & 29.12 & 0.75 & 7.37 & 9.81 & 0.67 \\
     & \makecell[r]{External\\latent\\field} & 25.55 & 29.14 & 0.75 & 7.22 & 9.82 & 0.69 \\
     & SA2 RE & 25.49 & 29.35 & 0.75 & 7.03 & 9.85 & 0.71 \\
     & SA3 RE & 25.43 & 29.41 & 0.75 & 7.00 & 9.81 & 0.71 \\
     & Benchmarking & 25.45 & 28.64 & 0.78 & 1.91 & 3.59 & 1.00 \\
    \hline\hline
\end{tabular}
\end{table}

% Model building: Stage 2.2
\begin{table}[H]
\centering
\caption{See description in \cref{table:model_building_s22}}
\label{table:model_building_s22}
\begin{tabular}{rr|rrr|rrr}
     &  & \multicolumn{3}{c}{SA4} & \multicolumn{3}{c}{Major-by-state} \\ 
    \cline{3-8}
     &  & MARB & MRRMSE & MIOP & MARB & MRRMSE & MIOP \\ 
    \hline\hline
    \multirow{7}{*}{\makecell[r]{Inadequate\\activity\\(leisure)}} & Intercept only & 4.32 & 4.41 & 0.71 & 1.72 & 1.85 & 0.66 \\
     & \makecell[r]{Fixed\\effects (FE)\\(non-varying)} & 3.72 & 3.95 & 0.77 & 1.01 & 1.38 & 0.83 \\
     & FE (varying) & 3.23 & 3.85 & 0.79 & 1.23 & 1.68 & 0.75 \\
     & \makecell[r]{External\\latent\\field} & 3.21 & 3.87 & 0.78 & 1.24 & 1.71 & 0.73 \\
     & \makecell[r]{SA2 random\\effect (RE)} & 3.20 & 3.88 & 0.79 & 1.23 & 1.72 & 0.71 \\
     & SA3 RE & 3.19 & 3.89 & 0.77 & 1.22 & 1.72 & 0.73 \\
     & Benchmarking & 3.10 & 3.62 & 0.81 & 0.27 & 0.58 & 1.00 \\ 
    \hline
    \multirow{7}{*}{\makecell[r]{Inadequate\\activity (all)}} & Intercept only & 4.19 & 4.27 & 0.78 & 1.52 & 1.71 & 0.73 \\
     & \makecell[r]{FE\\(non-varying)} & 3.76 & 4.01 & 0.80 & 1.12 & 1.49 & 0.83 \\
     & FE (varying) & 3.33 & 3.99 & 0.80 & 1.09 & 1.64 & 0.76 \\
     & \makecell[r]{External\\latent\\field} & 3.34 & 4.03 & 0.81 & 1.08 & 1.66 & 0.78 \\
     & SA2 RE & 3.34 & 4.04 & 0.80 & 1.08 & 1.68 & 0.78 \\
     & SA3 RE & 3.33 & 4.05 & 0.80 & 1.06 & 1.68 & 0.78 \\
     & Benchmarking & 3.17 & 3.71 & 0.84 & 0.27 & 0.62 & 1.00 \\ 
    \hline
    \multirow{7}{*}{\makecell[r]{Overweight/\\obese}} & Intercept only & 8.63 & 8.72 & 0.73 & 5.23 & 5.34 & 0.34 \\
     & \makecell[r]{FE\\(non-varying)} & 6.15 & 6.54 & 0.76 & 2.45 & 2.88 & 0.62 \\
     & FE (varying) & 5.13 & 6.12 & 0.83 & 2.49 & 3.12 & 0.60 \\
     & \makecell[r]{External\\latent\\field} & 5.00 & 6.08 & 0.84 & 2.21 & 3.00 & 0.70 \\
     & SA2 RE & 4.99 & 6.09 & 0.84 & 2.16 & 3.00 & 0.72 \\
     & SA3 RE & 4.97 & 6.12 & 0.84 & 2.15 & 3.00 & 0.72 \\
     & Benchmarking & 4.83 & 5.81 & 0.88 & 0.65 & 1.13 & 1.00 \\ 
    \hline
    \multirow{7}{*}{\makecell[r]{Risky\\waist\\circumference}} & Intercept only & 11.48 & 11.57 & 0.58 & 6.24 & 6.36 & 0.29 \\
     & \makecell[r]{FE\\(non-varying)} & 9.29 & 9.62 & 0.70 & 4.00 & 4.34 & 0.44 \\
     & FE (varying) & 7.80 & 8.72 & 0.70 & 4.35 & 4.83 & 0.43 \\
     & \makecell[r]{External\\latent\\field} & 7.68 & 8.70 & 0.71 & 4.33 & 4.90 & 0.52 \\
     & SA2 RE & 7.61 & 8.69 & 0.72 & 4.25 & 4.85 & 0.53 \\
     & SA3 RE & 7.58 & 8.71 & 0.70 & 4.27 & 4.88 & 0.53 \\
     & Benchmarking & 7.32 & 8.55 & 0.68 & 1.20 & 1.61 & 1.00 \\
    \hline\hline
\end{tabular}
\end{table}

%% NEW SECTION %% --------------------------------------------------------------------------------------------
\newpage
\pagestyle{fancy}
\section{Additional results}

\subsection{Smoking}

Similar to previous work in Australia \cite{RN607}, the prevalence of smoking shows strong spatial patterns (\cref{fig:or_smoking,fig:mapprev_smoking,fig:mapprevep_smoking}), with generally lower rates in major cities and less disadvantaged areas. That said, there are several instances where an area classed as a major city shows substantively high rates of smoking. These are predominately most disadvantaged areas in WA, ACT and NSW; some of which are industrial areas with small populations. The median of the point estimates for all very remote areas (0.33) is around 126\% higher than that for the non-very remote areas (0.14), with 75\% higher median CVs (30.4\% compared to 17.3\%). By improving the reach, we garner insights about these disadvantaged areas, albeit with higher entropy.   

\subsection{Alcohol}

The prevalence of risky alcohol consumption shows strong spatial patterns (\cref{fig:or_alcohol,fig:mapprev_alcohol,fig:mapprevep_alcohol}). The results suggest that less disadvantaged areas have higher proportions of risky alcohol consumption, which generally manifests in higher prevalence in major cities. This is supported by other Australian surveys \cite{RN631, RN514} and previous research \cite{RN677, RN113}. The effect of the inverse relationship between higher prevalence of risky alcohol consumption and lower socioeconomic disadvantage is mirrored in Figure 10 in the main paper, where the modelled estimate for the non-benchmarked areas is lower than the direct estimate. 

Although the figures suggests lower than average prevalence in more disadvantaged and remote areas (e.g. the middle of Australia), the estimates for these very remote regions (such as the Northern Territory) have greater uncertainty and thus these are not substantially different to the national average (see the Additional File). Furthermore, unlike other risk factors where prevalence estimates exhibit relative homogeneity within SES and remoteness groups (see \cref{fig:variance_irsdera}), for risky alcohol consumption the estimates exhibit far greater heterogeneity for more disadvantaged areas in major cities. 

\subsection{Diet}

Unlike the other risk factors, inadequate diet (\cref{fig:or_diet,fig:mapprev_diet,fig:mapprevep_diet}) tends to exhibit less dependence with SES and remoteness, which agrees with the broad findings from other Australian surveys \cite{RN631, RN645} (see \cref{supp:other_aus_data}). As a result, inadequate diet was one of the only risk factors for which the modelled estimate for the non-benchmarked areas was lower than the direct estimate. Unlike most of the other risk factor measures where healthy behaviour was generally reserved for major cities, areas with low prevalence of inadequate diet can be found outside major cities. For example some areas in Northern Queensland, Western New South Wales and Western Australia exhibit lower prevalence than the national average. This is supported by the new estimates for very remote areas. The median point estimates for these areas are the same as that for non-very remote areas, albeit with over three times the uncertainty (using CV). 

\subsection{Weight}

The three weight measures (overweight/obese, obese and risky waist circumference) exhibit significant spatial variation with lower proportions in less disadvantaged and urban areas (\cref{fig:or_obesity,fig:mapprev_obesity,fig:mapprevep_obesity,fig:or_overweight,fig:mapprev_overweight,fig:mapprevep_overweight,fig:or_waist_circum,fig:mapprev_waist_circum,fig:mapprevep_waist_circum}). Across all three measures, the prevalence was very strongly tied to remoteness with substantially lower prevalence almost exclusively occurring in major cities. Furthermore, the most notable differences in patterns between the estimates for obese and overweight/obese are found in major cities. 

Some low proportions for both obese and overweight/obese are found in the Northern Territory, which is largely composed of very remote and more disadvantaged areas. However, these areas have a high level of uncertainty and thus provide insufficient evidence of a meaningful difference. Across the three weight measures, the uncertainty (using CV) of estimates for very remote areas are, on average, over 2.5 times those for non-very remote areas.  

\subsection{Physical activity}

Similar to our other modelled risk factors and other Australian research \cite{RN204}, both activity variables (\cref{fig:or_activityleis,fig:mapprev_activityleis,fig:mapprevep_activityleis,fig:or_activityleiswkpl,fig:mapprev_activityleiswkpl,fig:mapprevep_activityleiswkpl}) exhibit high spatial variation, with lower prevalence of inadequate activity in major cities and least disadvantaged areas. One novelty of this work is estimates for inadequate activity (all) as well as inadequate activity (leisure). The most notable difference between the two was in non-remote areas. 

By improving the reach of the SHAA, the modelled risk factors provide insights into the spatial disparities of physical activity in very remote areas of Australia. These areas have, on average, 7\% higher prevalence but also over 2.5 times the uncertainty (using CV), compared with non-very remote areas. 

%According to the 2016 census, the areas with proportions of persons working as labourers above the 95\% percentile are 76\% and 62\% most disadvantaged and major cities or outer regional, respectively. 

%% NEW SECTION %% --------------------------------------------------------------------------------------------
\newpage
\section{Additional plots} \label{supp:add_plots}

In this section, we provide supplemental plots and maps to those provided in the main paper.

\begin{figure}[H]
    \centering
    \includegraphics[width=\textwidth]{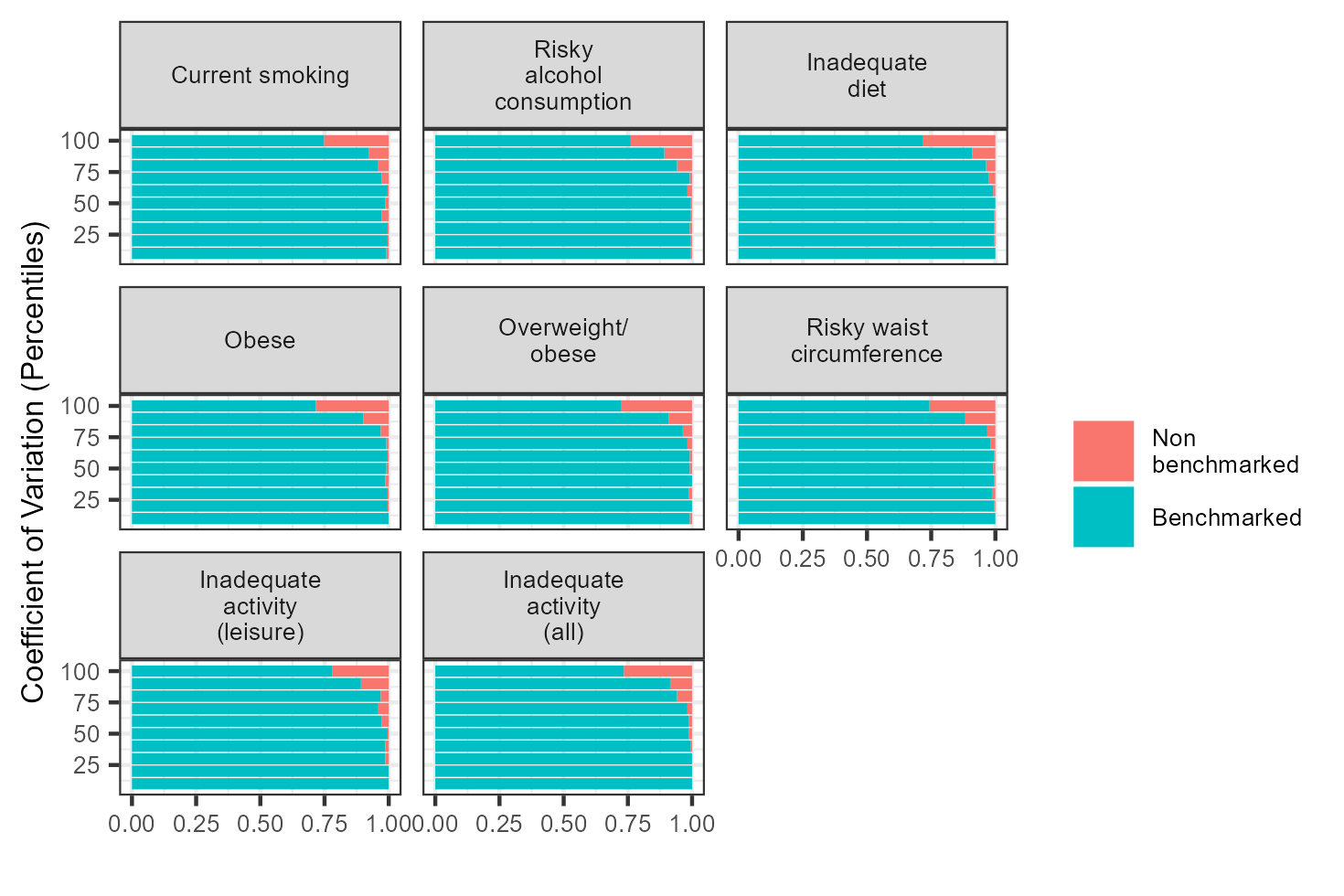}
    \caption{Distribution of coefficients of variation (CV) of the posterior prevalence estimates by risk factor and whether the area was benchmarked. The CVs are represented in percentiles with higher values denoting higher CVs and thus more uncertainty. The $x$-axis represents the proportion of the areas in each percentile that were benchmarked.}
    \label{fig:benchmark_CV}
\end{figure}

\begin{figure}[H]
    \centering
    \includegraphics[width=\textwidth]{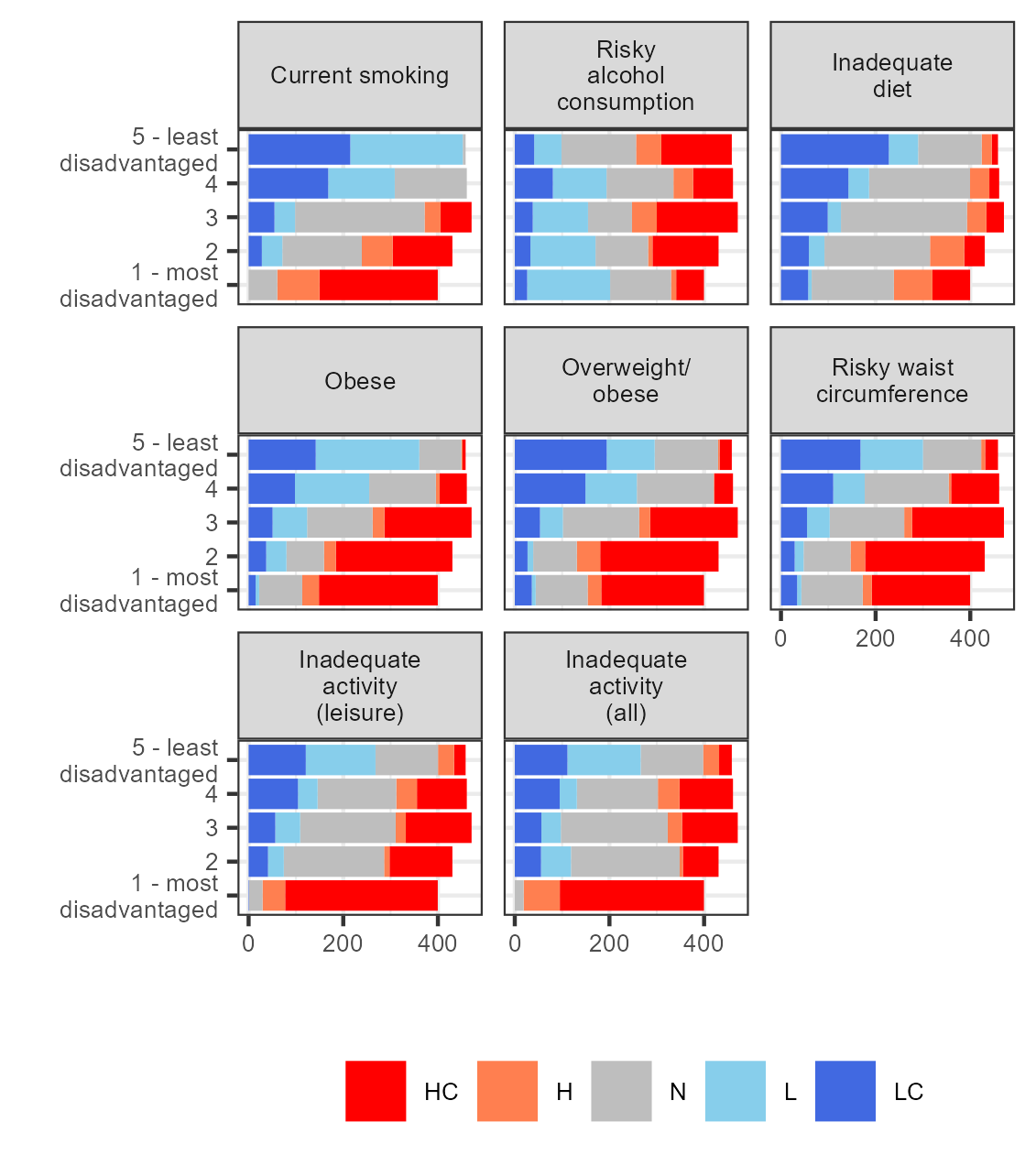}
    \caption{Distribution of the evidence classifications (HC, H, L, and LC) by SES index quintile (IRSD) and risk factor. The $x$-axis is the weighted number of SA2s using the 2017-18 ERP as weights.}
    \label{fig:ec_irsd_barchart_pw_cec}
\end{figure}

\newpage
\begin{landscape}

\begin{figure}[H]
    \centering
    \includegraphics[width=\textwidth]{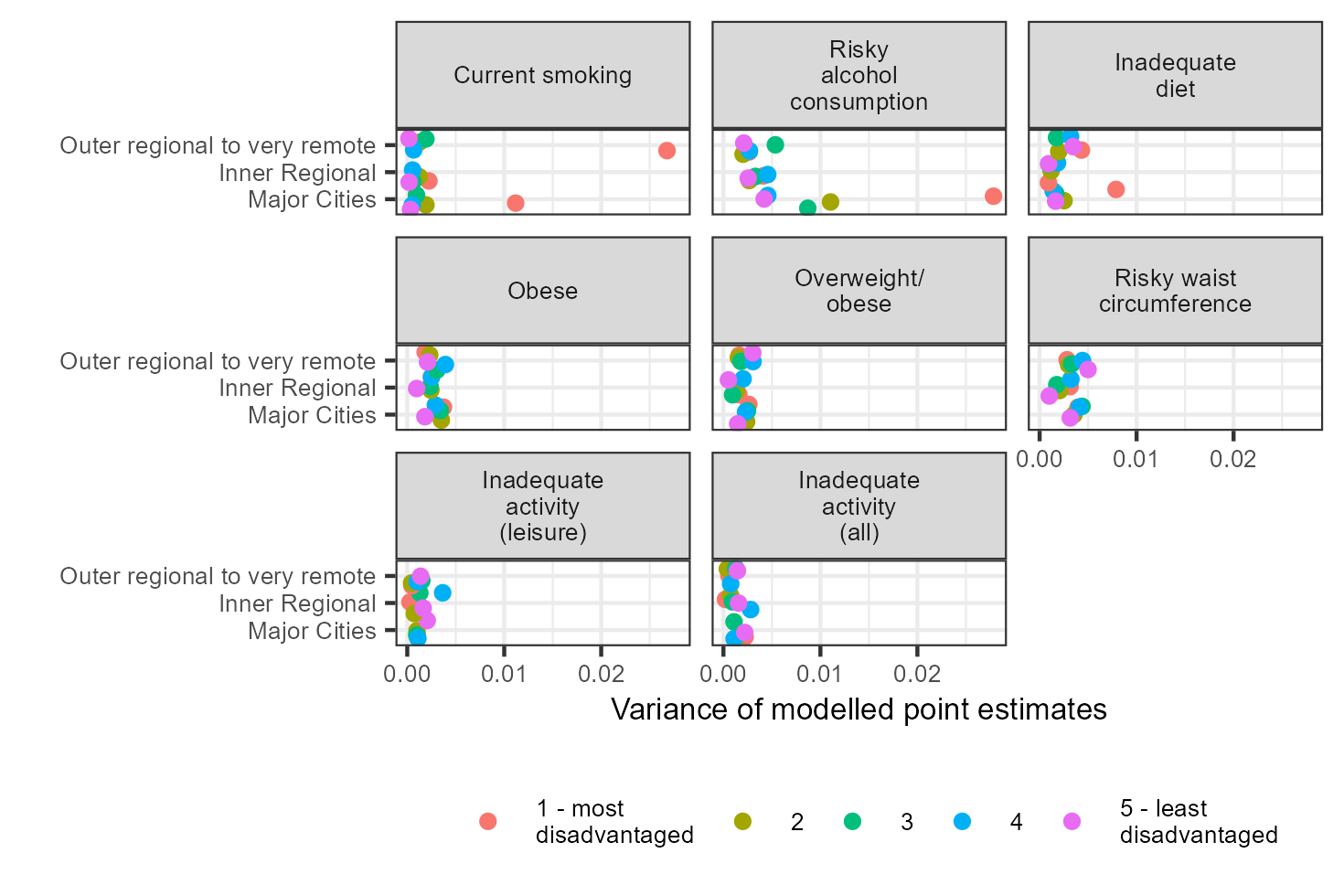}
    \caption{Summary of the variance of modelled point estimates (posterior medians) by SES index quintiles (IRSD), remoteness and risk factor. Points have been randomly jittered along the $y$-axis to avoid overlap. Each point represents the variance of all the point estimates for the specific risk factor, SES index quintile and remoteness category.}
    \label{fig:variance_irsdera}
\end{figure}

\begin{figure}[H]
     \begin{subfigure}[b]{0.7\textwidth}
         \includegraphics[width=\textwidth]{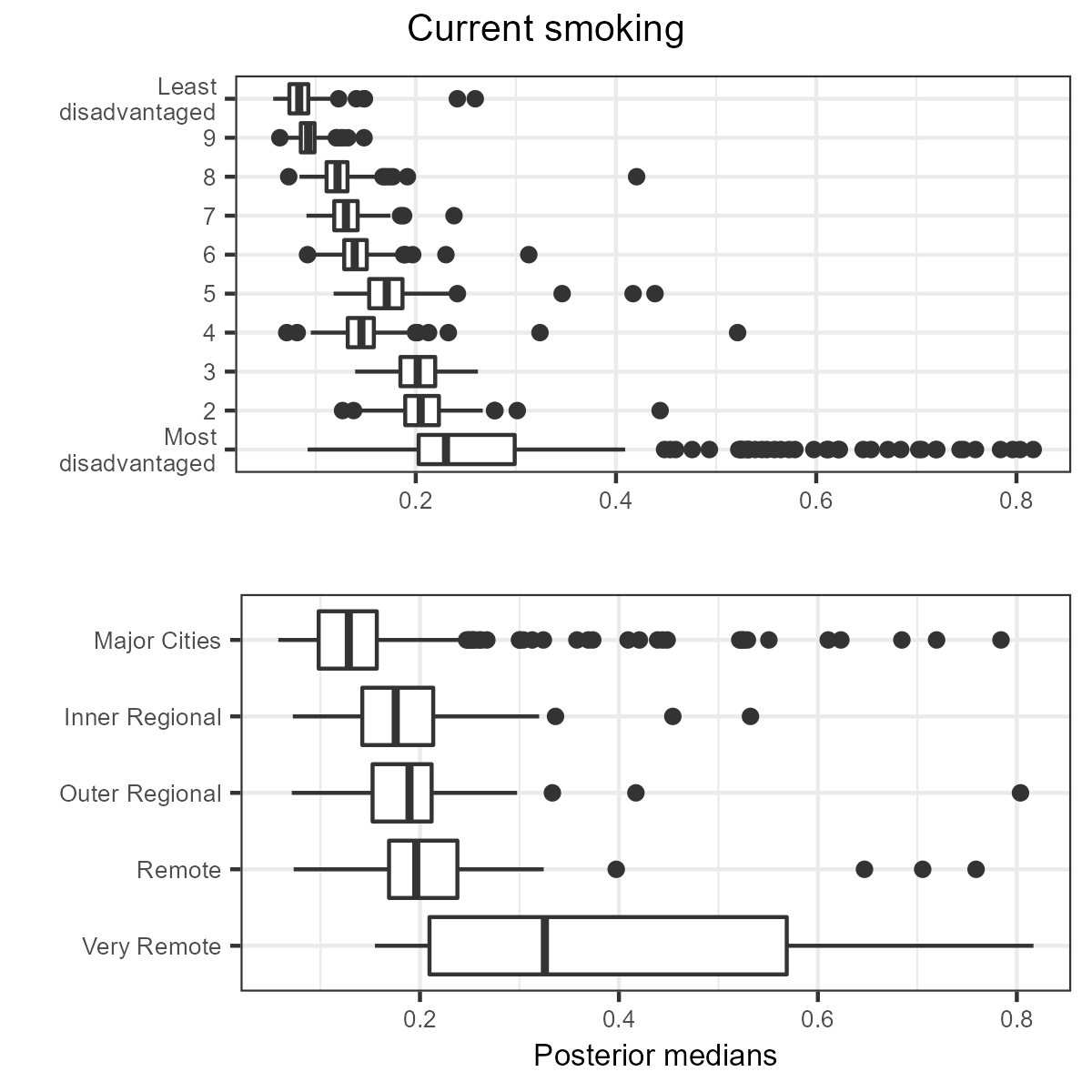}
         \caption{}
     \end{subfigure}
     \hspace{1em}
     \begin{subfigure}[b]{0.7\textwidth}
         \includegraphics[width=\textwidth]{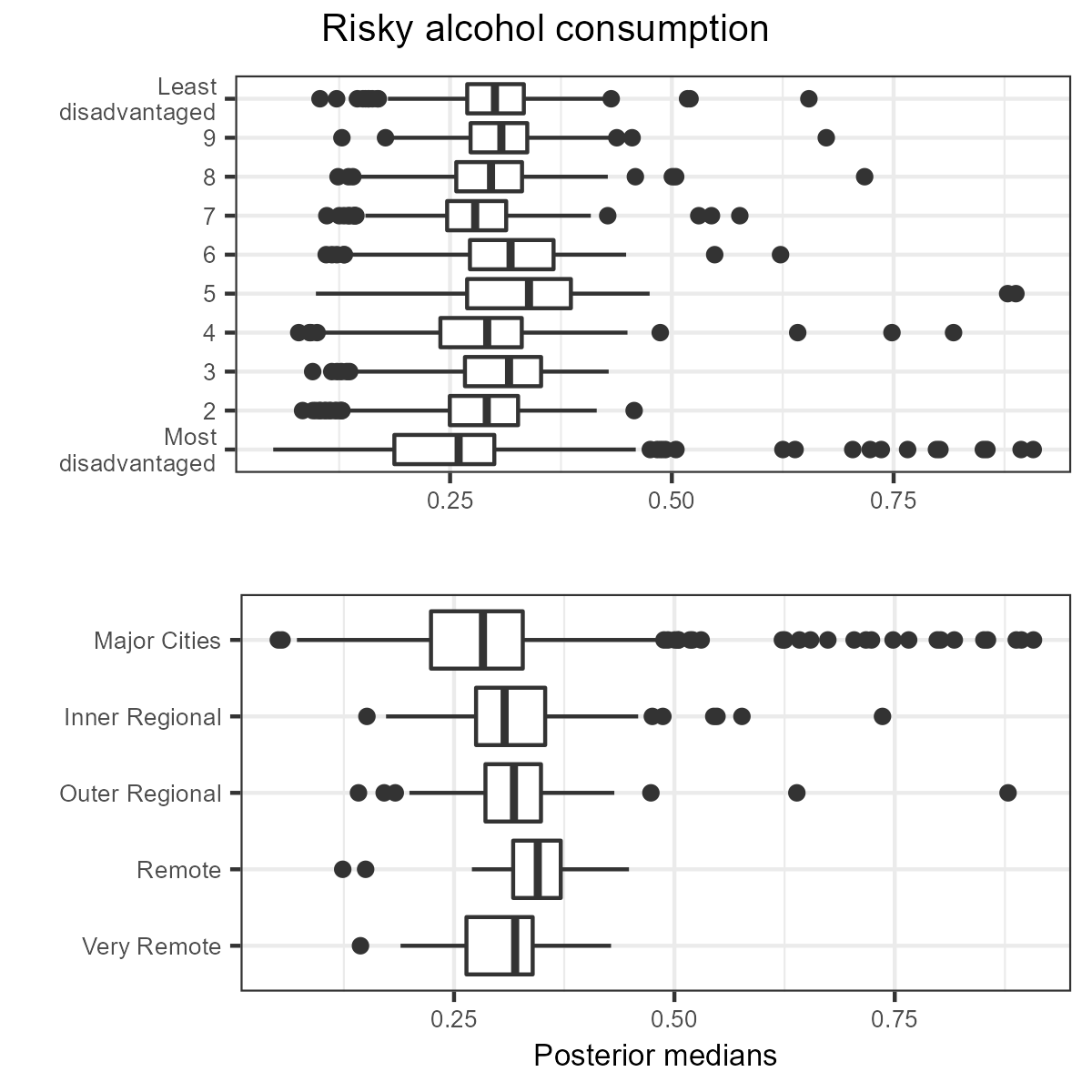}
         \caption{}
     \end{subfigure}
    \caption{\small Boxplots summarising the distribution of the modelled point estimates (posterior medians) by the SES index and remoteness.}
    \label{fig:seifa_irsd_1}
\end{figure}

\begin{figure}[H]
     \begin{subfigure}[b]{0.7\textwidth}
         \includegraphics[width=\textwidth]{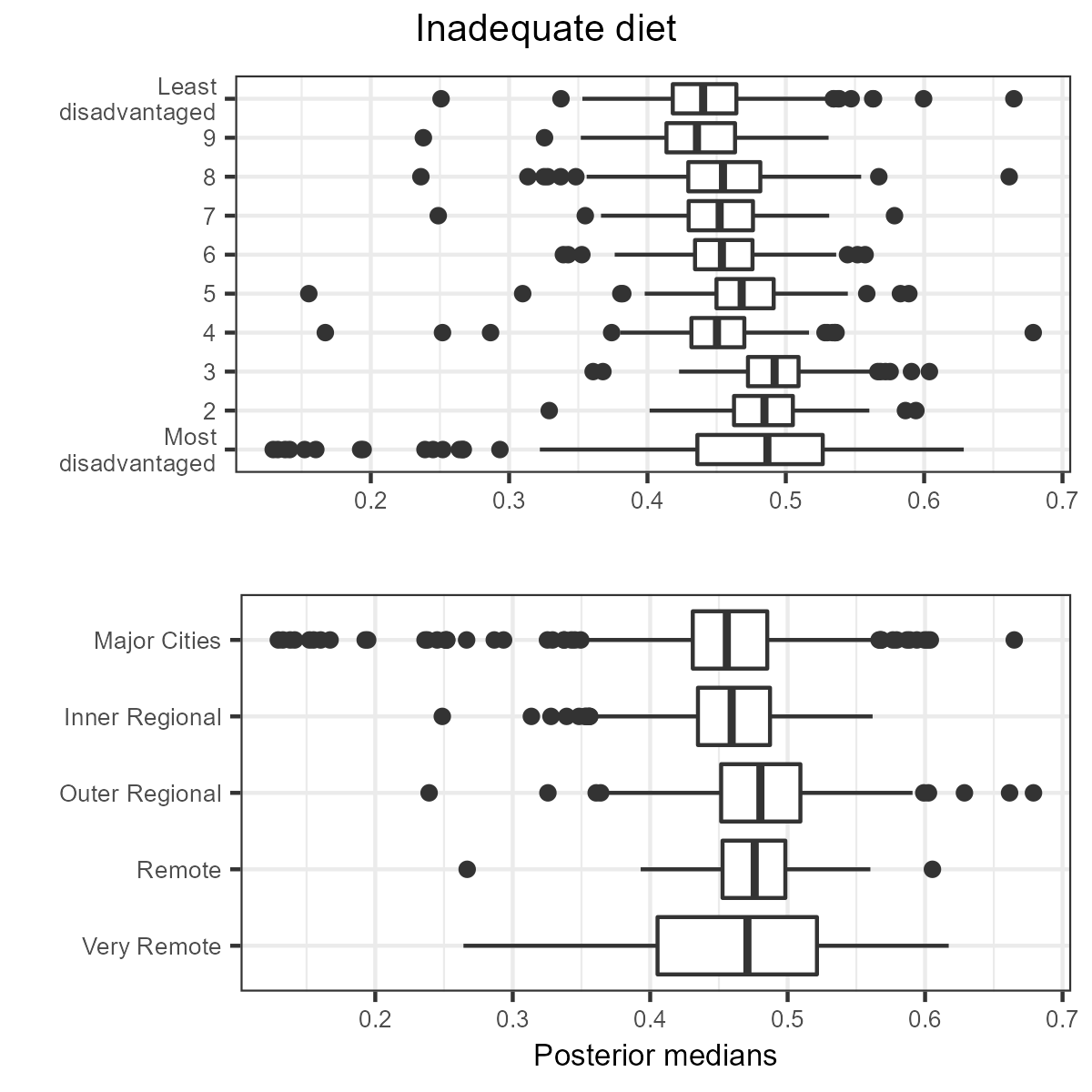}
         \caption{}
     \end{subfigure}
     \hspace{1em}
     \begin{subfigure}[b]{0.7\textwidth}
         \includegraphics[width=\textwidth]{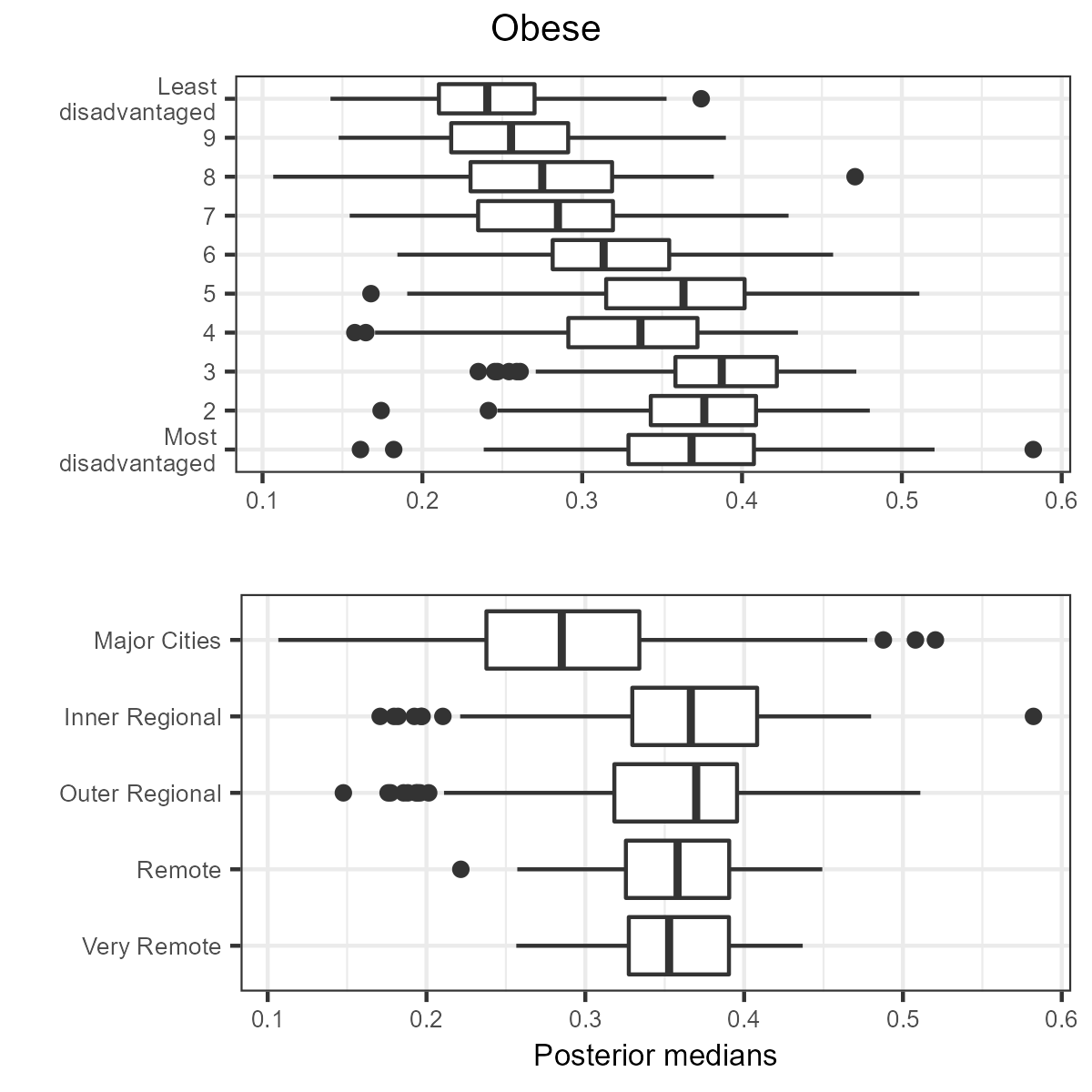}
         \caption{}
     \end{subfigure}
    \caption{\small Boxplots summarising the distribution of the modelled point estimates (posterior medians) by the SES index and remoteness.}
    \label{fig:seifa_irsd_2}
\end{figure}

\begin{figure}[H]
     \begin{subfigure}[b]{0.7\textwidth}
         \includegraphics[width=\textwidth]{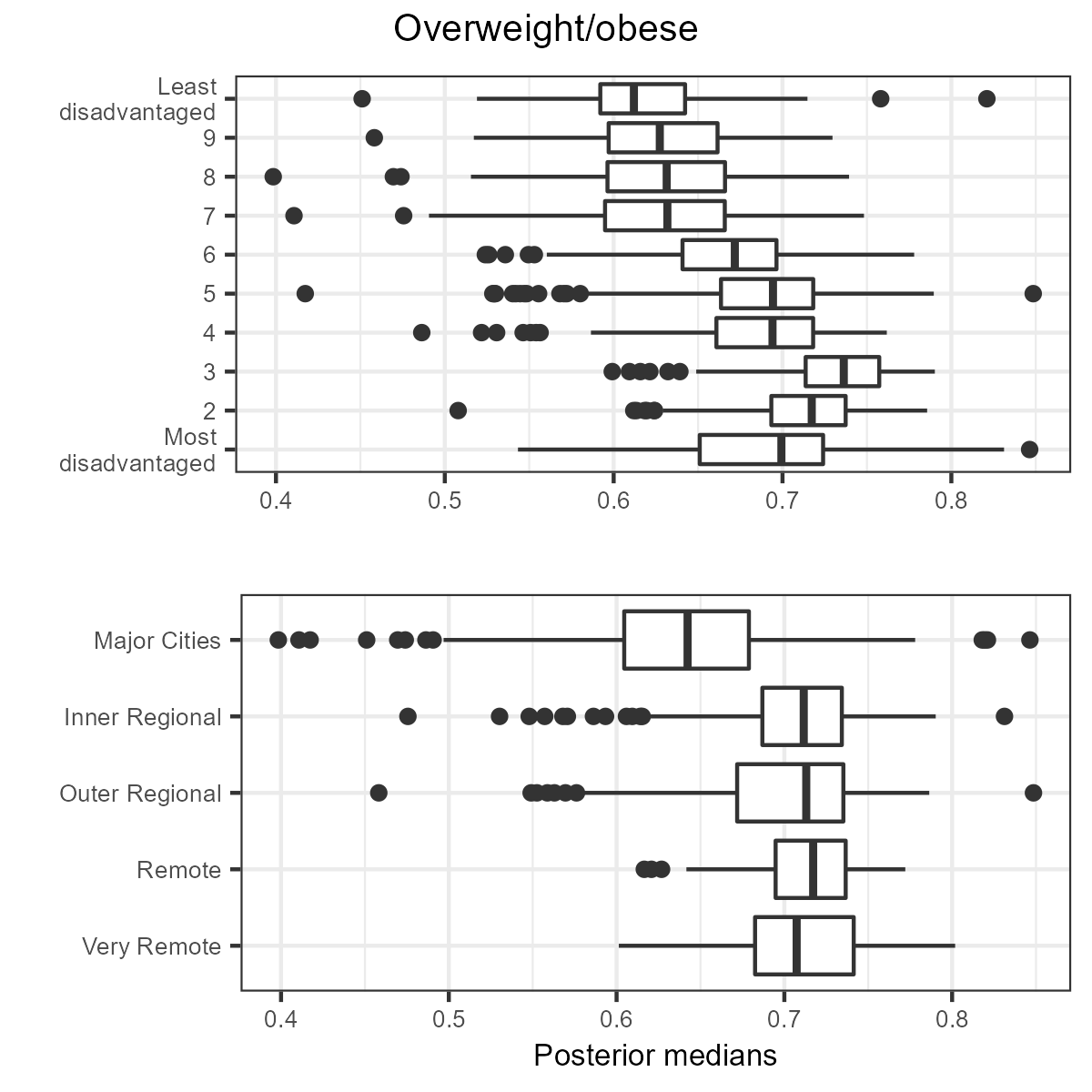}
         \caption{}
     \end{subfigure}
     \hspace{1em}
     \begin{subfigure}[b]{0.7\textwidth}
         \includegraphics[width=\textwidth]{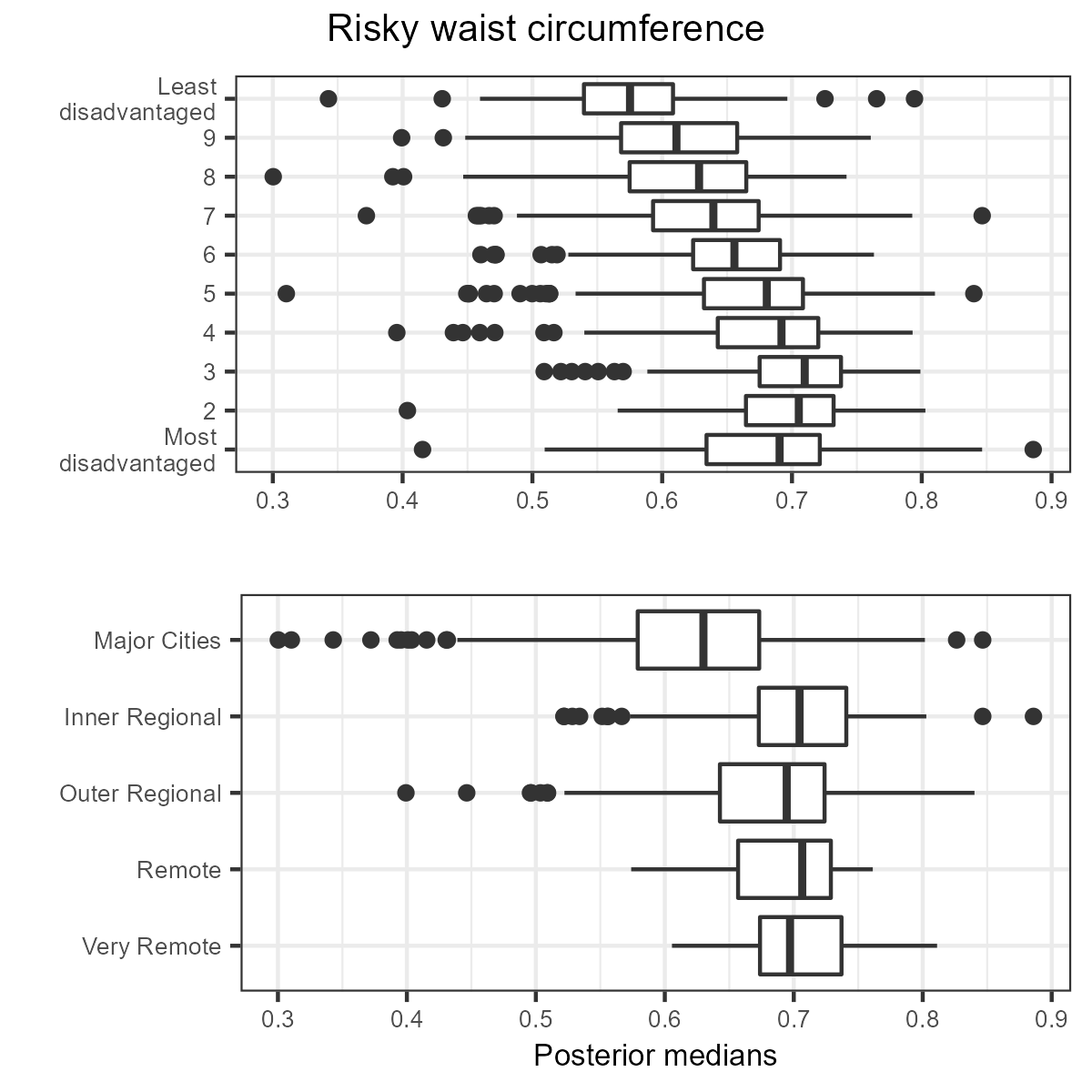}
         \caption{}
     \end{subfigure}
    \caption{\small Boxplots summarising the distribution of the modelled point estimates (posterior medians) by the SES index and remoteness.}
    \label{fig:seifa_irsd_3}
\end{figure}

\begin{figure}[H]
     \begin{subfigure}[b]{0.7\textwidth}
         \includegraphics[width=\textwidth]{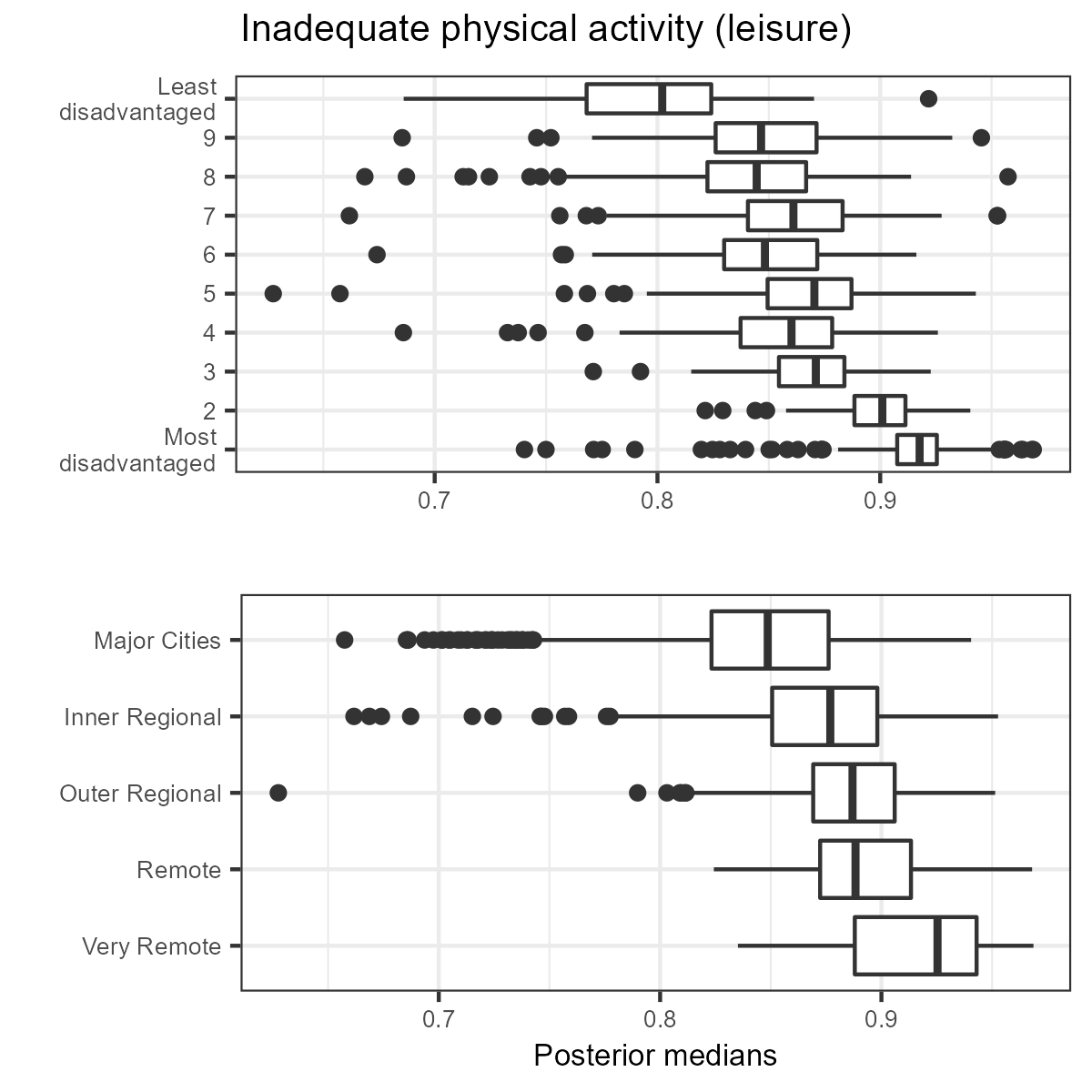}
         \caption{}
     \end{subfigure}
     \hspace{1em}
     \begin{subfigure}[b]{0.7\textwidth}
         \includegraphics[width=\textwidth]{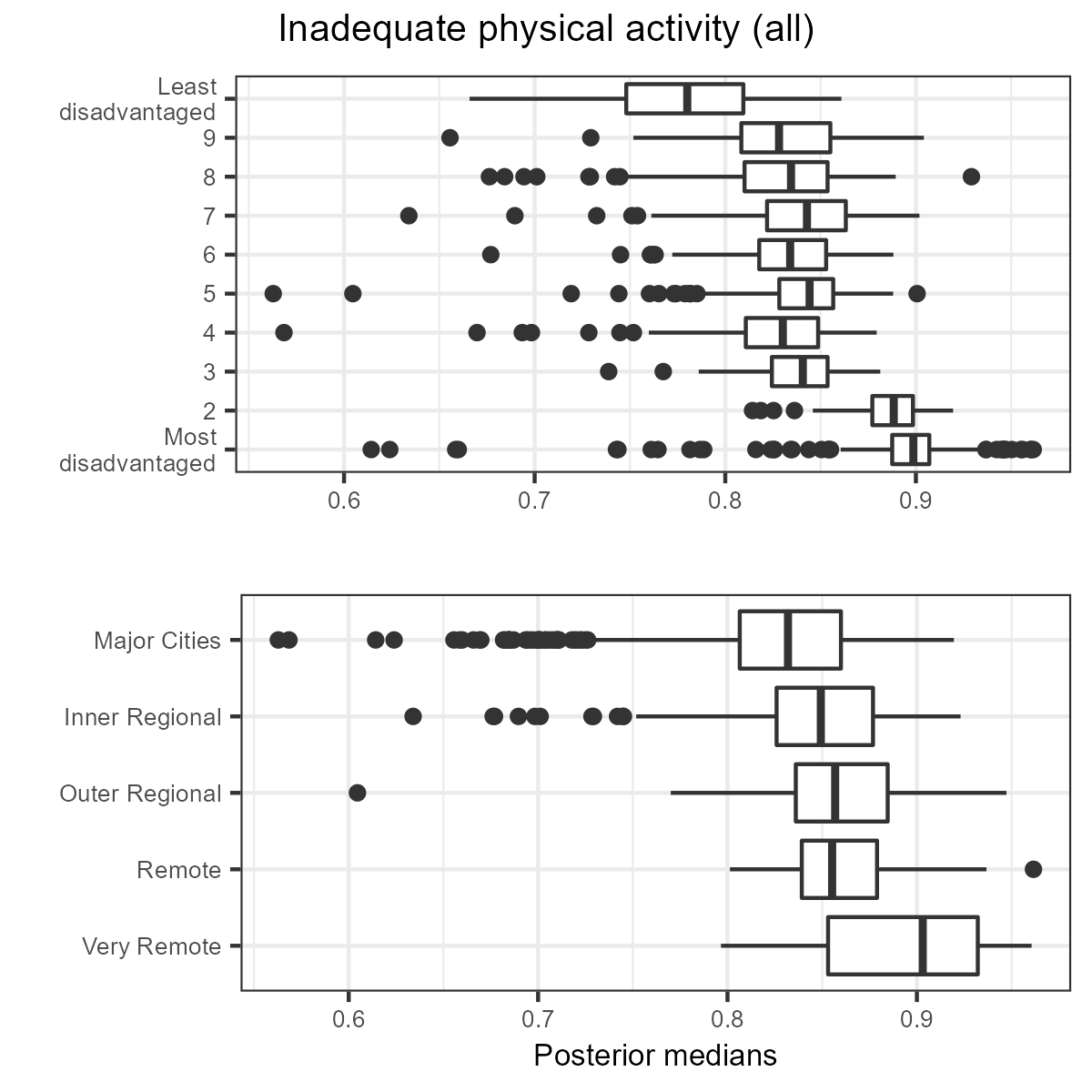}
         \caption{}
     \end{subfigure}
    \caption{\small Boxplots summarising the distribution of the modelled point estimates (posterior medians) by the SES index and remoteness.}
    \label{fig:seifa_irsd_4}
\end{figure}

%% NEW SECTION %% --------------------------------------------------------------------------------------------
\newpage
\section{Additional maps} \label{supp:add_maps}

\subsection{Socioeconomic status}

\begin{figure}[H]
    \centering
    \includegraphics[width=0.9\textwidth]{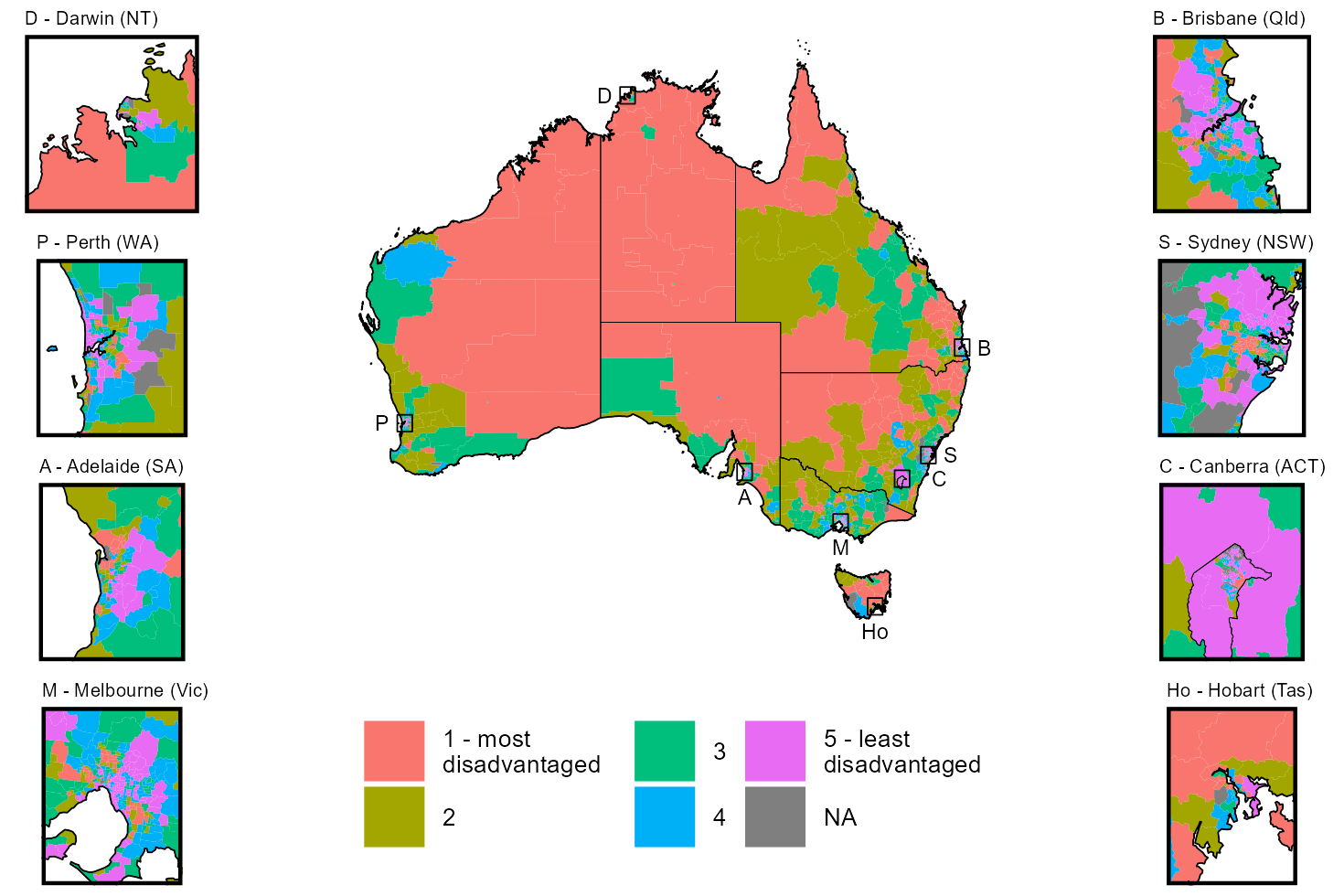}
    \caption{\small Choropleth maps displaying quintiles of the SES index for 2221 SA2s across Australia. The map includes insets for the eight capital cities for each state and territory, with black boxes on the main map indicating each insets' respective location. Gray areas were excluded from estimation due to the exclusion criteria described in Section 2.1 of the main paper. Black lines represent the boundaries of the eight states and territories of Australia.}
    \label{fig:map_irsd}
\end{figure}

\subsection{Current smoking}

\begin{figure}[H]
    \centering
    \includegraphics{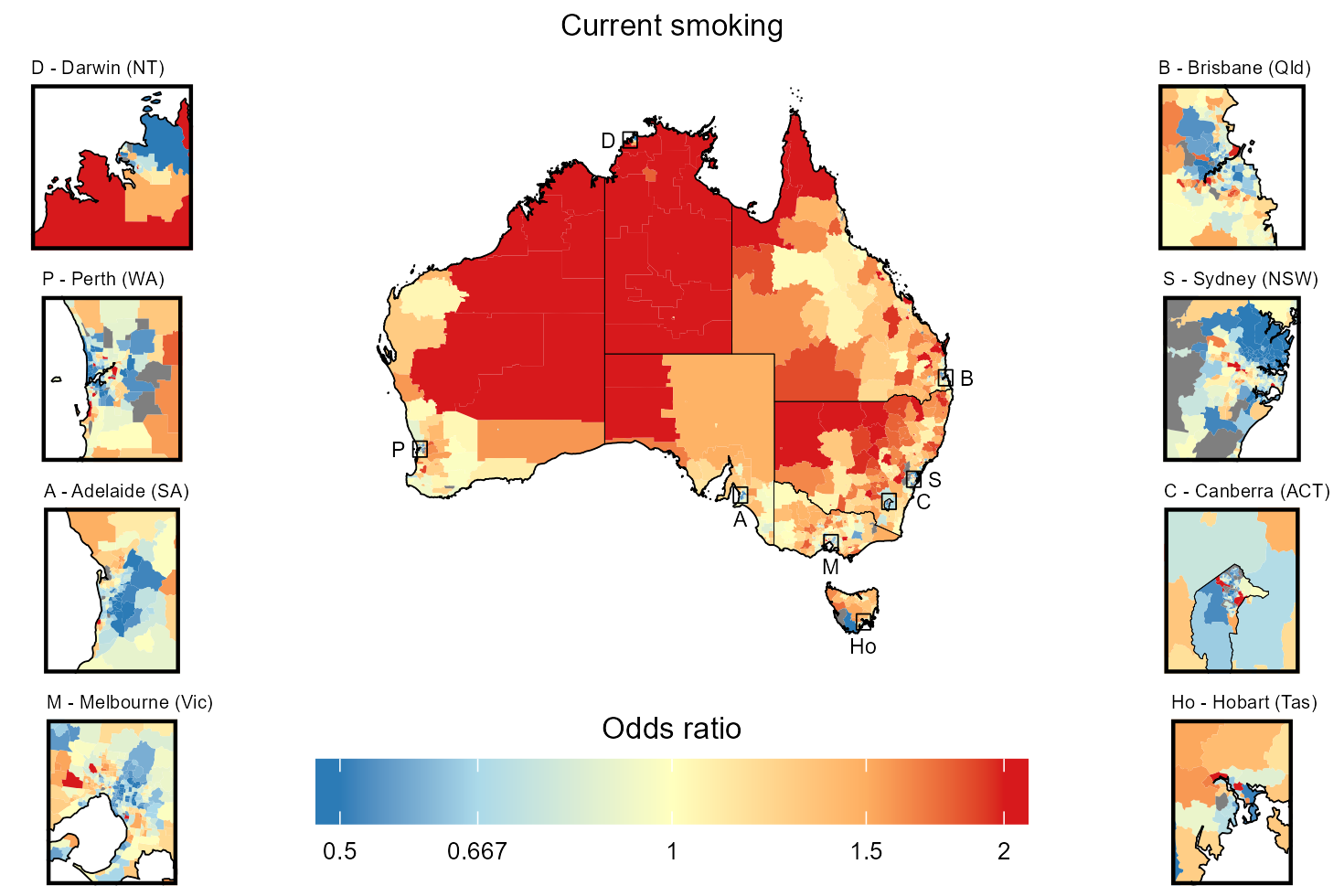}
    \caption{\small Choropleth maps displaying the modelled odds ratios (OR) for 2221 SA2s in Australia. ORs above 1 indicate that the prevalence is higher than the national average. The map includes insets for the eight capital cities for each state and territory, with black boxes on the main map indicating each insets' respective location. Note that some values are lower (or higher) than the range of color scales shown; for these values, the lowest (or highest) color is shown. White areas were excluded from estimation due to the exclusion criteria described in Section 2.1 of the main paper. Black lines represent the boundaries of the eight states and territories of Australia.}
    \label{fig:or_smoking}
\end{figure}

\begin{figure}[H]
     \begin{subfigure}[b]{0.7\textwidth}
         \includegraphics[width=\textwidth]{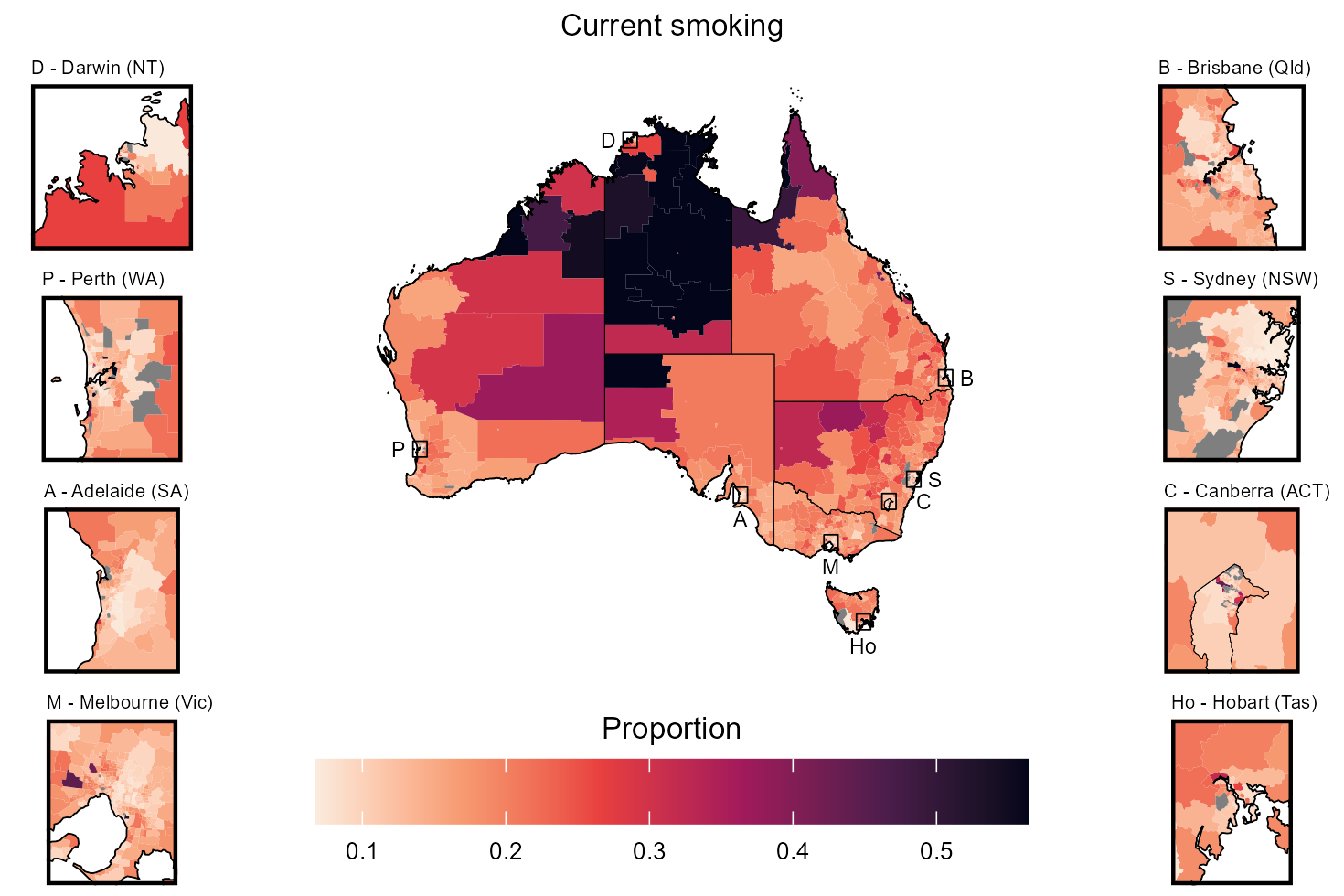}
         \caption{}
     \end{subfigure}
     \hspace{1em}
     \begin{subfigure}[b]{0.7\textwidth}
         \includegraphics[width=\textwidth]{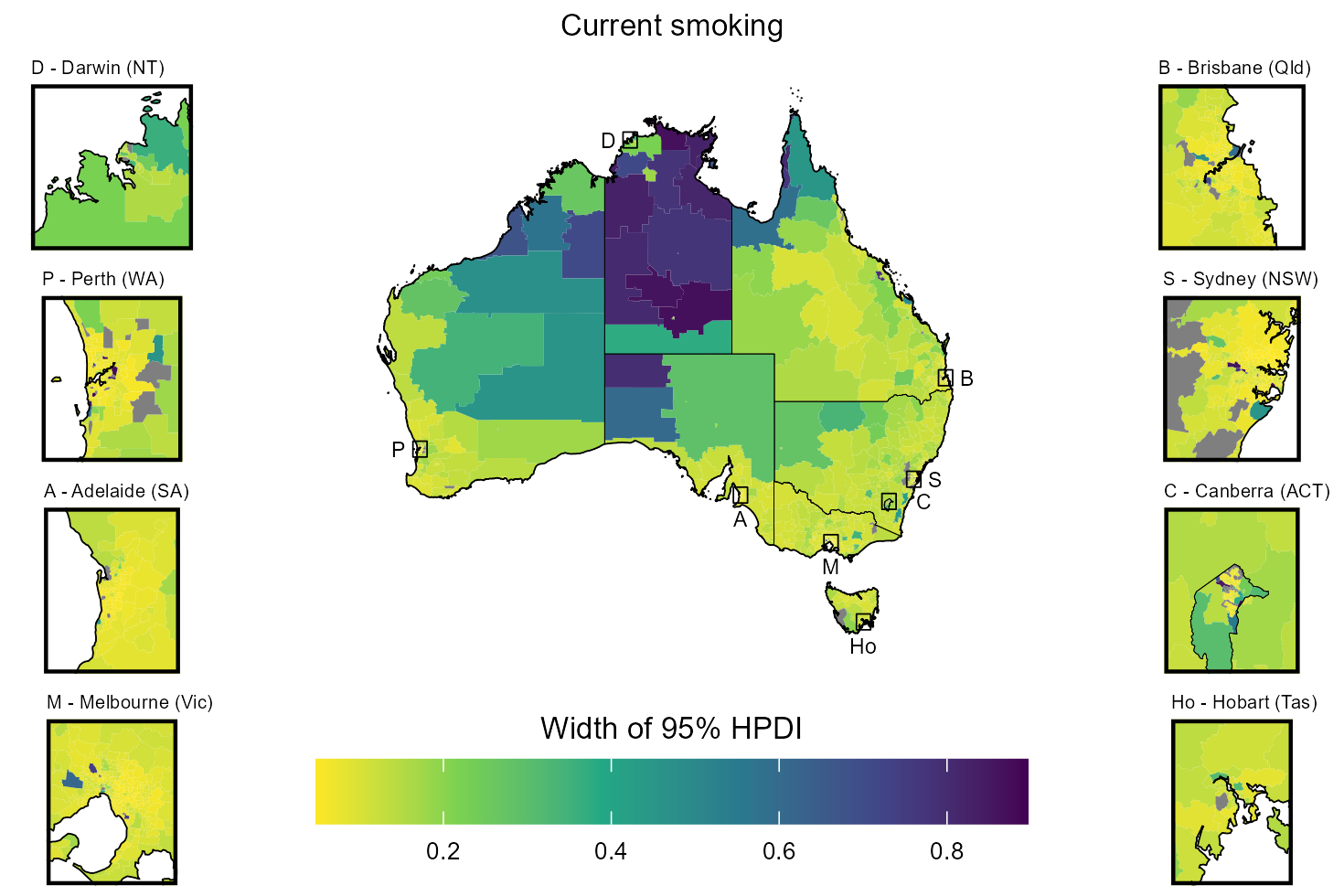}
         \caption{}
     \end{subfigure}
    \caption{\small Choropleth maps displaying the modelled proportion (a) and width of the 95\% HPDIs (b) for 2221 SA2s across Australia. The map includes insets for the eight capital cities for each state and territory, with black boxes on the main map indicating each insets' respective location. Note that prevalence values above the 99th and below the 1th percentile were assigned to the highest and lowest color, respectively. Gray areas were excluded from estimation due to the exclusion criteria described in the main paper. Black lines represent the boundaries of the eight states and territories of Australia.}
    \label{fig:mapprev_smoking}
\end{figure}

\newpage
\begin{figure}[H]
     \begin{subfigure}[b]{0.7\textwidth}
         \includegraphics[width=\textwidth]{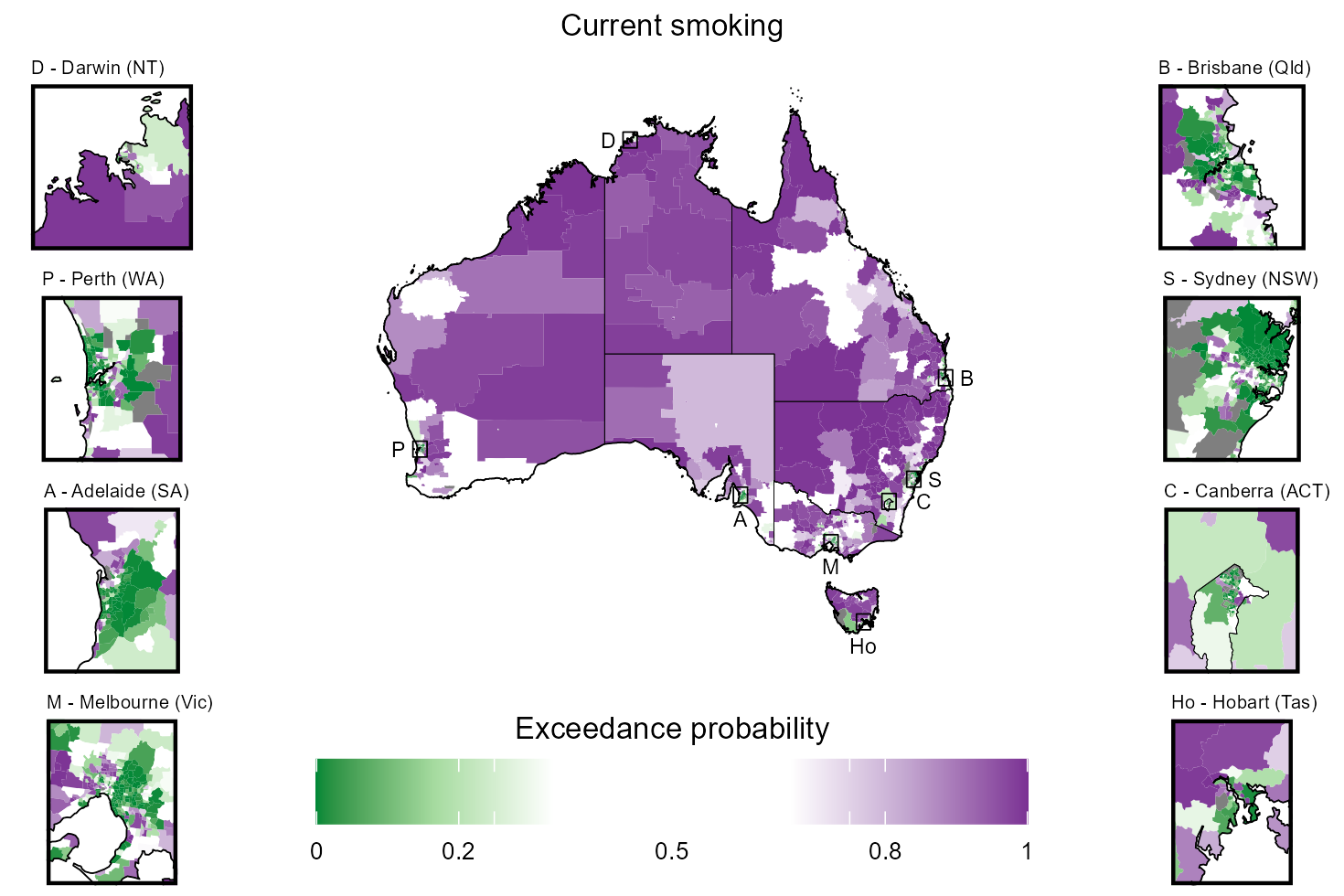}
         \caption{}
     \end{subfigure}
     \hspace{1em}
     \begin{subfigure}[b]{0.7\textwidth}
         \includegraphics[width=\textwidth]{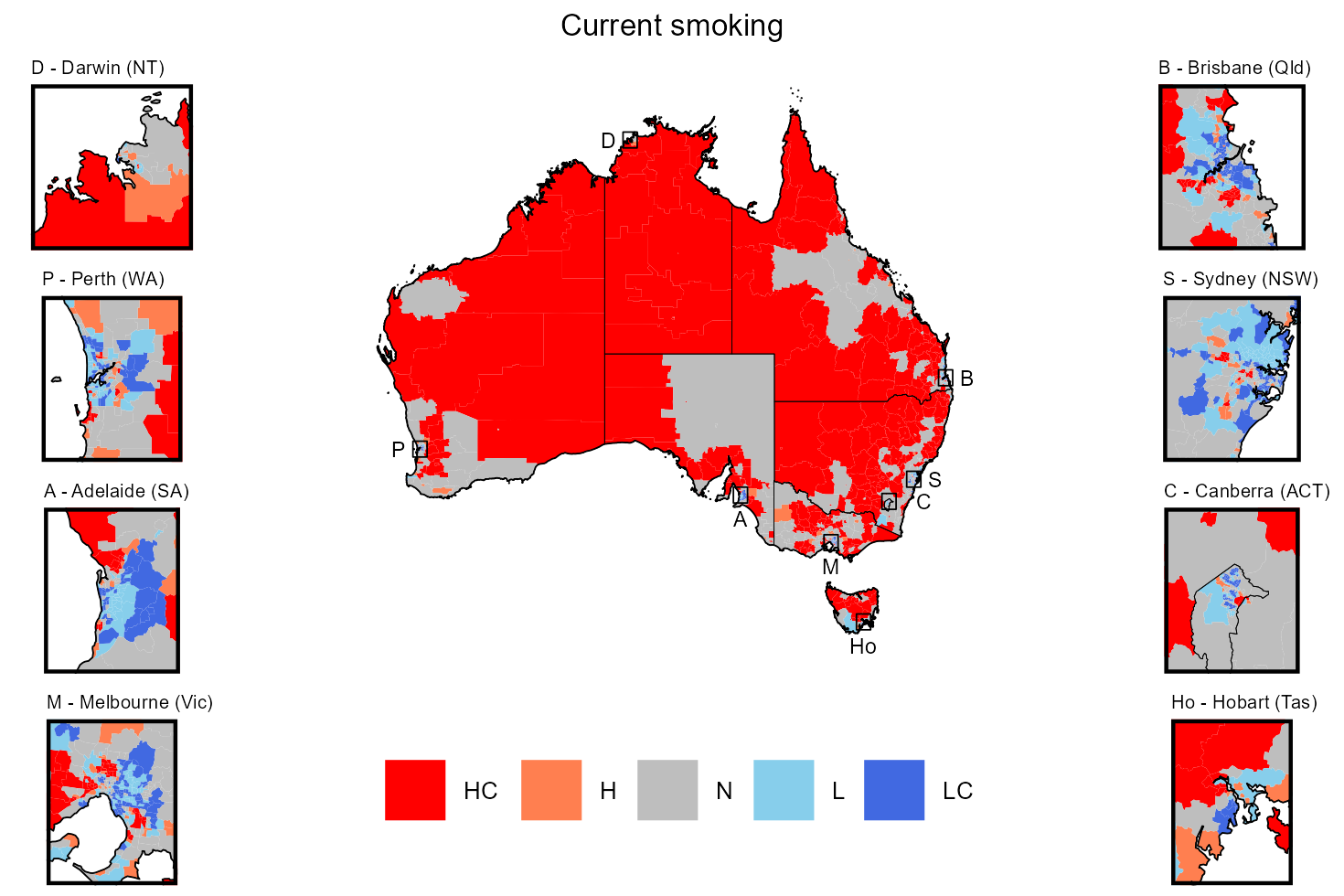}
         \caption{}
     \end{subfigure}
    \caption{\small Choropleth maps displaying the exceedance probabilities (a) and evidence classifications (b) for 2221 SA2s across Australia. The map includes insets for the eight capital cities for each state and territory, with black boxes on the main map indicating each insets' respective location. Gray areas were either excluded from estimation due to the exclusion criteria described in the main paper or were not classified according to one of the four categories. Black lines represent the boundaries of the eight states and territories of Australia.}
    \label{fig:mapprevep_smoking}
\end{figure}

\newpage
\subsection{Risky alcohol consumption}

\begin{figure}[H]
    \centering
    \includegraphics{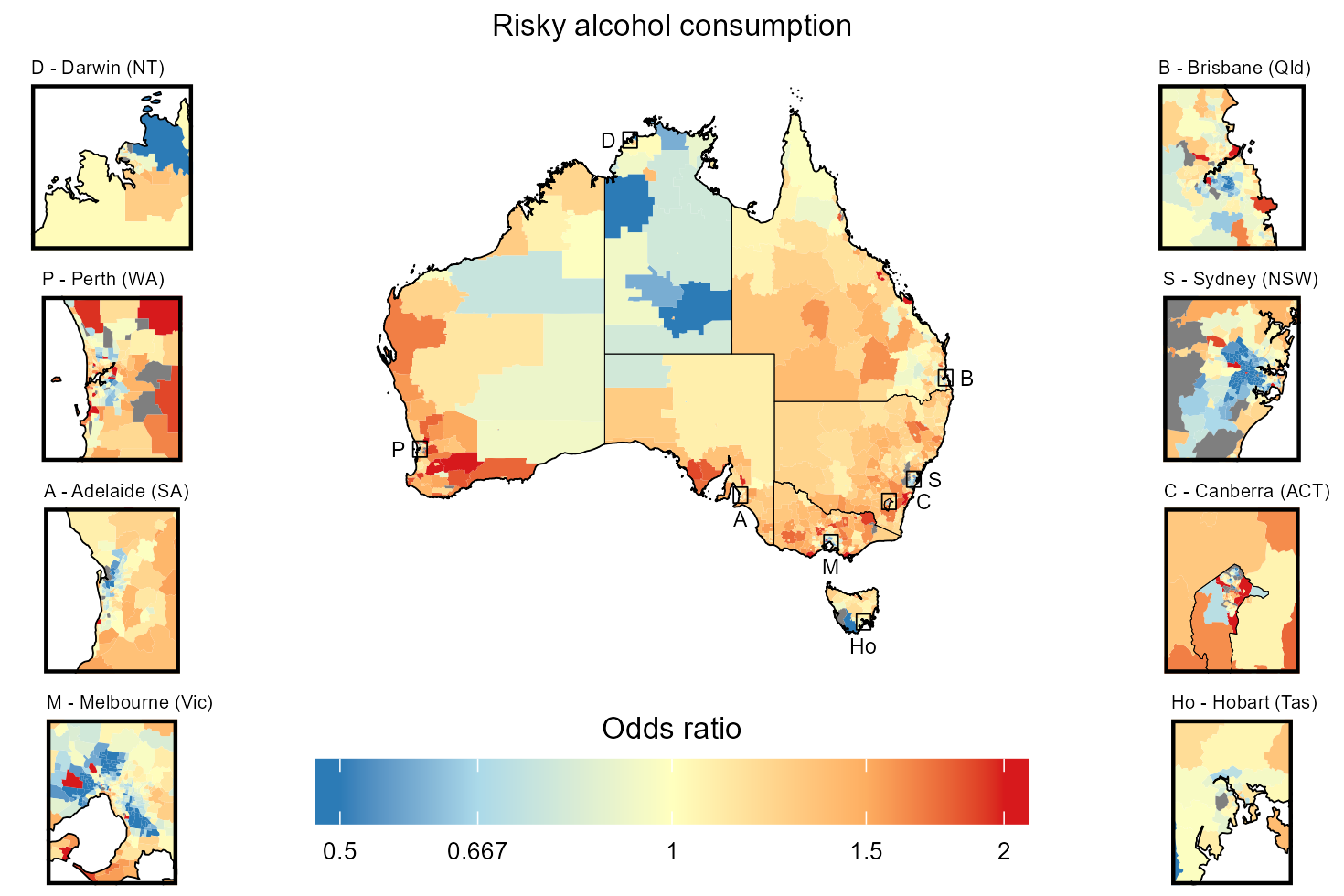}
    \caption{See caption for \cref{fig:or_smoking}}
    \label{fig:or_alcohol}
\end{figure}

\begin{figure}[H]
     \begin{subfigure}[b]{0.7\textwidth}
         \includegraphics[width=\textwidth]{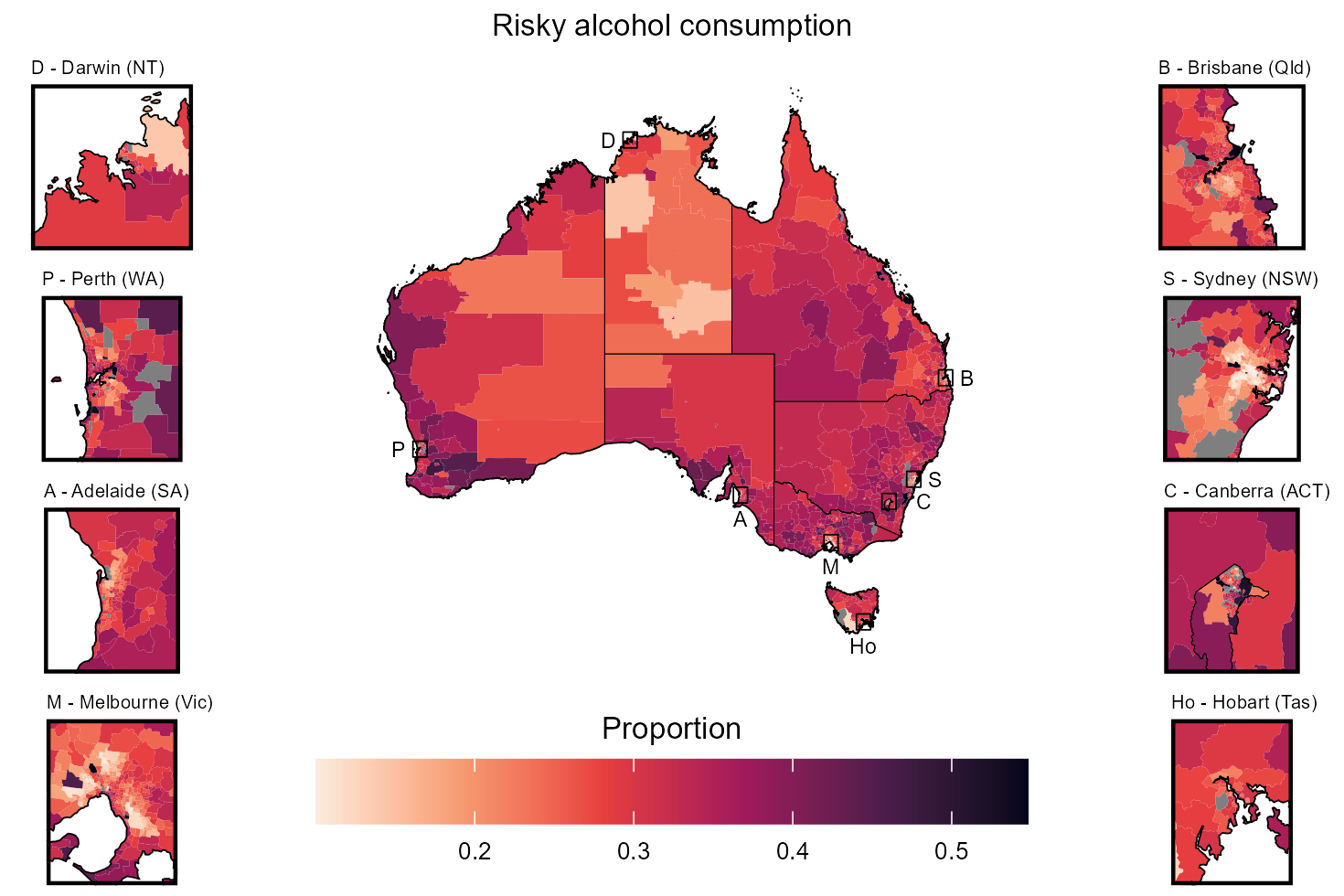}
         \caption{}
     \end{subfigure}
     \hspace{1em}
     \begin{subfigure}[b]{0.7\textwidth}
         \includegraphics[width=\textwidth]{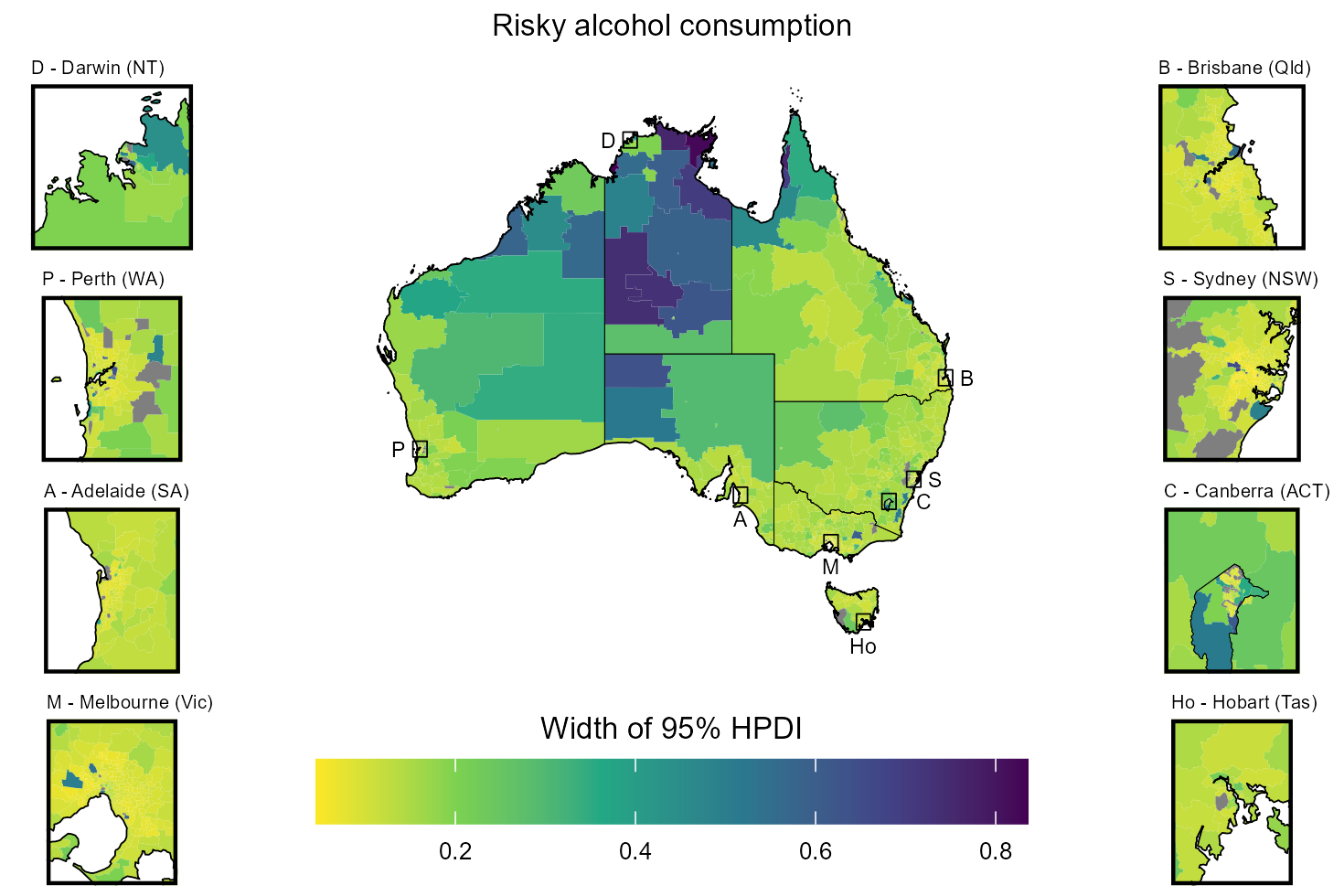}
         \caption{}
     \end{subfigure}
    \caption{\small See caption for \cref{fig:mapprev_smoking}}
    \label{fig:mapprev_alcohol}
\end{figure}

\newpage
\begin{figure}[H]
     \begin{subfigure}[b]{0.7\textwidth}
         \includegraphics[width=\textwidth]{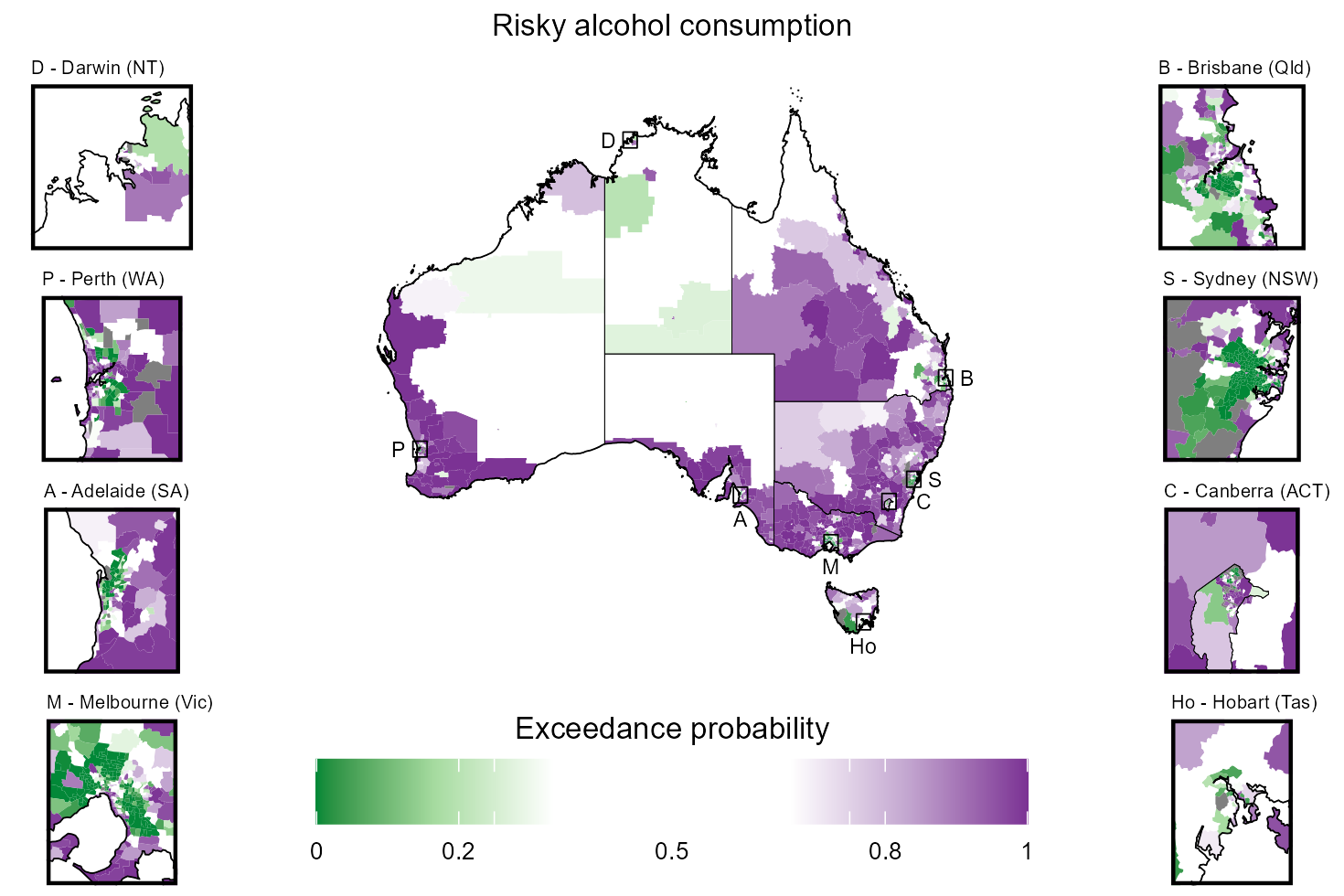}
         \caption{}
     \end{subfigure}
     \hspace{1em}
     \begin{subfigure}[b]{0.7\textwidth}
         \includegraphics[width=\textwidth]{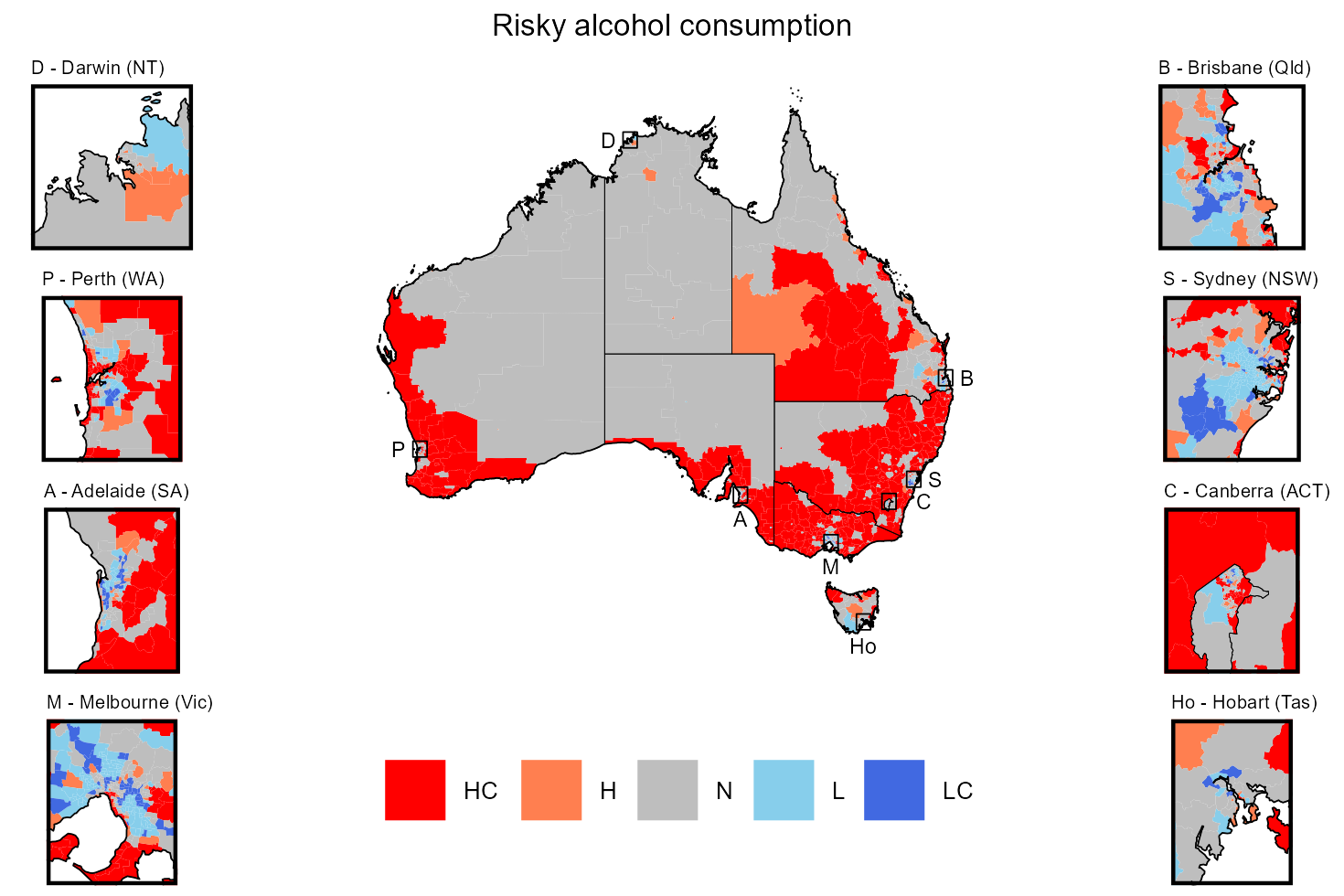}
         \caption{}
     \end{subfigure}
    \caption{\small See caption for \cref{fig:mapprevep_smoking}}
    \label{fig:mapprevep_alcohol}
\end{figure}

\subsection{Inadequate diet}

\begin{figure}[H]
    \centering
    \includegraphics{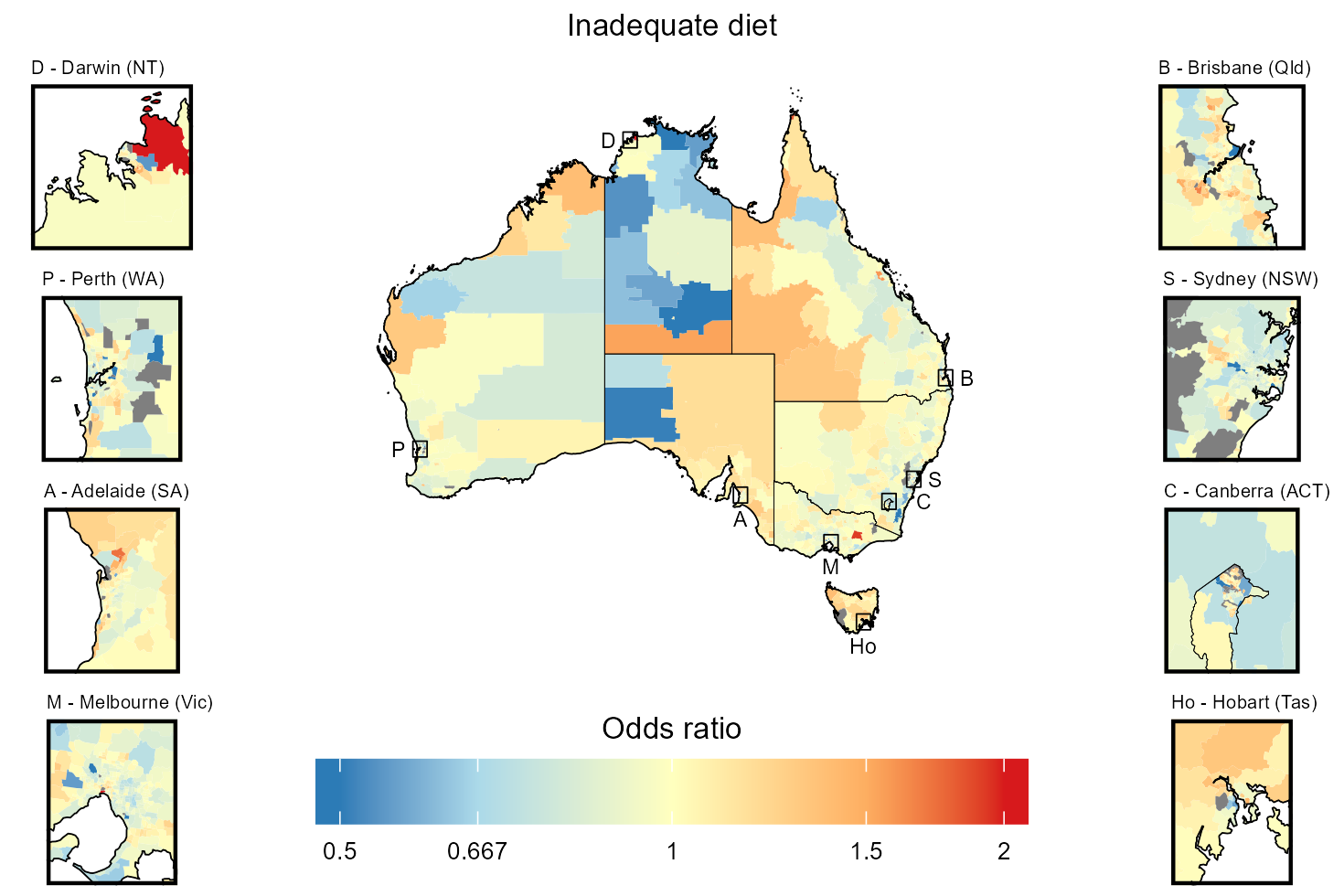}
    \caption{See caption for \cref{fig:or_smoking}}
    \label{fig:or_diet}
\end{figure}

\begin{figure}[H]
     \begin{subfigure}[b]{0.7\textwidth}
         \includegraphics[width=\textwidth]{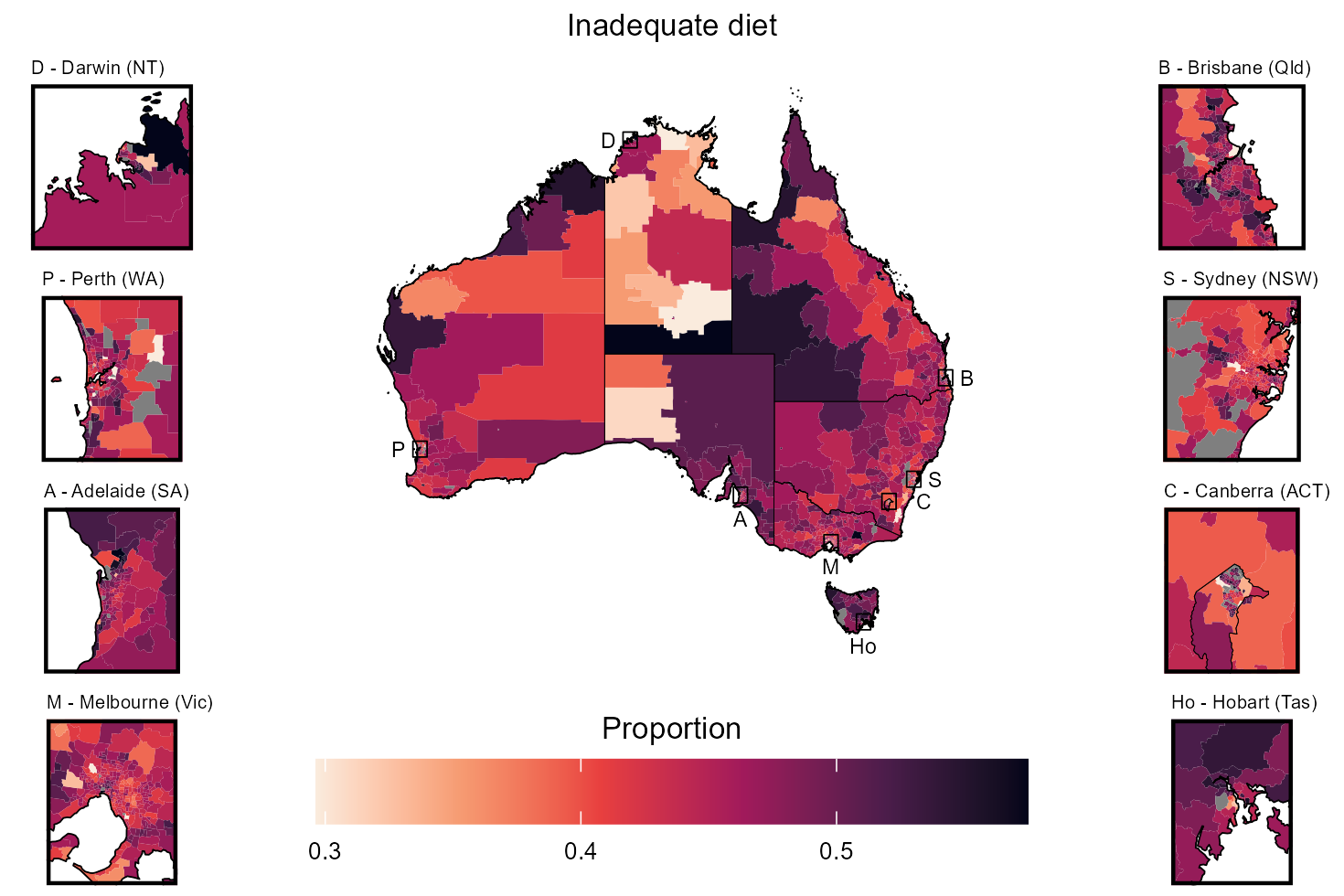}
         \caption{}
     \end{subfigure}
     \hspace{1em}
     \begin{subfigure}[b]{0.7\textwidth}
         \includegraphics[width=\textwidth]{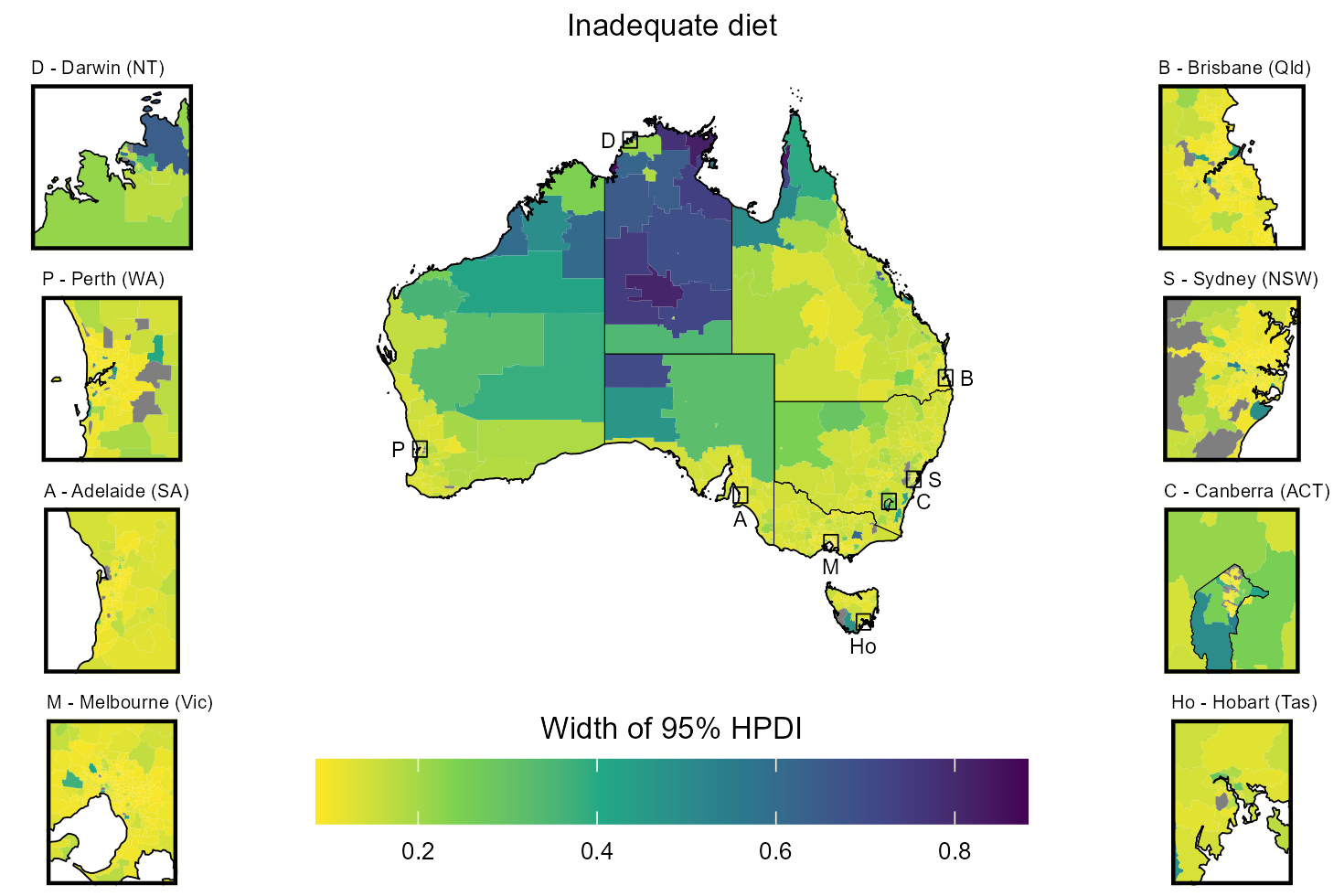}
         \caption{}
     \end{subfigure}
    \caption{\small See caption for \cref{fig:mapprev_smoking}.}
    \label{fig:mapprev_diet}
\end{figure}

\newpage
\begin{figure}[H]
     \begin{subfigure}[b]{0.7\textwidth}
         \includegraphics[width=\textwidth]{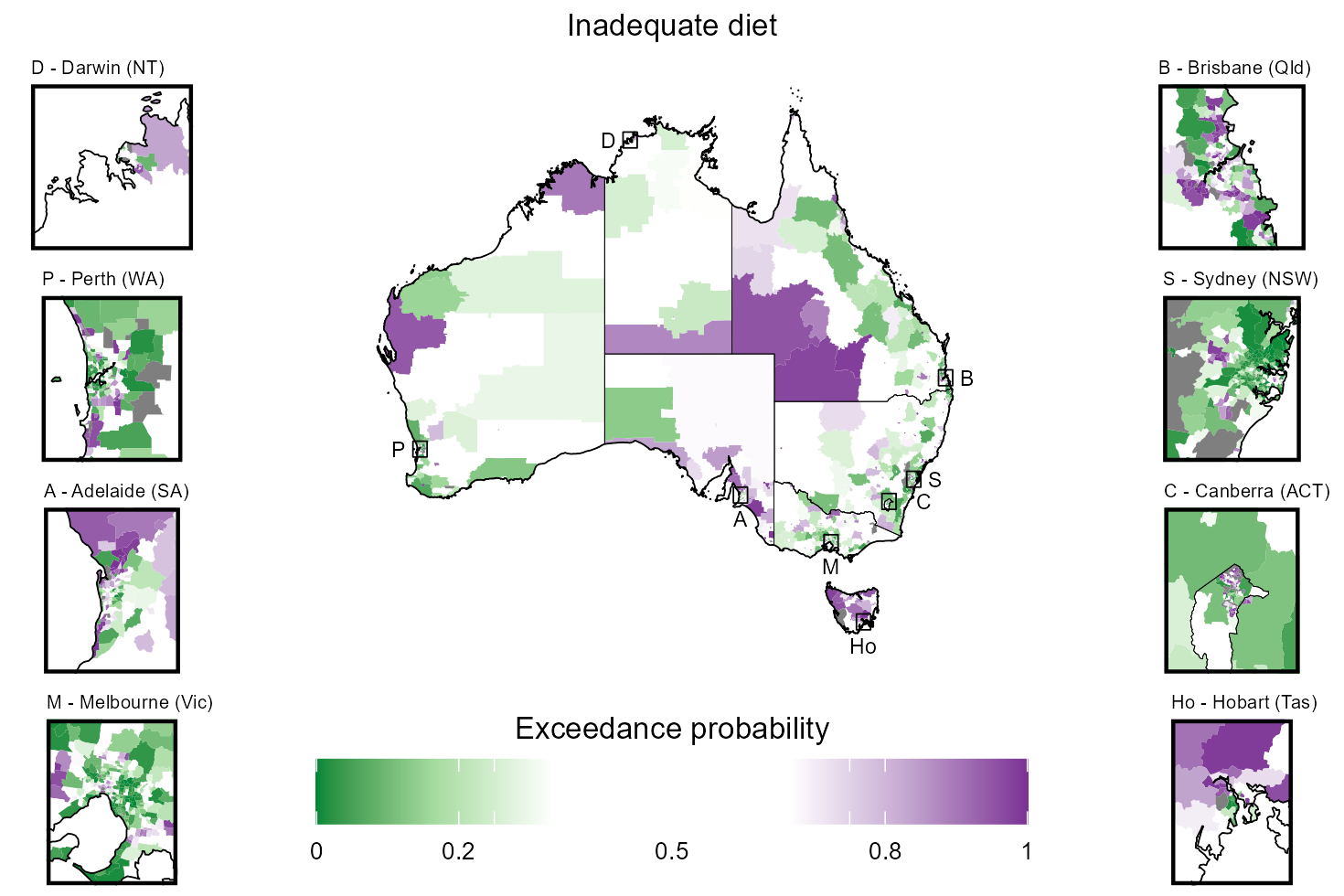}
         \caption{}
     \end{subfigure}
     \hspace{1em}
     \begin{subfigure}[b]{0.7\textwidth}
         \includegraphics[width=\textwidth]{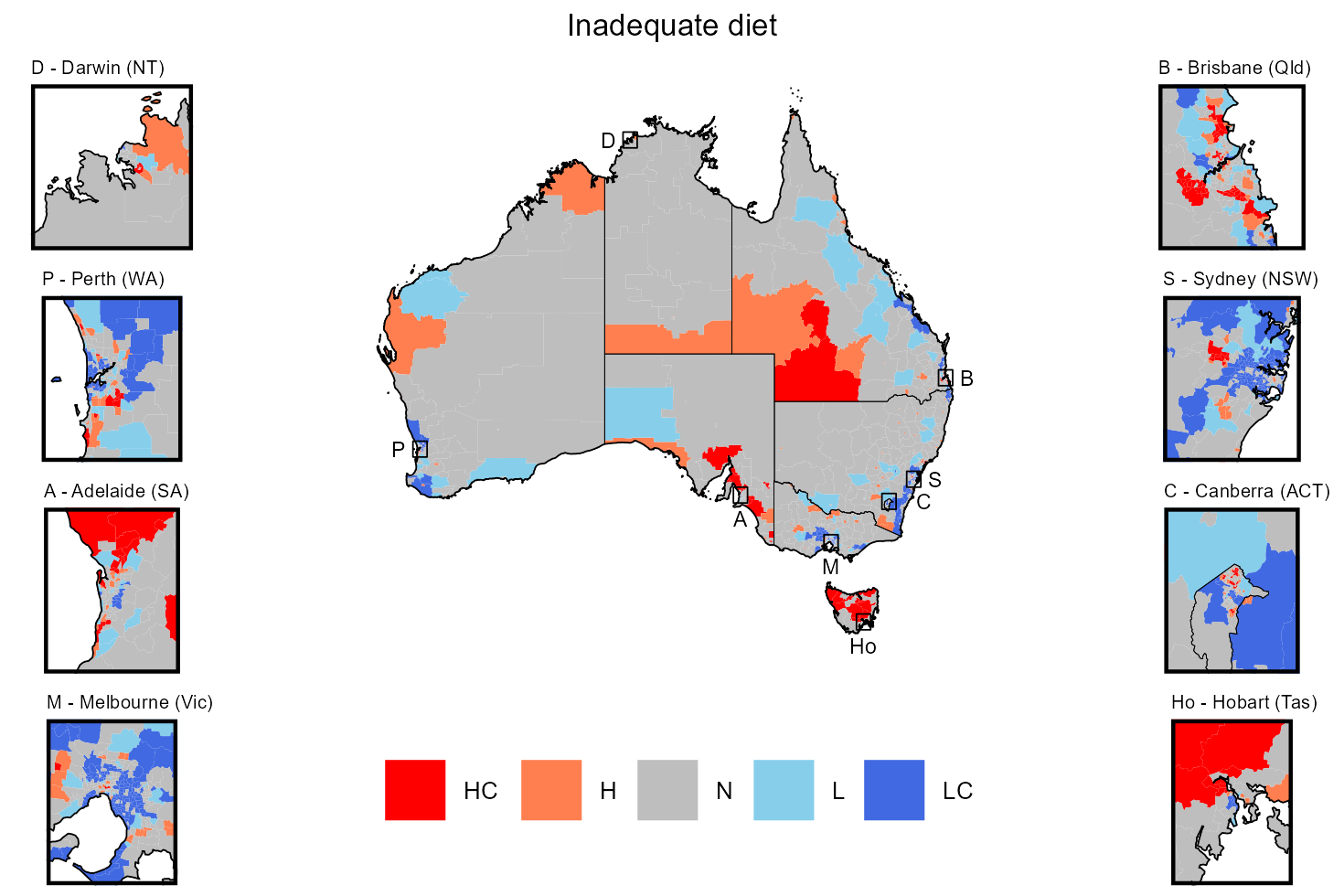}
         \caption{}
     \end{subfigure}
    \caption{\small See caption for \cref{fig:mapprevep_smoking}.}
    \label{fig:mapprevep_diet}
\end{figure}

\subsection{Obese}

\begin{figure}[H]
    \centering
    \includegraphics{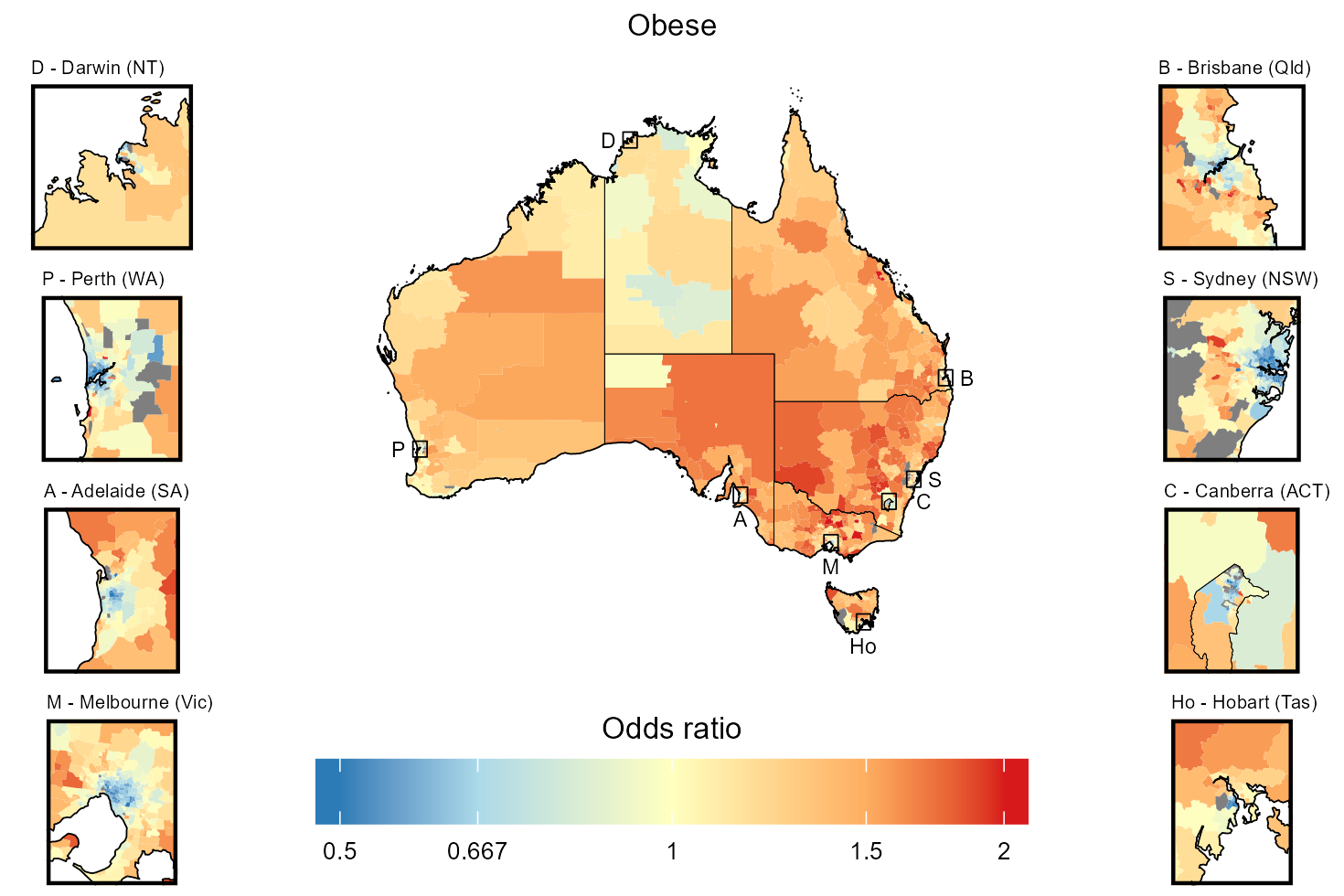}
    \caption{See caption for \cref{fig:or_smoking}}
    \label{fig:or_obesity}
\end{figure}

\begin{figure}[H]
     \begin{subfigure}[b]{0.7\textwidth}
         \includegraphics[width=\textwidth]{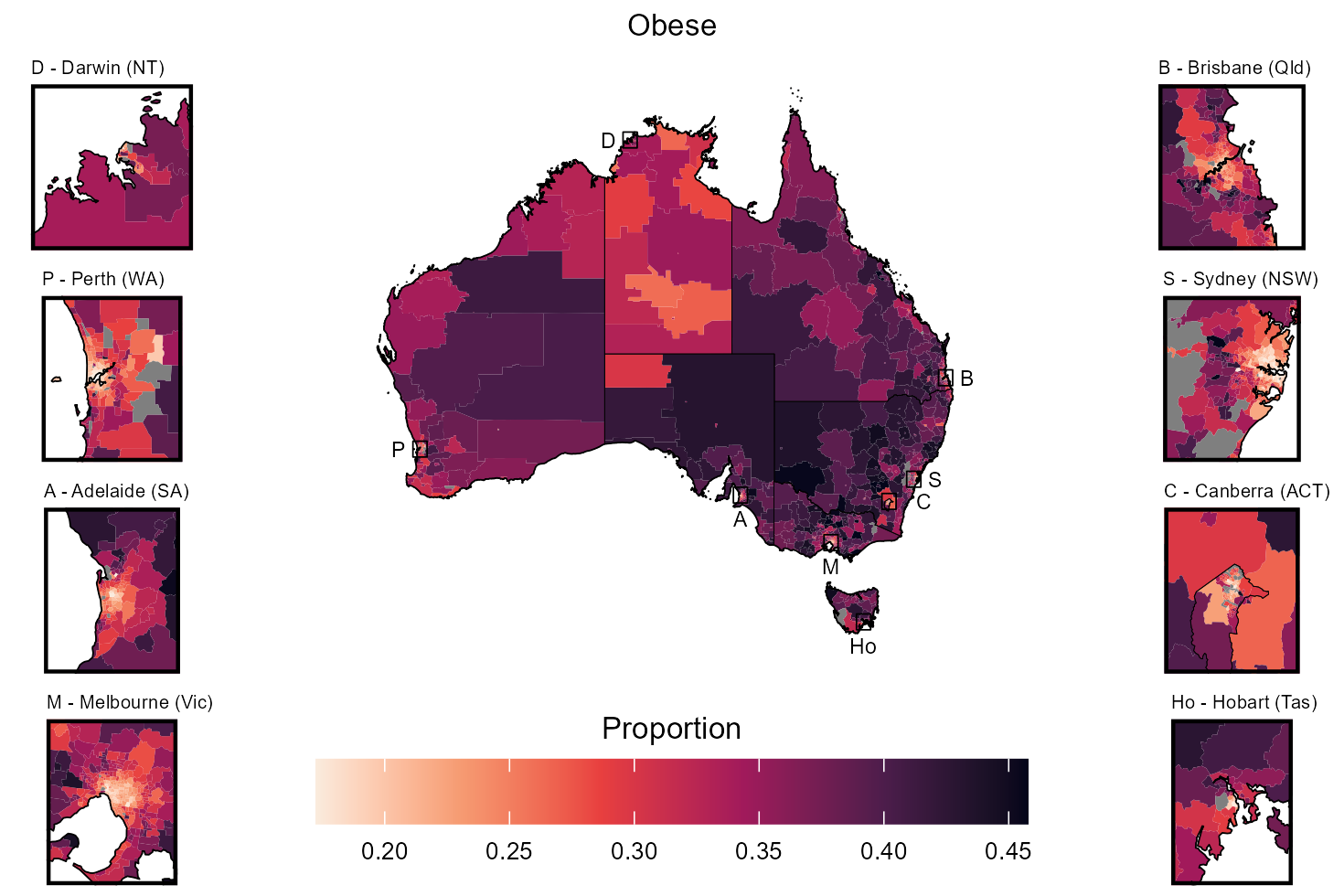}
         \caption{}
     \end{subfigure}
     \hspace{1em}
     \begin{subfigure}[b]{0.7\textwidth}
         \includegraphics[width=\textwidth]{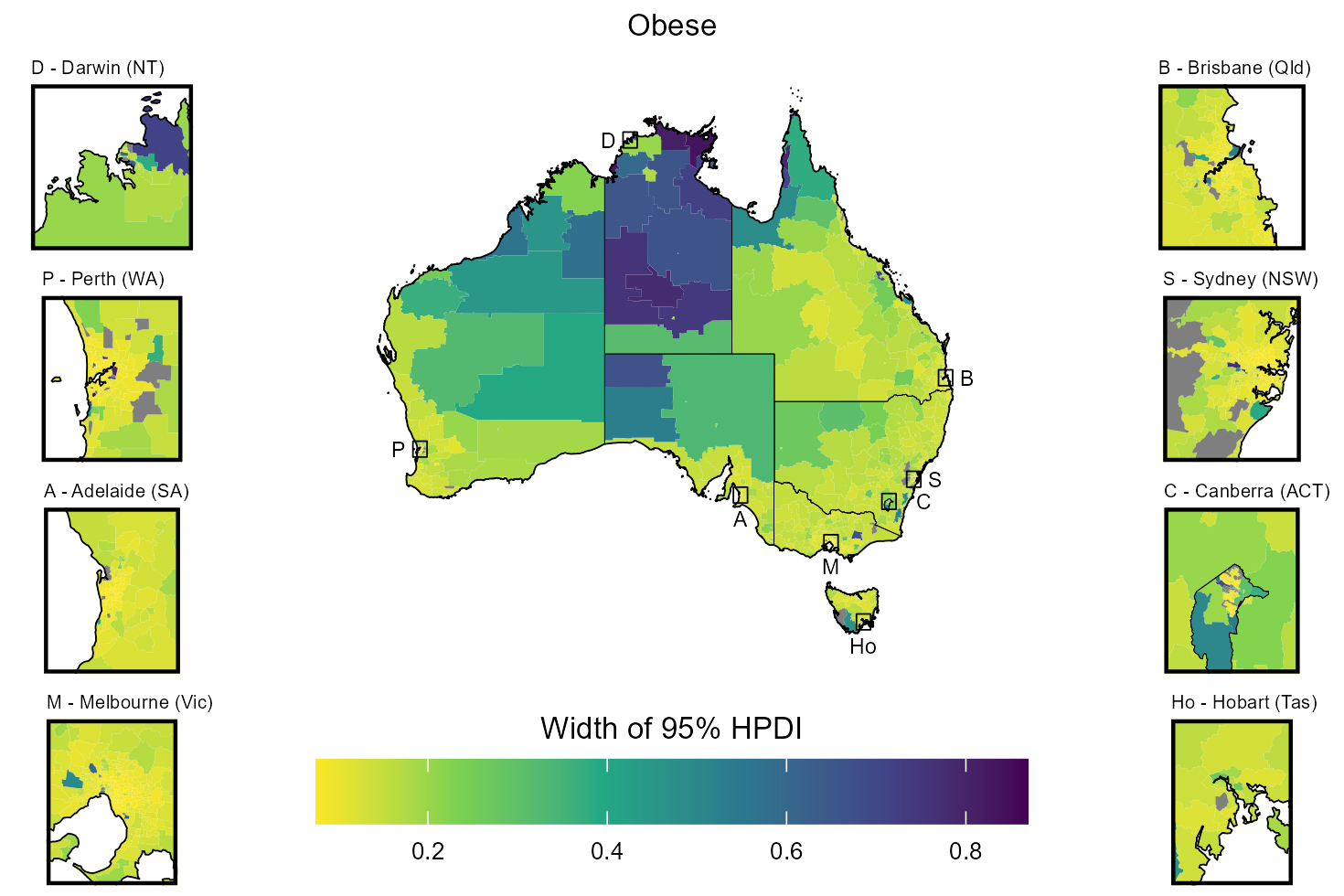}
         \caption{}
     \end{subfigure}
    \caption{\small See caption for \cref{fig:mapprev_smoking}.}
    \label{fig:mapprev_obesity}
\end{figure}

\newpage
\begin{figure}[H]
     \begin{subfigure}[b]{0.7\textwidth}
         \includegraphics[width=\textwidth]{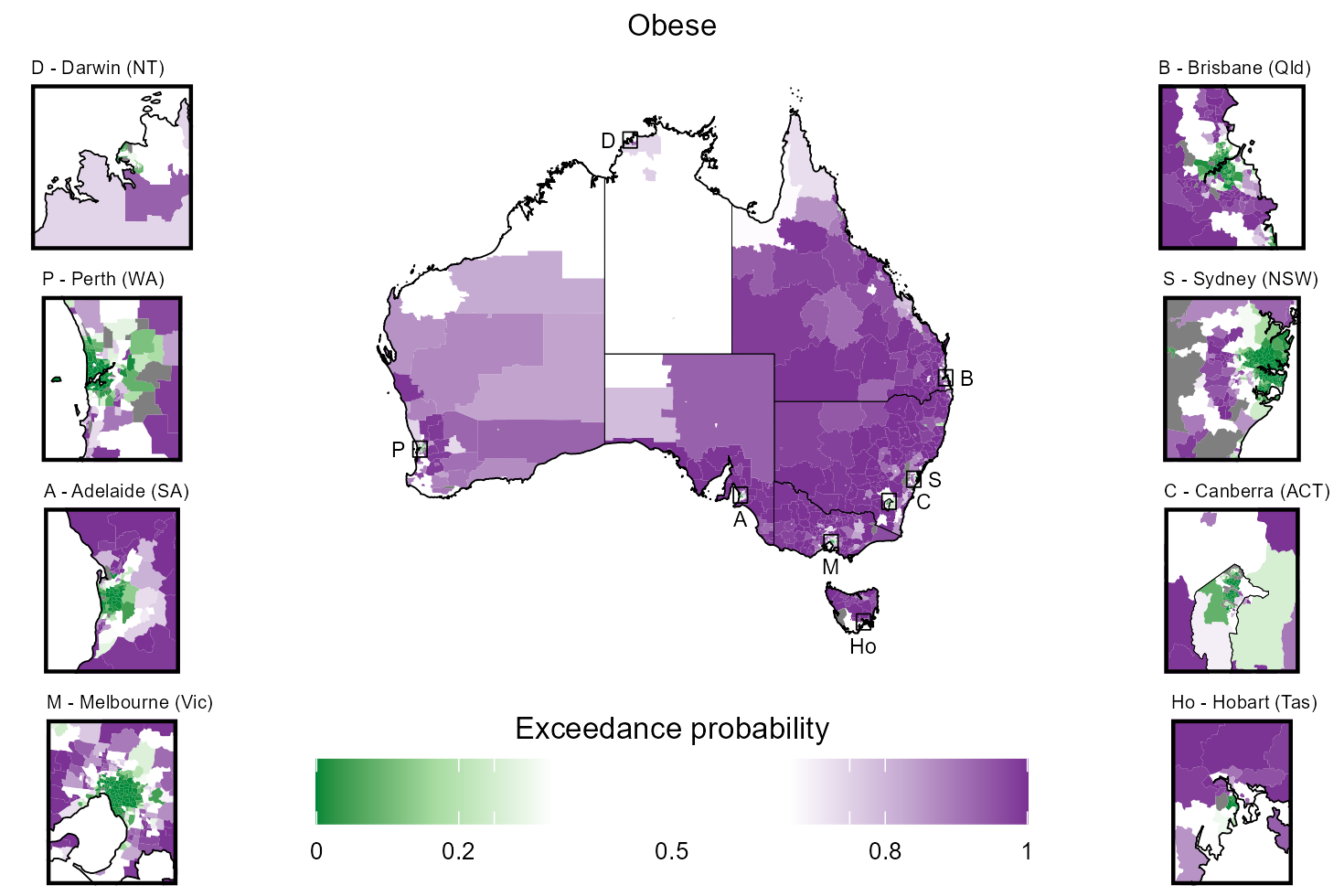}
         \caption{}
     \end{subfigure}
     \hspace{1em}
     \begin{subfigure}[b]{0.7\textwidth}
         \includegraphics[width=\textwidth]{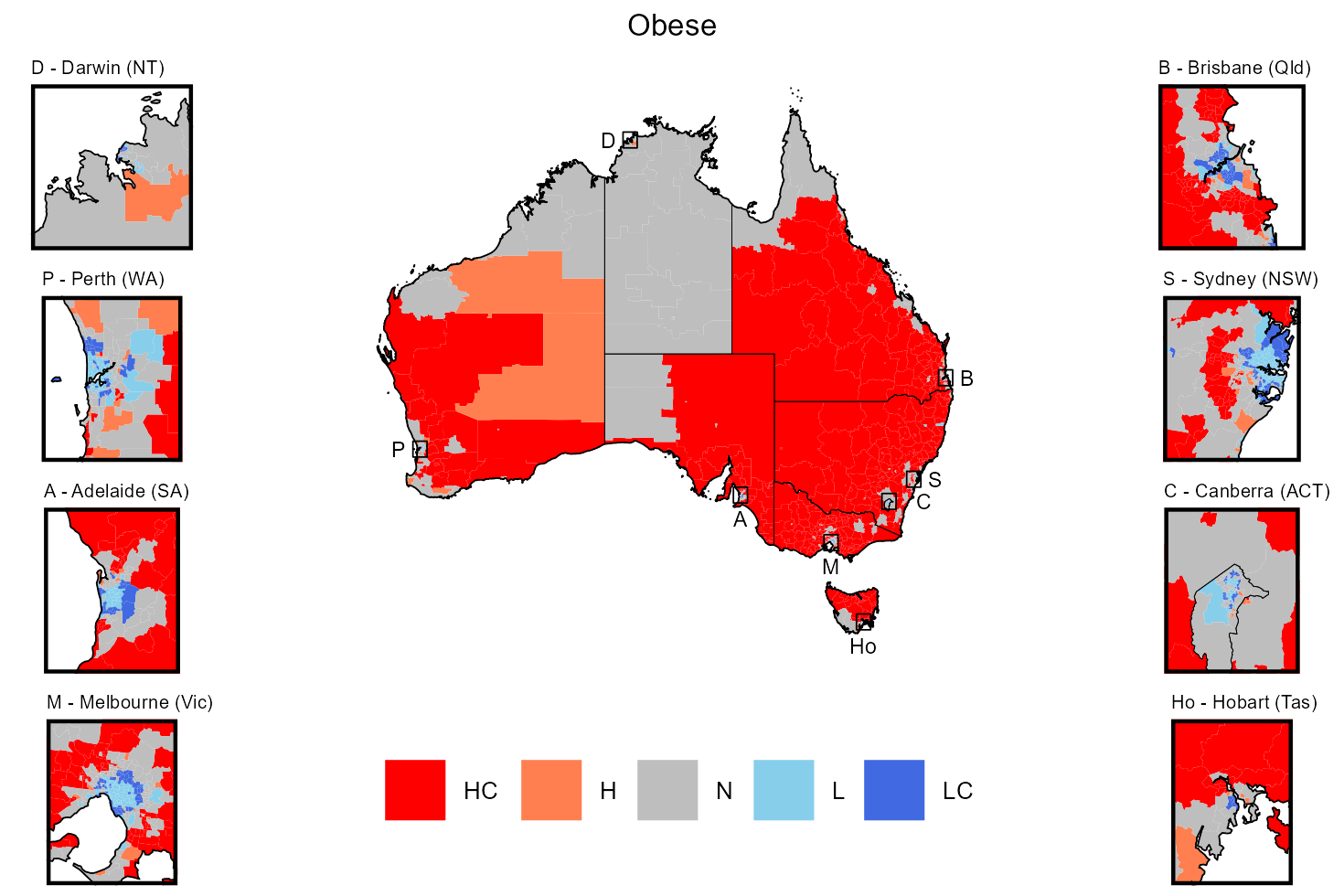}
         \caption{}
     \end{subfigure}
    \caption{\small See caption for \cref{fig:mapprevep_smoking}.}
    \label{fig:mapprevep_obesity}
\end{figure}

\subsection{Overweight}

\begin{figure}[H]
    \centering
    \includegraphics{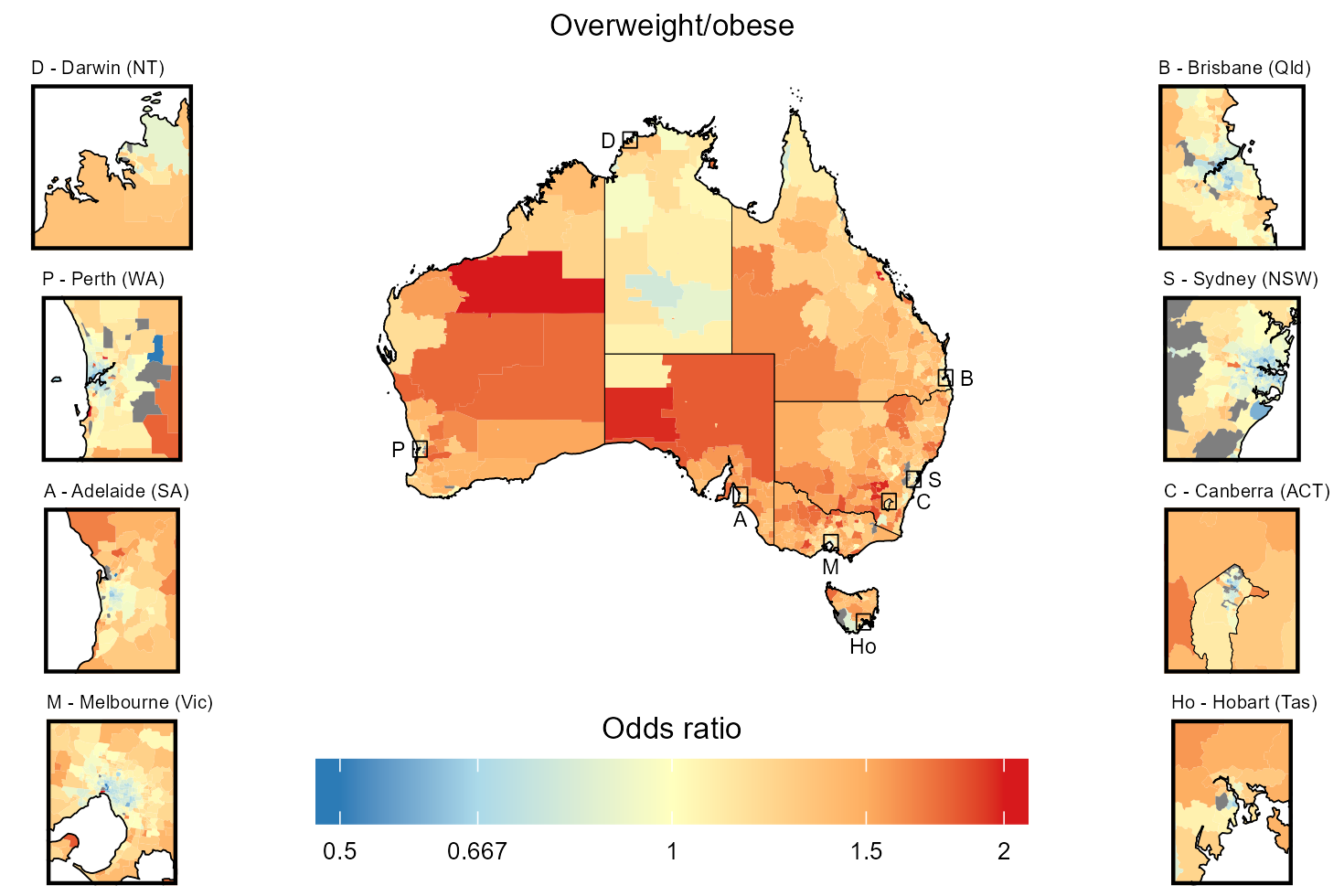}
    \caption{See caption for \cref{fig:or_smoking}}
    \label{fig:or_overweight}
\end{figure}

\begin{figure}[H]
     \begin{subfigure}[b]{0.7\textwidth}
         \includegraphics[width=\textwidth]{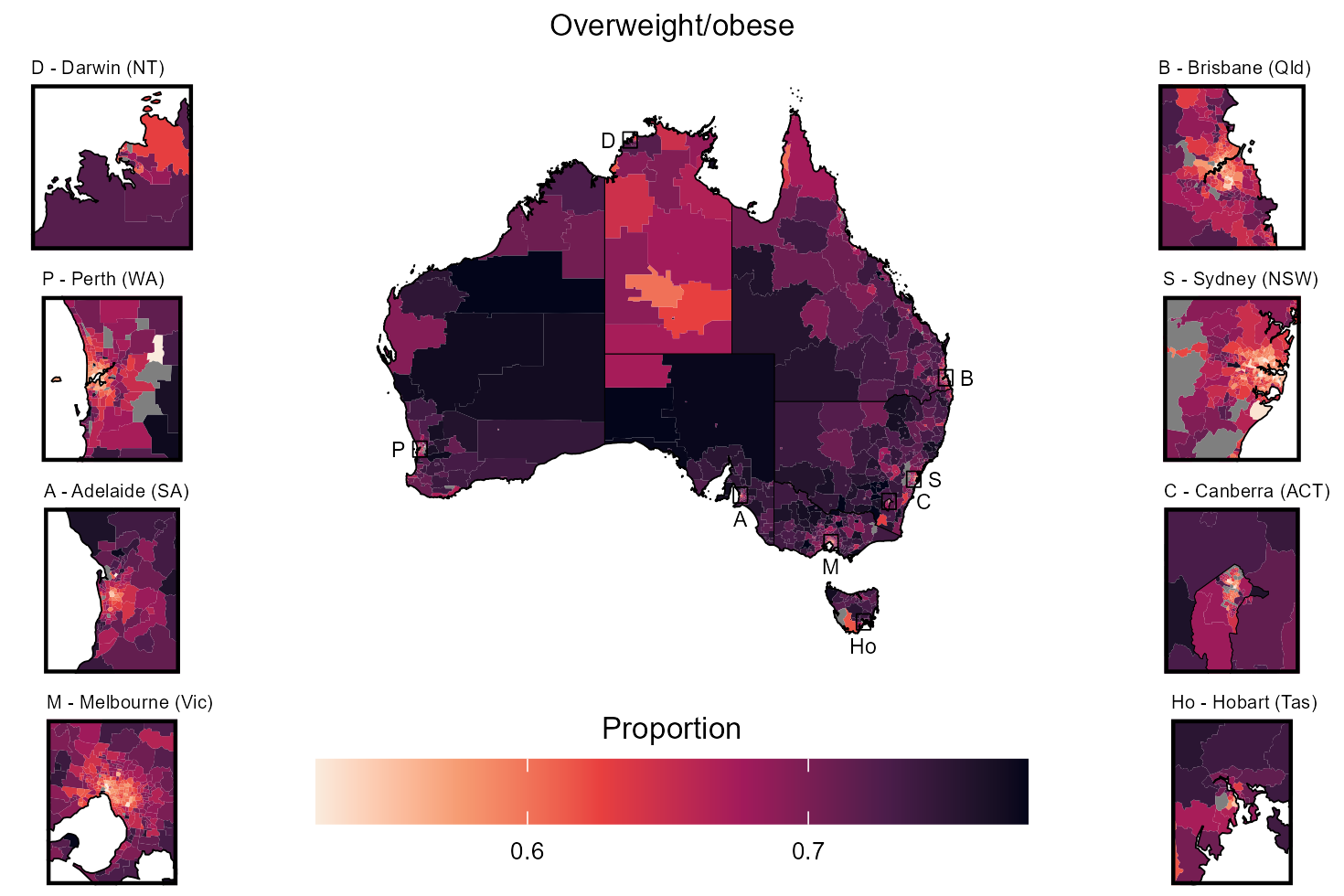}
         \caption{}
     \end{subfigure}
     \hspace{1em}
     \begin{subfigure}[b]{0.7\textwidth}
         \includegraphics[width=\textwidth]{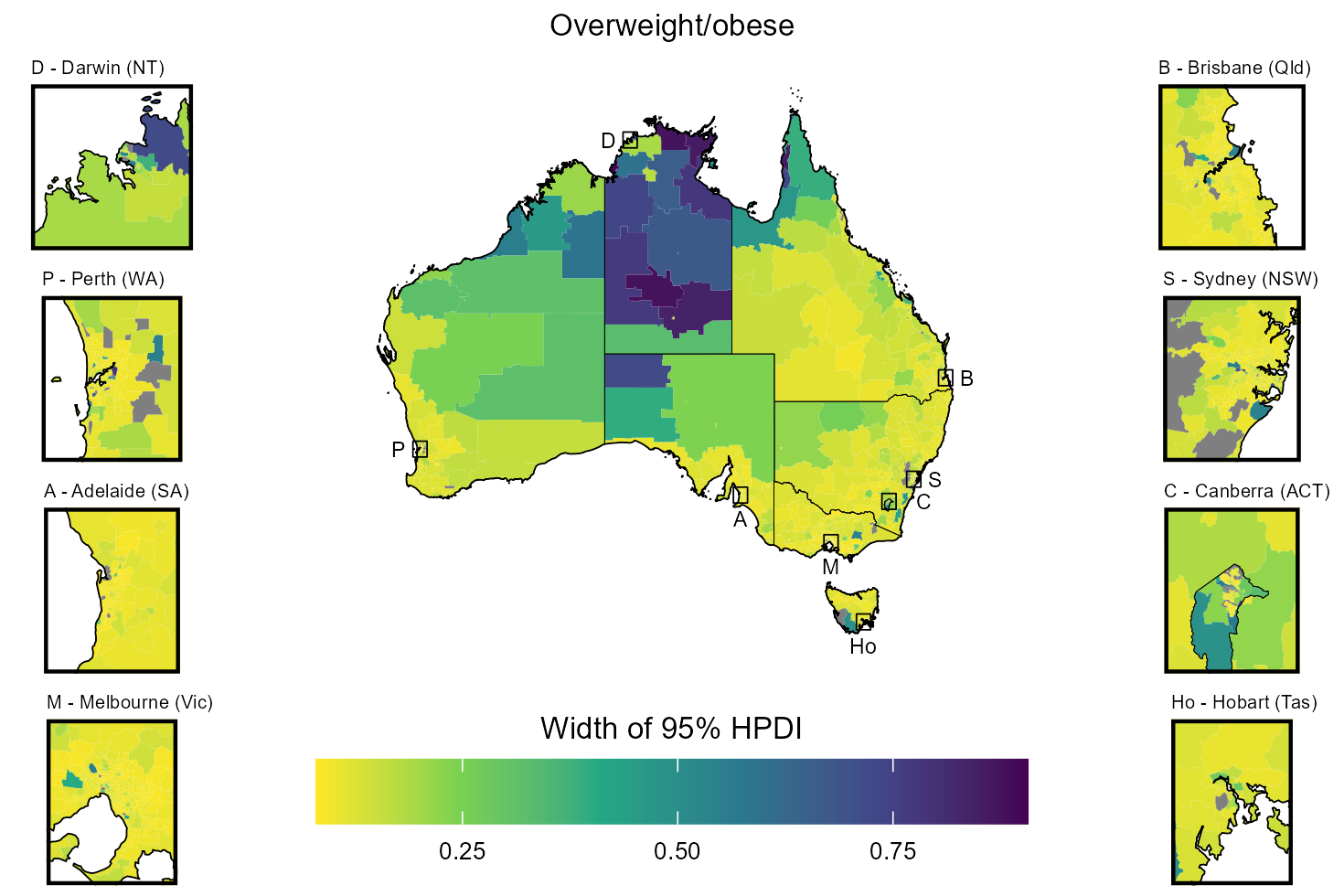}
         \caption{}
     \end{subfigure}
    \caption{\small See caption for \cref{fig:mapprev_smoking}.}
    \label{fig:mapprev_overweight}
\end{figure}

\newpage
\begin{figure}[H]
     \begin{subfigure}[b]{0.7\textwidth}
         \includegraphics[width=\textwidth]{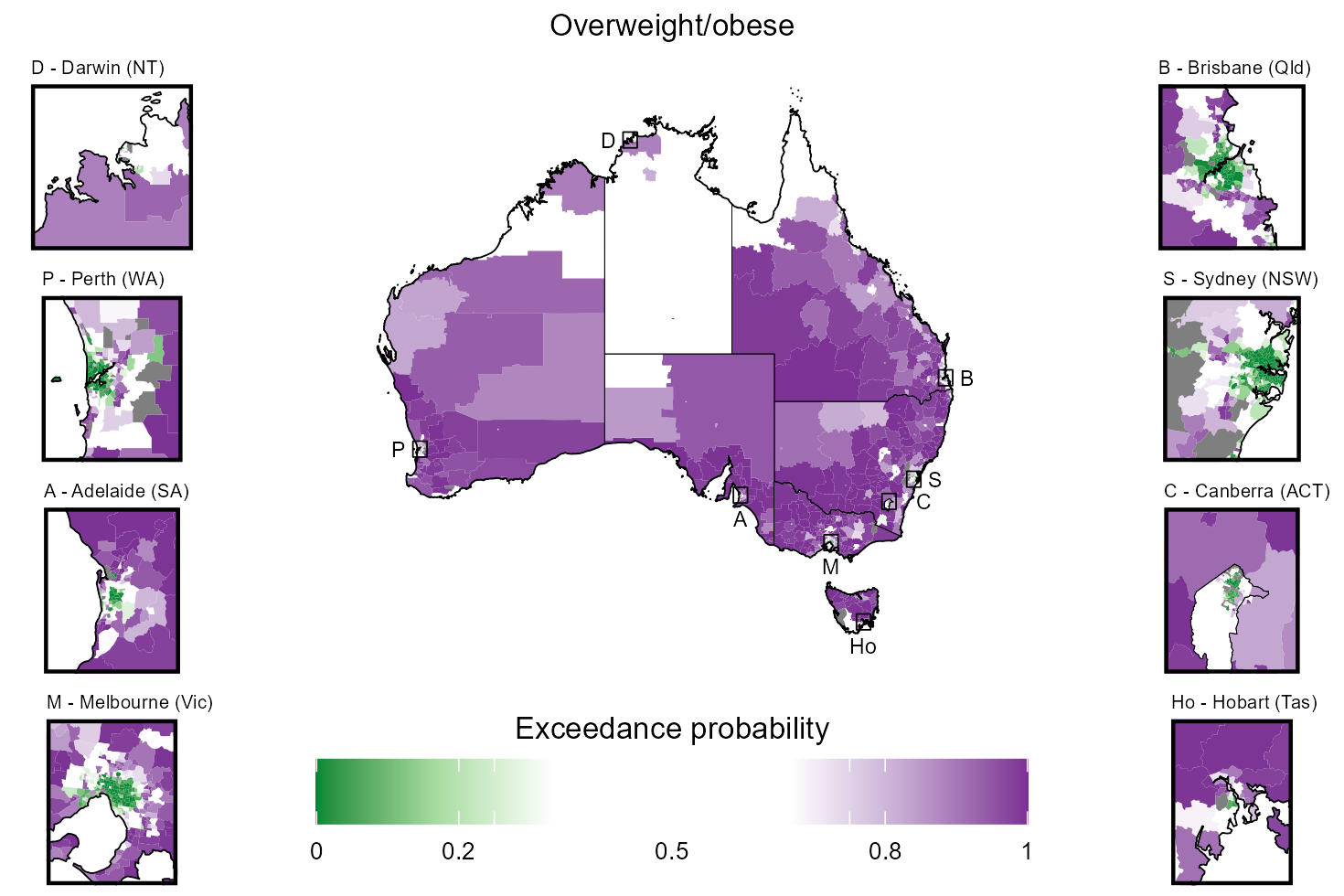}
         \caption{}
     \end{subfigure}
     \hspace{1em}
     \begin{subfigure}[b]{0.7\textwidth}
         \includegraphics[width=\textwidth]{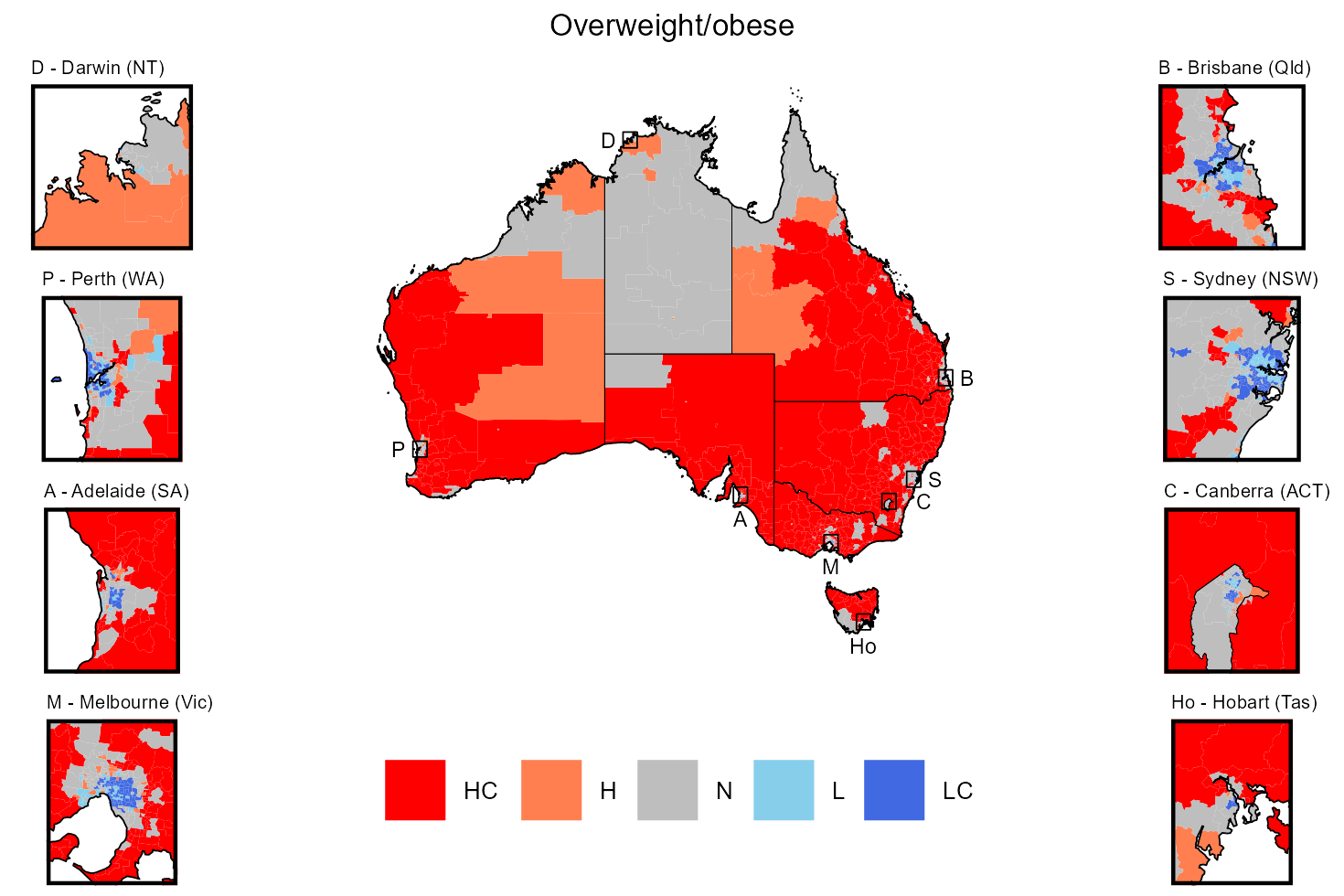}
         \caption{}
     \end{subfigure}
    \caption{\small See caption for \cref{fig:mapprevep_smoking}.}
    \label{fig:mapprevep_overweight}
\end{figure}

\subsection{Risky waist circumference}

\begin{figure}[H]
    \centering
    \includegraphics{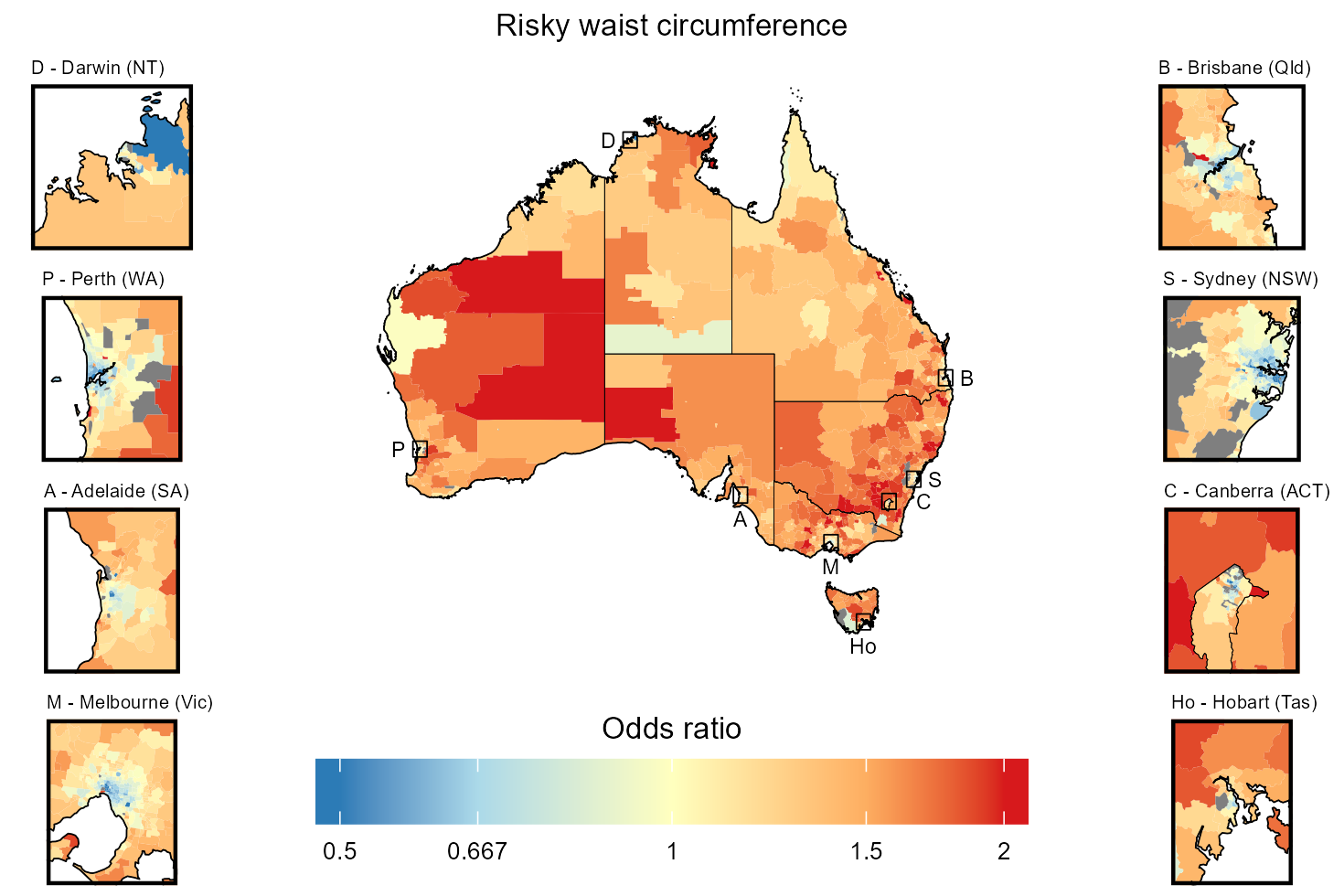}
    \caption{See caption for \cref{fig:or_smoking}}
    \label{fig:or_waist_circum}
\end{figure}

\begin{figure}[H]
     \begin{subfigure}[b]{0.7\textwidth}
         \includegraphics[width=\textwidth]{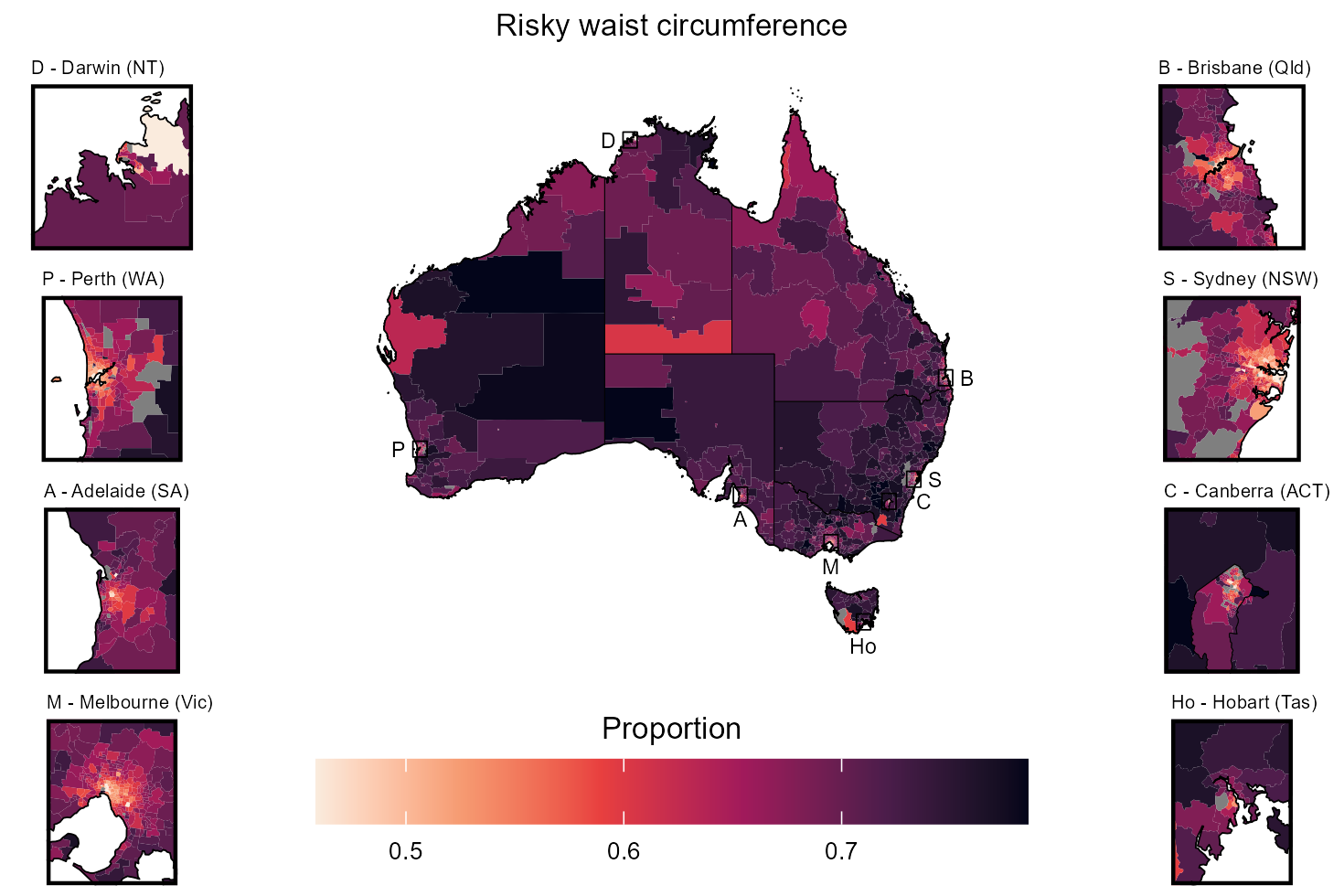}
         \caption{}
     \end{subfigure}
     \hspace{1em}
     \begin{subfigure}[b]{0.7\textwidth}
         \includegraphics[width=\textwidth]{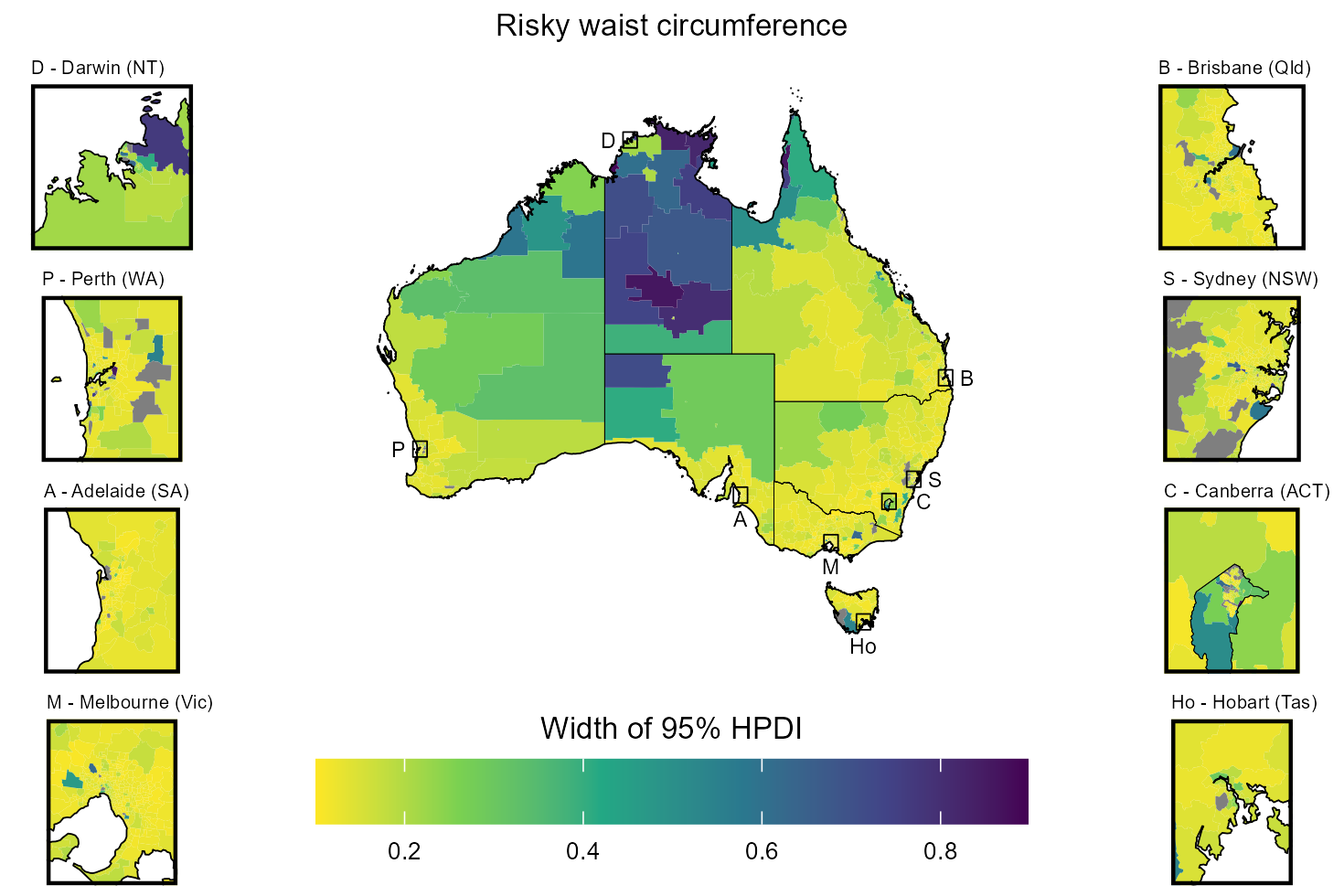}
         \caption{}
     \end{subfigure}
    \caption{\small See caption for \cref{fig:mapprev_smoking}.}
    \label{fig:mapprev_waist_circum}
\end{figure}

\newpage
\begin{figure}[H]
     \begin{subfigure}[b]{0.7\textwidth}
         \includegraphics[width=\textwidth]{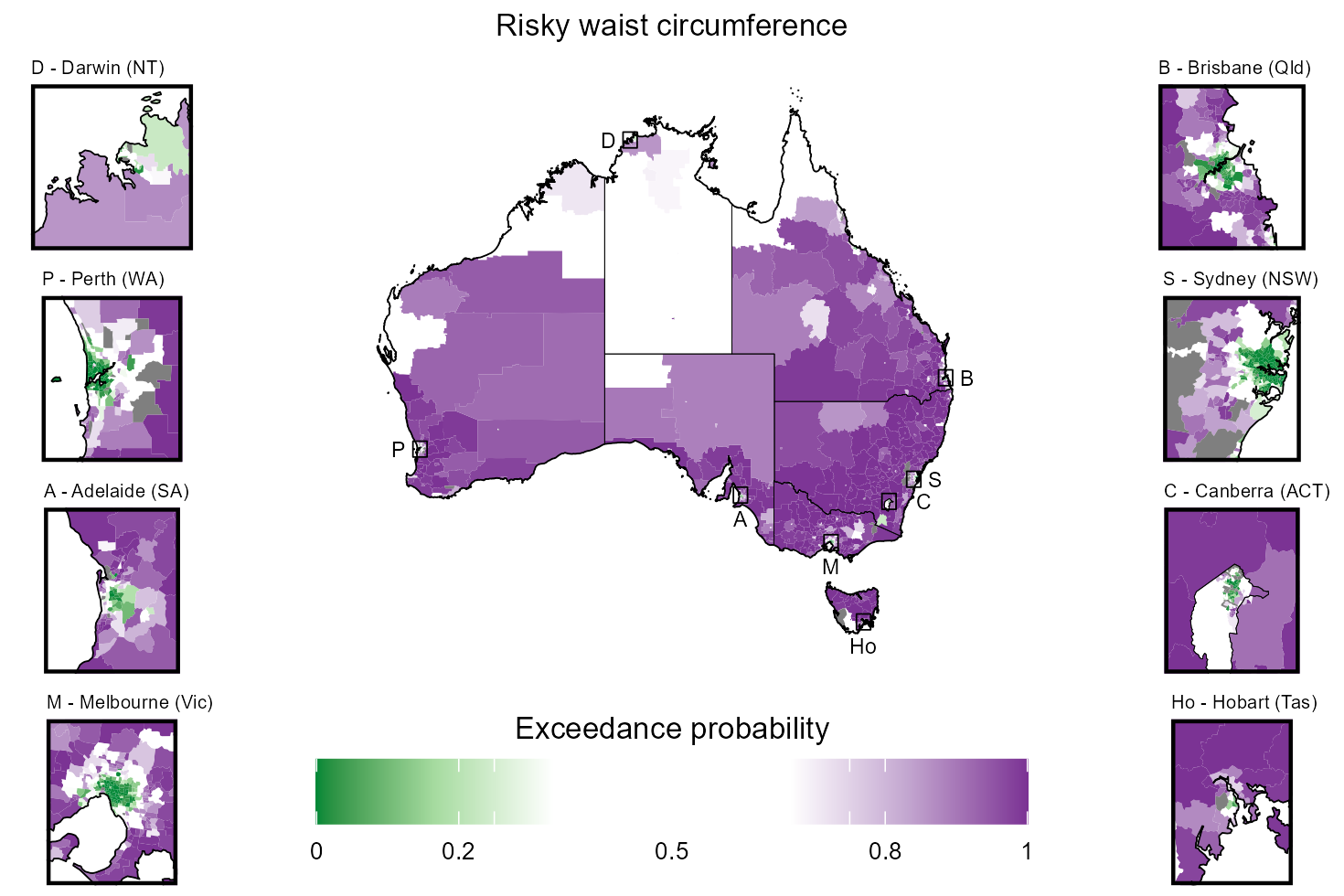}
         \caption{}
     \end{subfigure}
     \hspace{1em}
     \begin{subfigure}[b]{0.7\textwidth}
         \includegraphics[width=\textwidth]{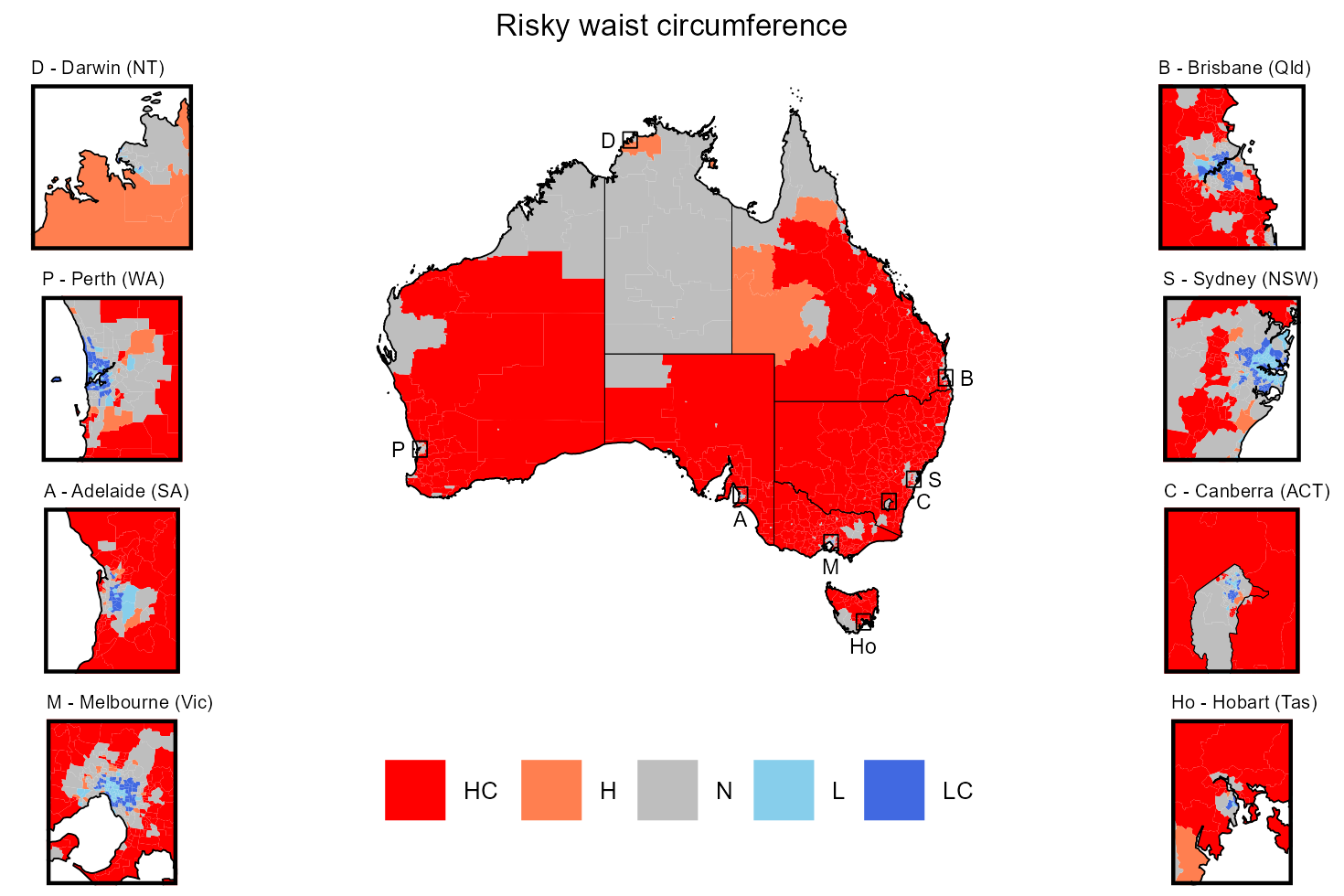}
         \caption{}
     \end{subfigure}
    \caption{\small See caption for \cref{fig:mapprevep_smoking}.}
    \label{fig:mapprevep_waist_circum}
\end{figure}

\subsection{Inadequate activity (leisure)}

\begin{figure}[H]
    \centering
    \includegraphics{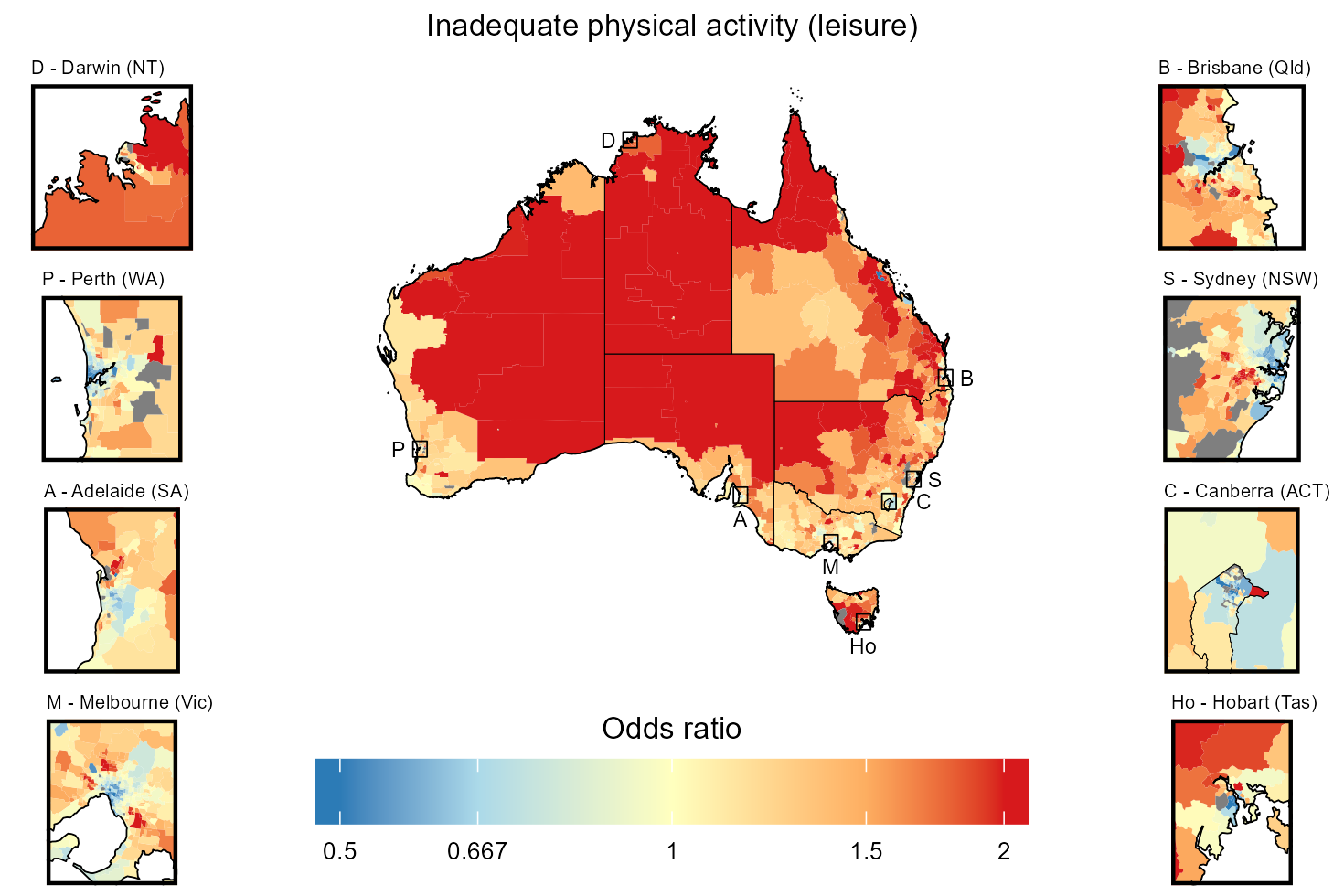}
    \caption{See caption for \cref{fig:or_smoking}}
    \label{fig:or_activityleis}
\end{figure}

\begin{figure}[H]
     \begin{subfigure}[b]{0.7\textwidth}
         \includegraphics[width=\textwidth]{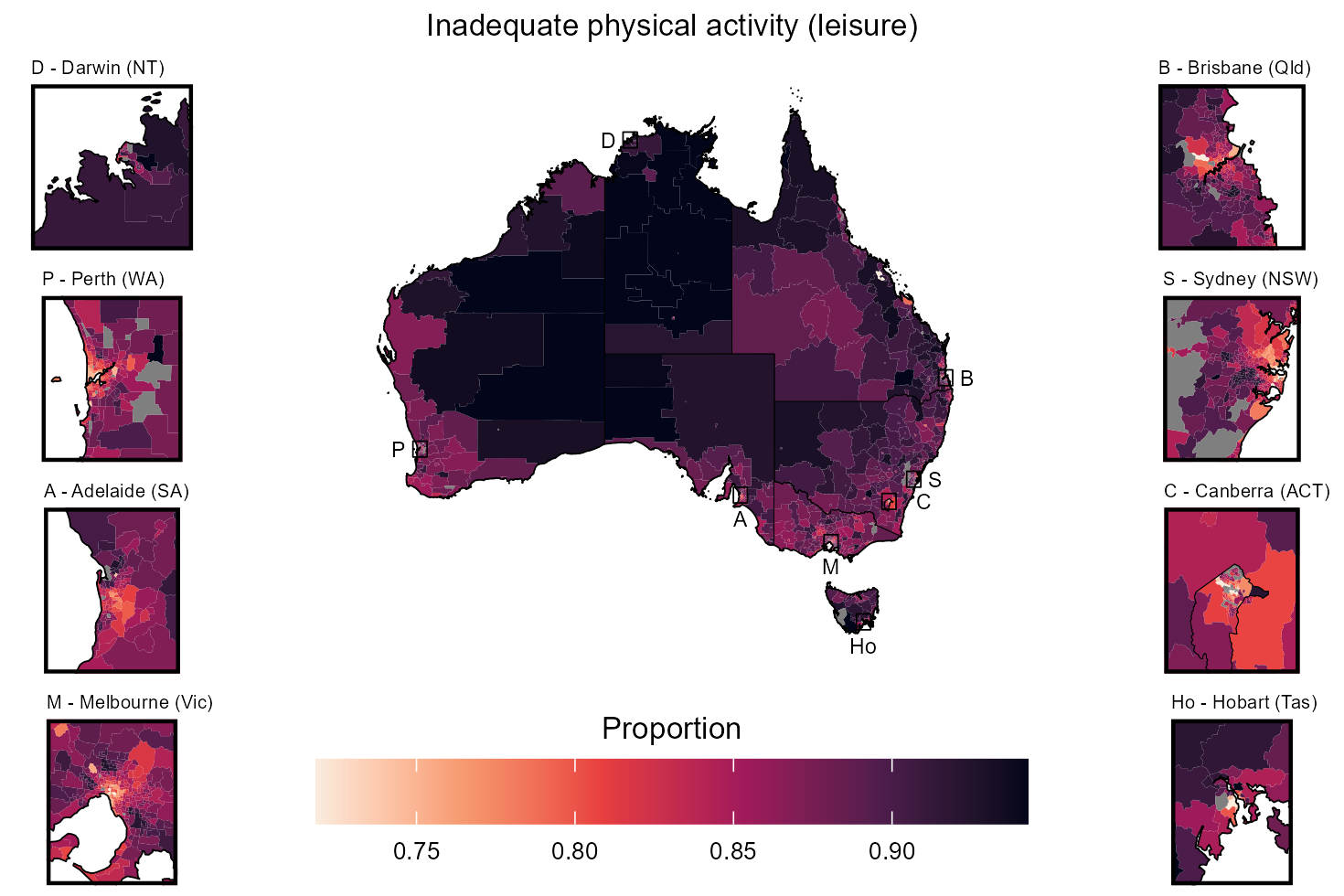}
         \caption{}
     \end{subfigure}
     \hspace{1em}
     \begin{subfigure}[b]{0.7\textwidth}
         \includegraphics[width=\textwidth]{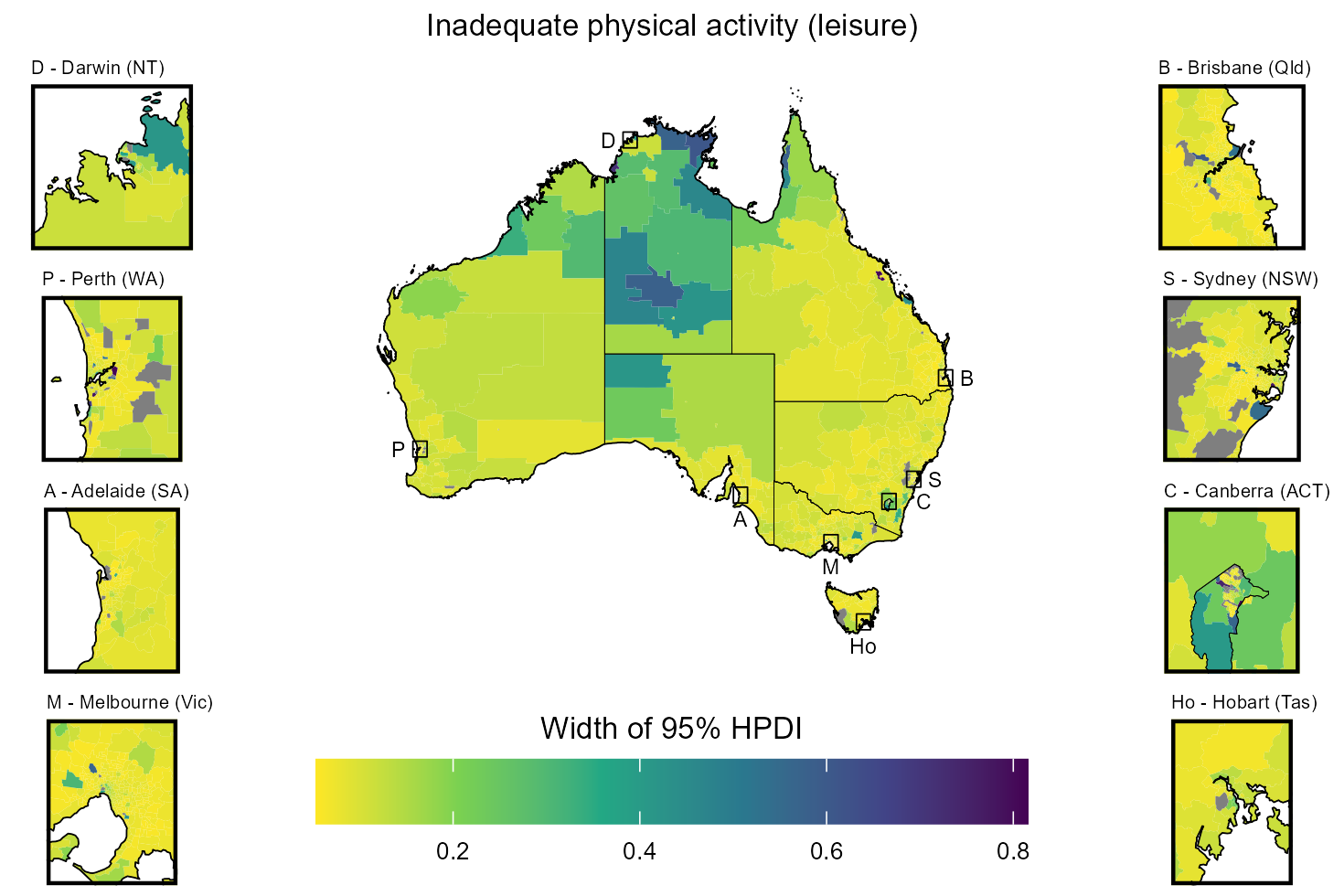}
         \caption{}
     \end{subfigure}
    \caption{\small See caption for \cref{fig:mapprev_smoking}.}
    \label{fig:mapprev_activityleis}
\end{figure}

\newpage
\begin{figure}[H]
     \begin{subfigure}[b]{0.7\textwidth}
         \includegraphics[width=\textwidth]{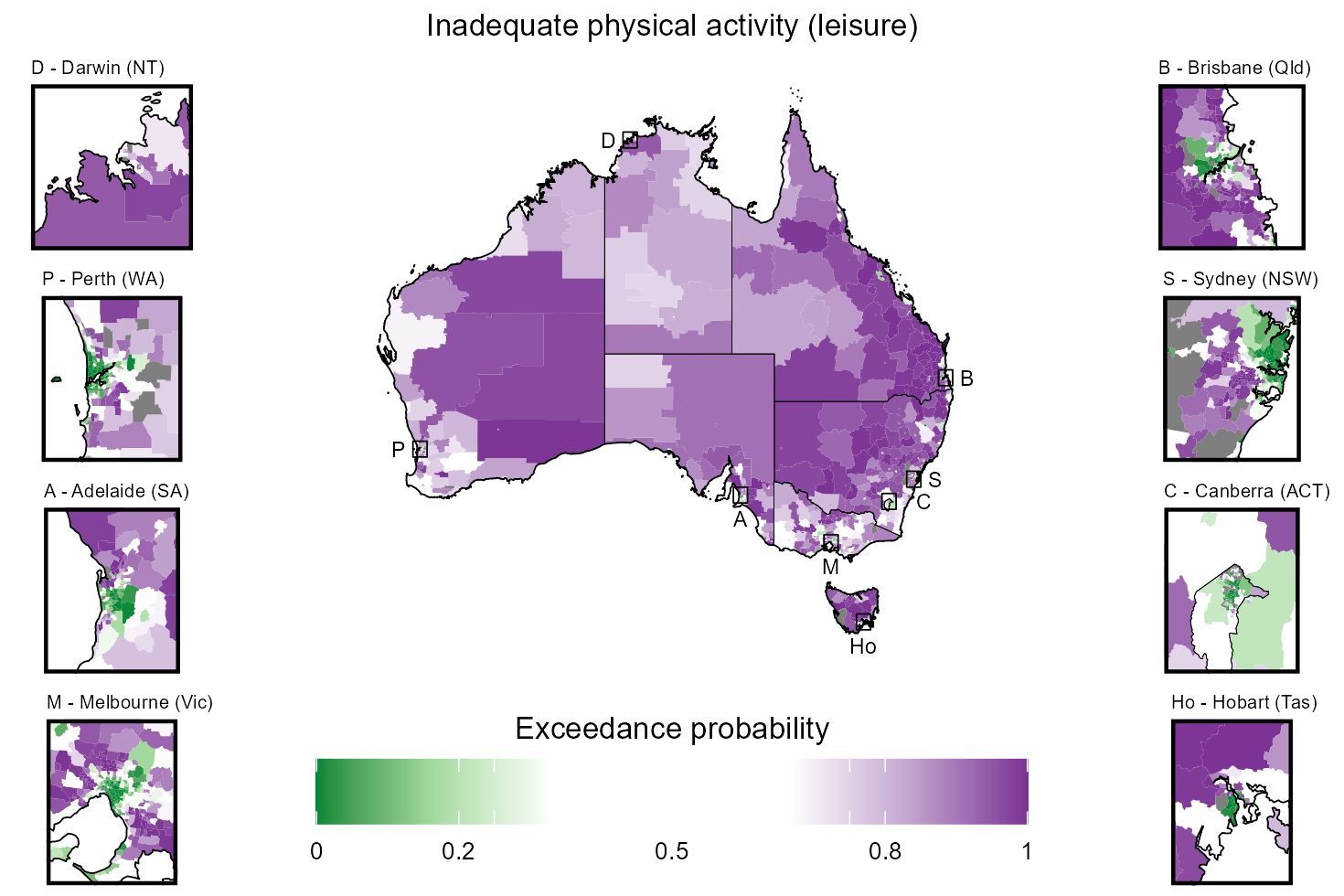}
         \caption{}
     \end{subfigure}
     \hspace{1em}
     \begin{subfigure}[b]{0.7\textwidth}
         \includegraphics[width=\textwidth]{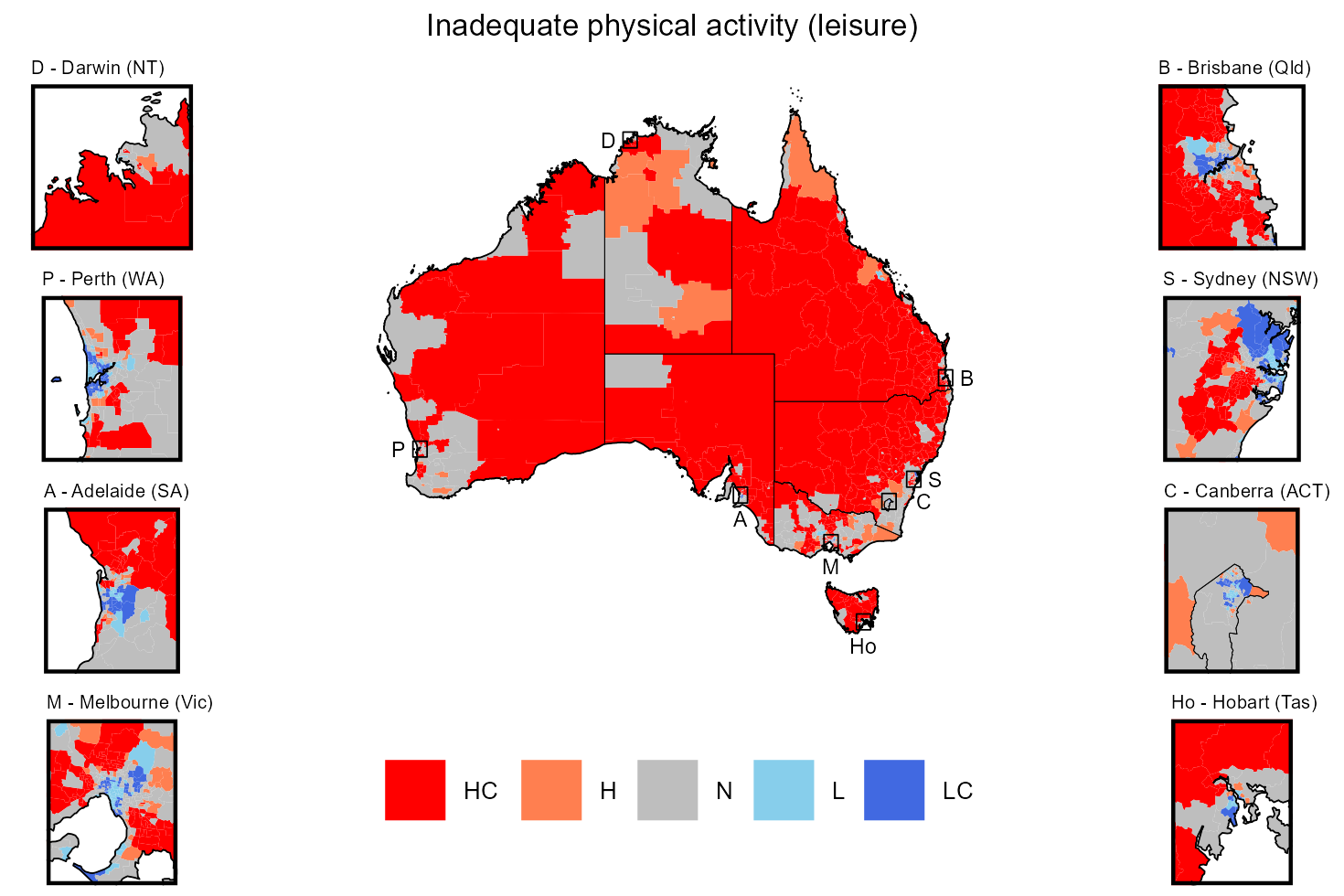}
         \caption{}
     \end{subfigure}
    \caption{\small See caption for \cref{fig:mapprevep_smoking}.}
    \label{fig:mapprevep_activityleis}
\end{figure}

\subsection{Inadequate activity (all)}

\begin{figure}[H]
    \centering
    \includegraphics{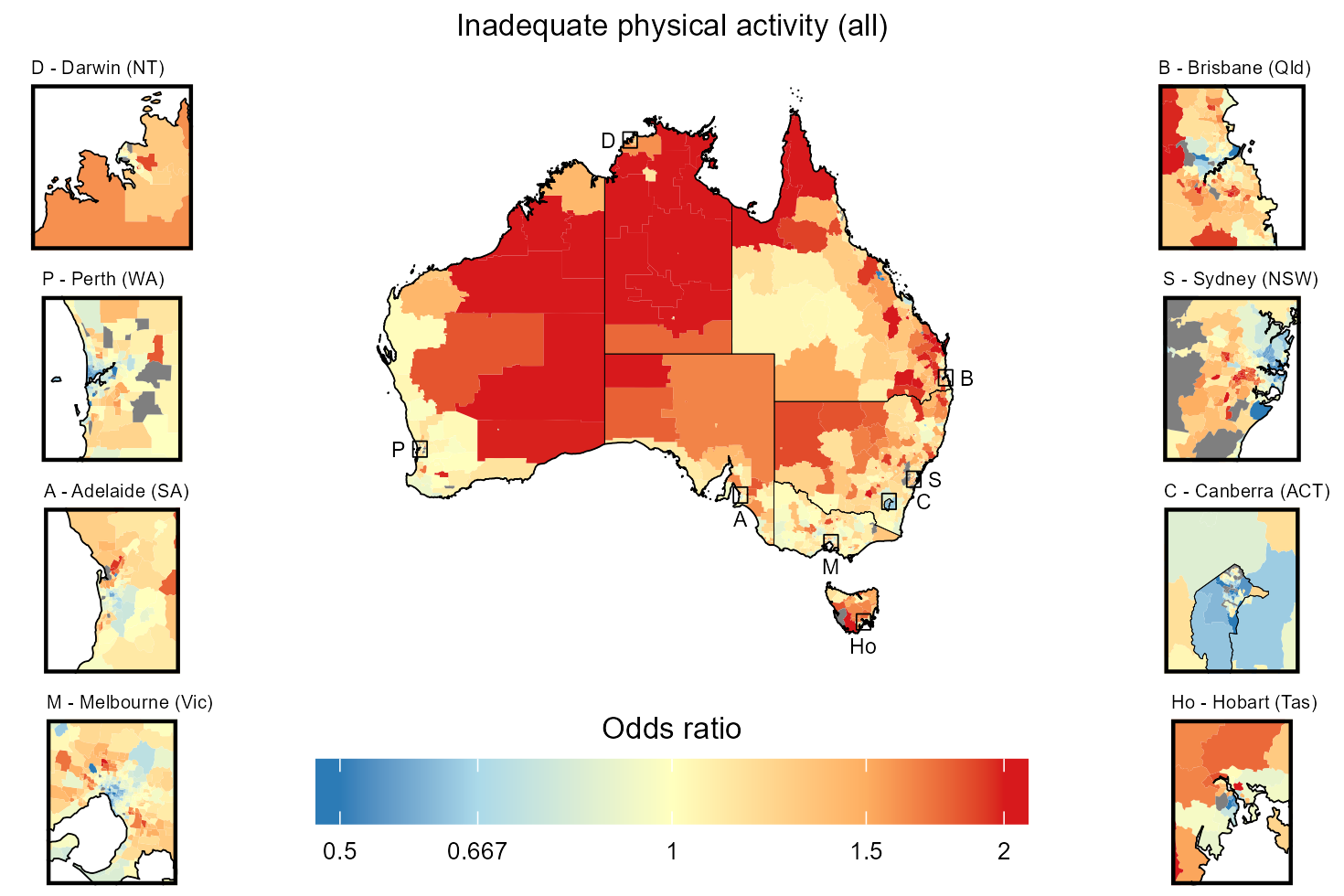}
    \caption{See caption for \cref{fig:or_smoking}}
    \label{fig:or_activityleiswkpl}
\end{figure}

\begin{figure}[H]
     \begin{subfigure}[b]{0.7\textwidth}
         \includegraphics[width=\textwidth]{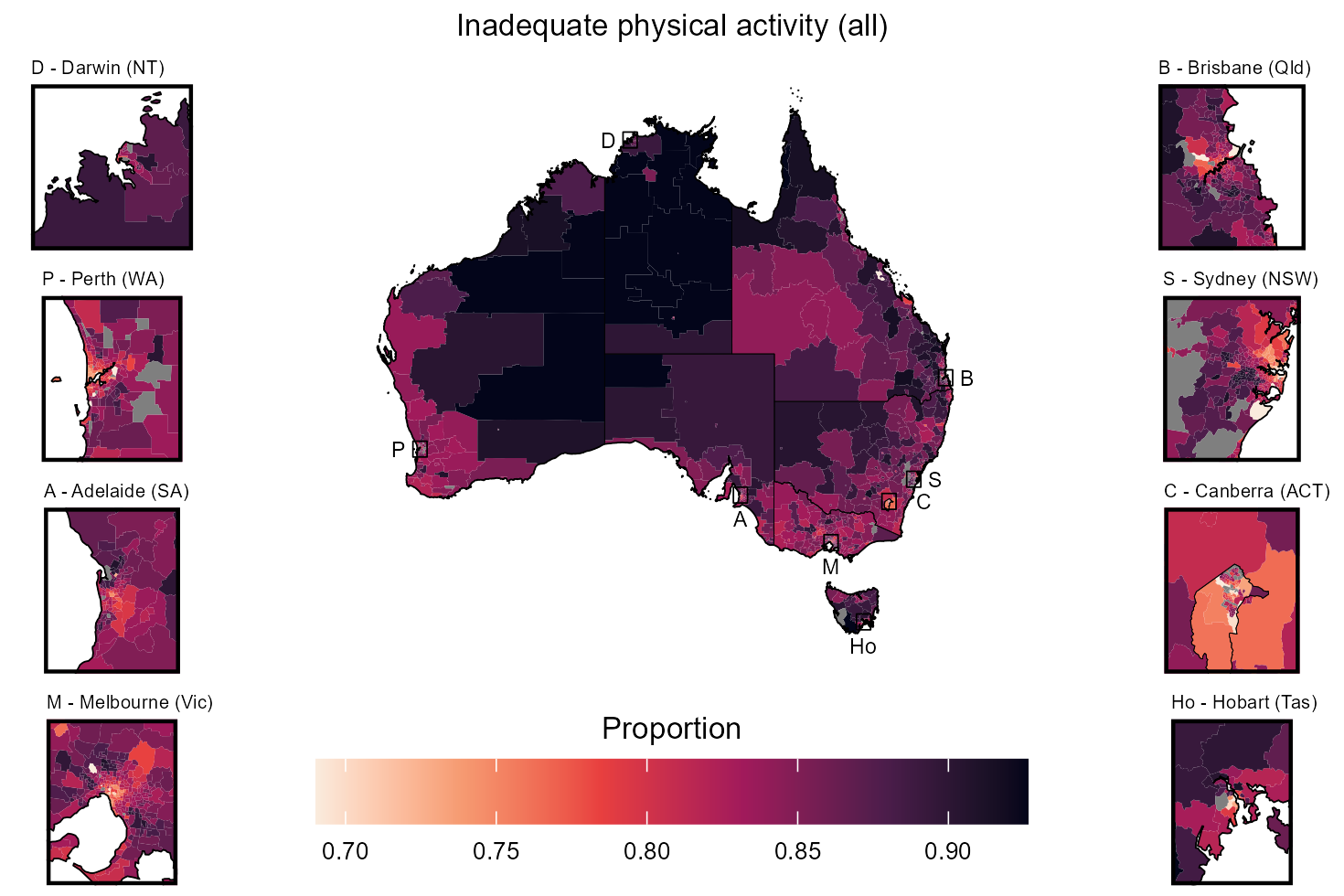}
         \caption{}
     \end{subfigure}
     \hspace{1em}
     \begin{subfigure}[b]{0.7\textwidth}
         \includegraphics[width=\textwidth]{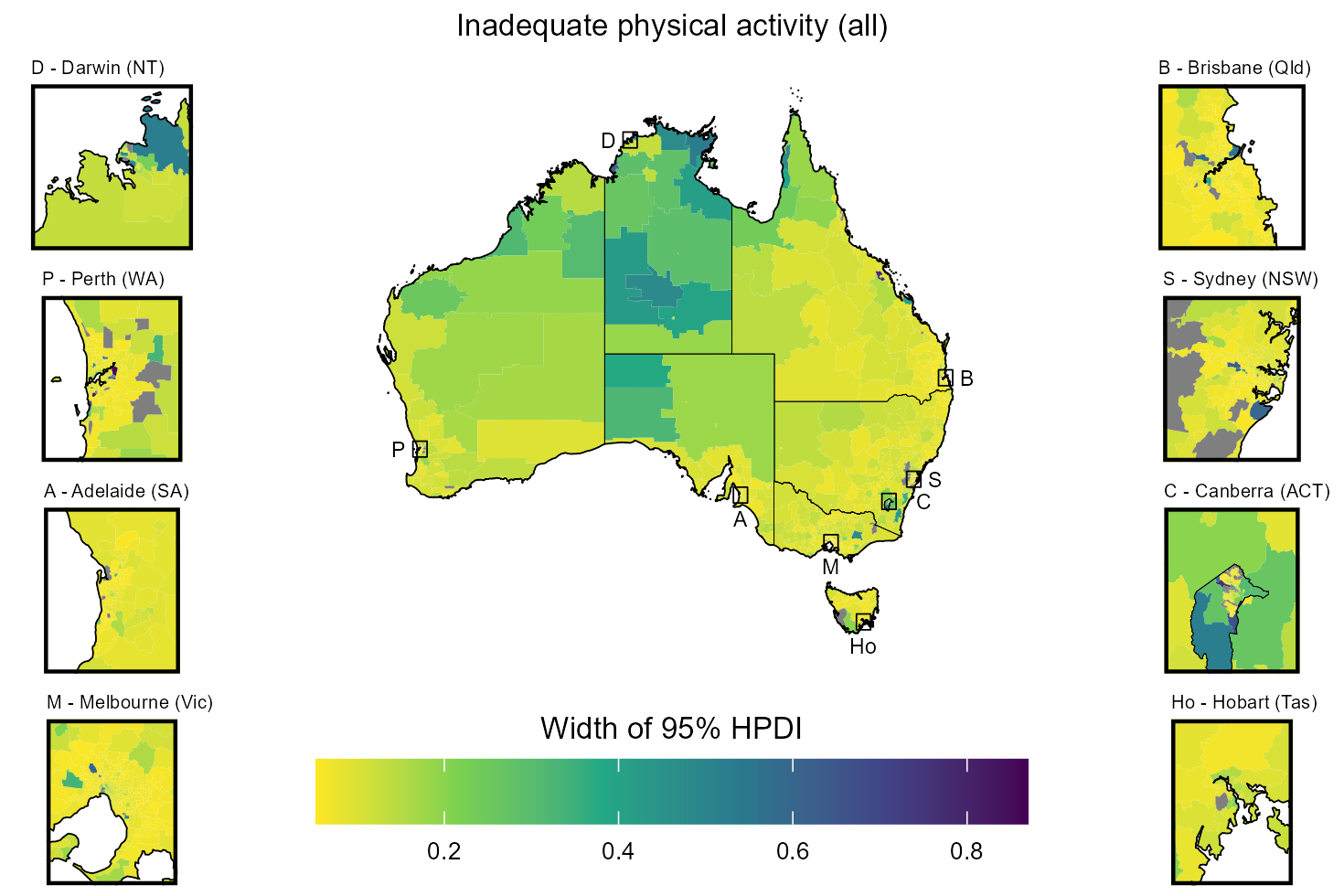}
         \caption{}
     \end{subfigure}
    \caption{\small See caption for \cref{fig:mapprev_smoking}.}
    \label{fig:mapprev_activityleiswkpl}
\end{figure}

\newpage
\begin{figure}[H]
     \begin{subfigure}[b]{0.7\textwidth}
         \includegraphics[width=\textwidth]{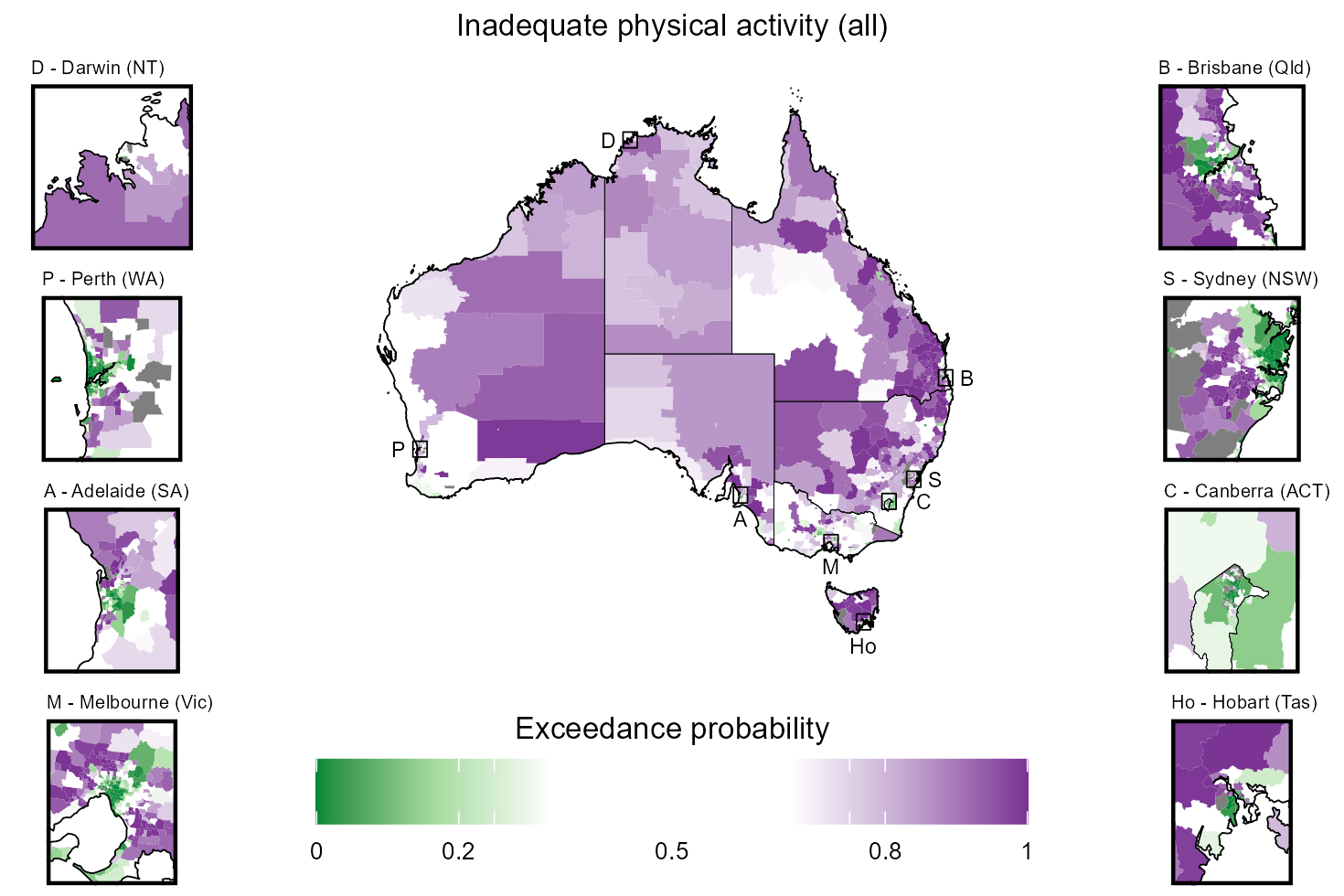}
         \caption{}
     \end{subfigure}
     \hspace{1em}
     \begin{subfigure}[b]{0.7\textwidth}
         \includegraphics[width=\textwidth]{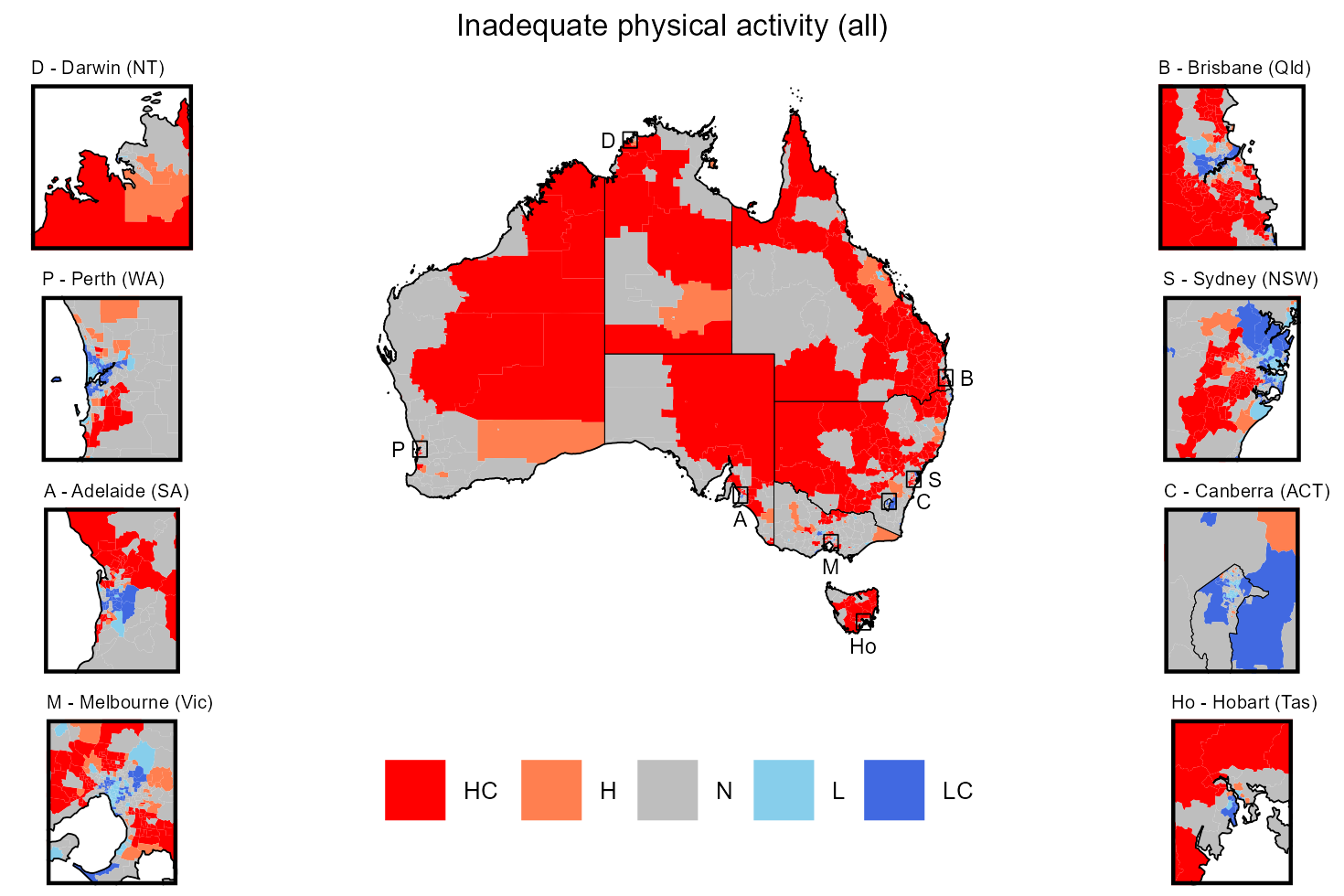}
         \caption{}
     \end{subfigure}
    \caption{\small See caption for \cref{fig:mapprevep_smoking}.}
    \label{fig:mapprevep_activityleiswkpl}
\end{figure}

\end{landscape}

\section{Abbreviations}%% if any
ABS: Australian Bureau of Statistics; ACA: Australian Cancer Atlas; ACT: Australian Capital Territory; AIHW: Australian Institute of Health and Welfare; ALC: Area linear comparison; ASGS: Australian Statistical Geography Standard; BMI: Body Mass Index; DH: Demographic-health; DOH: Department of Health; EP: Exceedance probability; HPDI: Highest posterior density interval; ICAR: Intrinsic conditional autoregressive; IID: Independent and identically distributed; IOP: Interval overlap probability; IQR: Interquartile range; IRSD: Index of Relative Socio-Economic Disadvantage; MARB: Mean absolute relative bias; MCMC: Markov Chain monte carlo; MIOP: Mean interval overlap probability; MrP: Multilevel regression and poststratification; MRRMSE: Mean relative root mean squared error; NHMRC: National Health and Medical Research Council; NHS: National Health Survey; NORF: Non-outcome risk factor; NSW: New South Wales; NT: Northern Territory; PC: Principal component; PHN: Primary health network; QLD: Queensland; RR: Rate ratio; SA: South Australia; SA2: Statistical area level 2; SA3: Statistical area level 3; SA4: Statistical area level 4; SEIFA: Socio-Economic Indexes for Areas; SES: Socioeconomic status; SHAA: Social Health Atlases of Australia; SR: Smoothing ratio; TAS: Tasmania; TSLN: Two-stage logistic-normal; VIC: Victoria; WA: Western Australia

\newpage
\bibliographystyle{unsrtnat}
\bibliography{ref}  %%% Uncomment this line and comment out the ``thebibliography'' section below to use the external .bib file (using bibtex) .